\documentclass[twocolumn]{aastex61}

\usepackage[utf8]{inputenc}
\DeclareUnicodeCharacter{2212}{-}
\usepackage{graphicx}
\graphicspath{{./figures/}{./}}
\usepackage{amsmath}
\usepackage{enumitem}
\setenumerate{itemsep=0mm}

\usepackage{soul} 
\usepackage{amsmath}
\usepackage{amssymb}
\usepackage{xspace}
\usepackage{xifthen}

\definecolor{forestgreen}{HTML}{228B22}
\definecolor{urlblue}{HTML}{000000}

\newcommand{\ie}{i.e.\xspace}
\newcommand{\eg}{e.g.\xspace}

\newcommand{\CHECK}[1]{{#1}}

\newcommand{\NKNOWN}{\CHECK{four}\xspace}
\newcommand{\NSTREAMS}{\CHECK{eleven}\xspace}

\newcommand{\NGLOB}{\CHECK{four}\xspace}

\mathchardef\mhyphen="2D

\newcommand{\roughly}{\ensuremath{ {\sim}\,} }
\newcommand{\gtr}{\ensuremath{ {>}\,} }

\newlength{\dhatheight}

\newcommand{\code}[1]{\texttt{#1}\xspace}

\newcommand{\unit}[1]{\ensuremath{\mathrm{\,#1}}\xspace}
\newcommand{\yr}{\unit{yr}}
\newcommand{\Gyr}{\unit{Gyr}}
\newcommand{\Myr}{\unit{Myr}}

\newcommand{\degree}{\ensuremath{{}^{\circ}}\xspace}
\newcommand{\degrees}{\degree}
\newcommand{\mas}{\unit{mas}}
\newcommand{\amin}{\unit{arcmin}}
\newcommand{\asec}{\unit{arcsec}}

\newcommand{\km}{\unit{km}}
\newcommand{\kms}{\km \second^{-1}}
\newcommand{\pc}{\unit{pc}}
\newcommand{\kpc}{\unit{kpc}}
\newcommand{\second}{\unit{s}}

\newcommand{\Msolar}{\unit{M_\odot}}
\newcommand{\Msun}{\unit{M_\odot}}

\newcommand{\Lsun}{\unit{L_\odot}}

\newcommand{\Dsun}{\unit{D_\odot}}
\newcommand{\Rgc}{\ensuremath{R_{GC}}\xspace}

\newcommand{\magn}{\unit{mag}}
\newcommand{\mmag}{\unit{mmag}}

\newcommand{\secref}[1]{Section~\ref{sec:#1}}

\newcommand{\tabref}[1]{Table~\ref{tab:#1}}

\newcommand{\figref}[1]{Figure~\ref{fig:#1}}

\newcommand{\eqnref}[1]{Equation~\eqref{eqn:#1}}

\newcommand{\bandvar}[2][]{%
  \ifthenelse{\isempty{#1}}{\var{#2}}{\var{#2\_#1}}%
}

\newcommand{\LCDM}{\ensuremath{\rm \Lambda CDM}\xspace}
\newcommand{\modulus}{\ensuremath{m - M}\xspace}
\newcommand{\mM}{\modulus}
\newcommand{\ra}{{\ensuremath{\alpha_{2000}}}\xspace}
\newcommand{\dec}{{\ensuremath{\delta_{2000}}}\xspace}
\newcommand{\age}{{\ensuremath{\tau}}\xspace}

\newcommand{\feh}{{\ensuremath{\rm [Fe/H]}}\xspace}

\newcommand{\ngmix}{\code{ngmix}}

\newcommand{\SExtractor}{\code{SExtractor}}
\newcommand{\sextractor}{\SExtractor}

\newcommand{\HEALPix}{\code{HEALPix}}
\newcommand{\healpix}{\HEALPix}

\newcommand{\mangle}{\code{mangle}}
\newcommand{\emcee}{\code{emcee}}
\newcommand{\ugali}{\code{ugali}}
\newcommand{\var}[1]{\ensuremath{\texttt{\MakeUppercase{#1}}}\xspace}
\newcommand{\nside}{\code{nside}}

\providecommand\physrep{\ref@jnl{Phys.~Rep.}}%
\providecommand\apjs{\ref@jnl{ApJS}}%
\providecommand{\jcap}{\ref@jnl{JCAP}}%

\begin{document}

\title{Stellar Streams Discovered in the Dark Energy Survey}


\author{N.~Shipp}
\affiliation{Kavli Institute for Cosmological Physics, University of Chicago, Chicago, IL 60637, USA}
\affiliation{Department of Astronomy and Astrophysics, The University of Chicago, Chicago IL 60637, USA}
\altaffiliation{LSSTC Data Science Fellow}
\author{A.~Drlica-Wagner}
\affiliation{Fermi National Accelerator Laboratory, P. O. Box 500, Batavia, IL 60510, USA}
\author{E.~Balbinot}
\affiliation{Department of Physics, University of Surrey, Guildford GU2 7XH, UK}
\author{P.~Ferguson}
\affiliation{George P. and Cynthia Woods Mitchell Institute for Fundamental Physics and Astronomy, and Department of Physics and Astronomy, Texas A\&M University, College Station, TX 77843,  USA}
\author{D.~Erkal}
\affiliation{Department of Physics, University of Surrey, Guildford GU2 7XH, UK}
\affiliation{Institute of Astronomy, University of Cambridge, Madingley Road, Cambridge CB3 0HA, UK}
\author{T.~S.~Li}
\affiliation{Fermi National Accelerator Laboratory, P. O. Box 500, Batavia, IL 60510, USA}
\author{K.~Bechtol}
\affiliation{LSST, 933 North Cherry Avenue, Tucson, AZ 85721, USA}
\author{V.~Belokurov}
\affiliation{Institute of Astronomy, University of Cambridge, Madingley Road, Cambridge CB3 0HA, UK}
\author{B.~Buncher}
\affiliation{Fermi National Accelerator Laboratory, P. O. Box 500, Batavia, IL 60510, USA}
\author{D.~Carollo}
\affiliation{ARC Centre of Excellence for All-sky Astrophysics (CAASTRO)}
\affiliation{INAF - Osservatorio Astrofisico di Torino, Pino Torinese, Italy}
\author{M.~Carrasco~Kind}
\affiliation{Department of Astronomy, University of Illinois at Urbana-Champaign, 1002 W. Green Street, Urbana, IL 61801, USA}
\affiliation{National Center for Supercomputing Applications, 1205 West Clark St., Urbana, IL 61801, USA}
\author{K.~Kuehn}
\affiliation{Australian Astronomical Observatory, North Ryde, NSW 2113, Australia}
\author{J.~L.~Marshall}
\affiliation{George P. and Cynthia Woods Mitchell Institute for Fundamental Physics and Astronomy, and Department of Physics and Astronomy, Texas A\&M University, College Station, TX 77843,  USA}
\author{A. B.~Pace}
\affiliation{George P. and Cynthia Woods Mitchell Institute for Fundamental Physics and Astronomy, and Department of Physics and Astronomy, Texas A\&M University, College Station, TX 77843,  USA}
\author{E.~S.~Rykoff}
\affiliation{Kavli Institute for Particle Astrophysics \& Cosmology, P. O. Box 2450, Stanford University, Stanford, CA 94305, USA}
\affiliation{SLAC National Accelerator Laboratory, Menlo Park, CA 94025, USA}
\author{I.~Sevilla-Noarbe}
\affiliation{Centro de Investigaciones Energ\'eticas, Medioambientales y Tecnol\'ogicas (CIEMAT), Madrid, Spain}
\author{E.~Sheldon}
\affiliation{Brookhaven National Laboratory, Bldg 510, Upton, NY 11973, USA}
\author{L.~Strigari}
\affiliation{George P. and Cynthia Woods Mitchell Institute for Fundamental Physics and Astronomy, and Department of Physics and Astronomy, Texas A\&M University, College Station, TX 77843,  USA}
\author{A.~K.~Vivas}
\affiliation{Cerro Tololo Inter-American Observatory, National Optical Astronomy Observatory, Casilla 603, La Serena, Chile}
\author{B.~Yanny}
\affiliation{Fermi National Accelerator Laboratory, P. O. Box 500, Batavia, IL 60510, USA}
\author{A.~Zenteno}
\affiliation{Cerro Tololo Inter-American Observatory, National Optical Astronomy Observatory, Casilla 603, La Serena, Chile}
\author{T.~M.~C.~Abbott}
\affiliation{Cerro Tololo Inter-American Observatory, National Optical Astronomy Observatory, Casilla 603, La Serena, Chile}
\author{F.~B.~Abdalla}
\affiliation{Department of Physics \& Astronomy, University College London, Gower Street, London, WC1E 6BT, UK}
\affiliation{Department of Physics and Electronics, Rhodes University, PO Box 94, Grahamstown, 6140, South Africa}
\author{S.~Allam}
\affiliation{Fermi National Accelerator Laboratory, P. O. Box 500, Batavia, IL 60510, USA}
\author{S.~Avila}
\affiliation{Institute of Cosmology \& Gravitation, University of Portsmouth, Portsmouth, PO1 3FX, UK}
\affiliation{Instituto de Fisica Teorica UAM/CSIC, Universidad Autonoma de Madrid, 28049 Madrid, Spain}
\author{E.~Bertin}
\affiliation{CNRS, UMR 7095, Institut d'Astrophysique de Paris, F-75014, Paris, France}
\affiliation{Sorbonne Universit\'es, UPMC Univ Paris 06, UMR 7095, Institut d'Astrophysique de Paris, F-75014, Paris, France}
\author{D.~Brooks}
\affiliation{Department of Physics \& Astronomy, University College London, Gower Street, London, WC1E 6BT, UK}
\author{D.~L.~Burke}
\affiliation{Kavli Institute for Particle Astrophysics \& Cosmology, P. O. Box 2450, Stanford University, Stanford, CA 94305, USA}
\affiliation{SLAC National Accelerator Laboratory, Menlo Park, CA 94025, USA}
\author{J.~Carretero}
\affiliation{Institut de F\'{\i}sica d'Altes Energies (IFAE), The Barcelona Institute of Science and Technology, Campus UAB, 08193 Bellaterra (Barcelona) Spain}
\author{F.~J.~Castander}
\affiliation{Institut d'Estudis Espacials de Catalunya (IEEC), 08193 Barcelona, Spain}
\affiliation{Institute of Space Sciences (ICE, CSIC),  Campus UAB, Carrer de Can Magrans, s/n,  08193 Barcelona, Spain}
\author{R.~Cawthon}
\affiliation{Kavli Institute for Cosmological Physics, University of Chicago, Chicago, IL 60637, USA}
\author{M.~Crocce}
\affiliation{Institut d'Estudis Espacials de Catalunya (IEEC), 08193 Barcelona, Spain}
\affiliation{Institute of Space Sciences (ICE, CSIC),  Campus UAB, Carrer de Can Magrans, s/n,  08193 Barcelona, Spain}
\author{C.~E.~Cunha}
\affiliation{Kavli Institute for Particle Astrophysics \& Cosmology, P. O. Box 2450, Stanford University, Stanford, CA 94305, USA}
\author{C.~B.~D'Andrea}
\affiliation{Department of Physics and Astronomy, University of Pennsylvania, Philadelphia, PA 19104, USA}
\author{L.~N.~da Costa}
\affiliation{Laborat\'orio Interinstitucional de e-Astronomia - LIneA, Rua Gal. Jos\'e Cristino 77, Rio de Janeiro, RJ - 20921-400, Brazil}
\affiliation{Observat\'orio Nacional, Rua Gal. Jos\'e Cristino 77, Rio de Janeiro, RJ - 20921-400, Brazil}
\author{C.~Davis}
\affiliation{Kavli Institute for Particle Astrophysics \& Cosmology, P. O. Box 2450, Stanford University, Stanford, CA 94305, USA}
\author{J.~De~Vicente}
\affiliation{Centro de Investigaciones Energ\'eticas, Medioambientales y Tecnol\'ogicas (CIEMAT), Madrid, Spain}
\author{S.~Desai}
\affiliation{Department of Physics, IIT Hyderabad, Kandi, Telangana 502285, India}
\author{H.~T.~Diehl}
\affiliation{Fermi National Accelerator Laboratory, P. O. Box 500, Batavia, IL 60510, USA}
\author{P.~Doel}
\affiliation{Department of Physics \& Astronomy, University College London, Gower Street, London, WC1E 6BT, UK}
\author{A.~E.~Evrard}
\affiliation{Department of Astronomy, University of Michigan, Ann Arbor, MI 48109, USA}
\affiliation{Department of Physics, University of Michigan, Ann Arbor, MI 48109, USA}
\author{B.~Flaugher}
\affiliation{Fermi National Accelerator Laboratory, P. O. Box 500, Batavia, IL 60510, USA}
\author{P.~Fosalba}
\affiliation{Institut d'Estudis Espacials de Catalunya (IEEC), 08193 Barcelona, Spain}
\affiliation{Institute of Space Sciences (ICE, CSIC),  Campus UAB, Carrer de Can Magrans, s/n,  08193 Barcelona, Spain}
\author{J.~Frieman}
\affiliation{Fermi National Accelerator Laboratory, P. O. Box 500, Batavia, IL 60510, USA}
\affiliation{Kavli Institute for Cosmological Physics, University of Chicago, Chicago, IL 60637, USA}
\author{J.~Garc\'ia-Bellido}
\affiliation{Instituto de Fisica Teorica UAM/CSIC, Universidad Autonoma de Madrid, 28049 Madrid, Spain}
\author{E.~Gaztanaga}
\affiliation{Institut d'Estudis Espacials de Catalunya (IEEC), 08193 Barcelona, Spain}
\affiliation{Institute of Space Sciences (ICE, CSIC),  Campus UAB, Carrer de Can Magrans, s/n,  08193 Barcelona, Spain}
\author{D.~W.~Gerdes}
\affiliation{Department of Astronomy, University of Michigan, Ann Arbor, MI 48109, USA}
\affiliation{Department of Physics, University of Michigan, Ann Arbor, MI 48109, USA}
\author{D.~Gruen}
\affiliation{Kavli Institute for Particle Astrophysics \& Cosmology, P. O. Box 2450, Stanford University, Stanford, CA 94305, USA}
\affiliation{SLAC National Accelerator Laboratory, Menlo Park, CA 94025, USA}
\author{R.~A.~Gruendl}
\affiliation{Department of Astronomy, University of Illinois at Urbana-Champaign, 1002 W. Green Street, Urbana, IL 61801, USA}
\affiliation{National Center for Supercomputing Applications, 1205 West Clark St., Urbana, IL 61801, USA}
\author{J.~Gschwend}
\affiliation{Laborat\'orio Interinstitucional de e-Astronomia - LIneA, Rua Gal. Jos\'e Cristino 77, Rio de Janeiro, RJ - 20921-400, Brazil}
\affiliation{Observat\'orio Nacional, Rua Gal. Jos\'e Cristino 77, Rio de Janeiro, RJ - 20921-400, Brazil}
\author{G.~Gutierrez}
\affiliation{Fermi National Accelerator Laboratory, P. O. Box 500, Batavia, IL 60510, USA}
\author{B.~Hoyle}
\affiliation{Max Planck Institute for Extraterrestrial Physics, Giessenbachstrasse, 85748 Garching, Germany}
\affiliation{Universit\"ats-Sternwarte, Fakult\"at f\"ur Physik, Ludwig-Maximilians Universit\"at M\"unchen, Scheinerstr. 1, 81679 M\"unchen, Germany}
\author{D.~J.~James}
\affiliation{Harvard-Smithsonian Center for Astrophysics, MS-42, 60 Garden Street, Cambridge, MA 02138, USA}
\author{M.~D.~Johnson}
\affiliation{National Center for Supercomputing Applications, 1205 West Clark St., Urbana, IL 61801, USA}
\author{E.~Krause}
\affiliation{Department of Astronomy/Steward Observatory, 933 North Cherry Avenue, Tucson, AZ 85721-0065, USA}
\affiliation{Jet Propulsion Laboratory, California Institute of Technology, 4800 Oak Grove Dr., Pasadena, CA 91109, USA}
\author{N.~Kuropatkin}
\affiliation{Fermi National Accelerator Laboratory, P. O. Box 500, Batavia, IL 60510, USA}
\author{O.~Lahav}
\affiliation{Department of Physics \& Astronomy, University College London, Gower Street, London, WC1E 6BT, UK}
\author{H.~Lin}
\affiliation{Fermi National Accelerator Laboratory, P. O. Box 500, Batavia, IL 60510, USA}
\author{M.~A.~G.~Maia}
\affiliation{Laborat\'orio Interinstitucional de e-Astronomia - LIneA, Rua Gal. Jos\'e Cristino 77, Rio de Janeiro, RJ - 20921-400, Brazil}
\affiliation{Observat\'orio Nacional, Rua Gal. Jos\'e Cristino 77, Rio de Janeiro, RJ - 20921-400, Brazil}
\author{M.~March}
\affiliation{Department of Physics and Astronomy, University of Pennsylvania, Philadelphia, PA 19104, USA}
\author{P.~Martini}
\affiliation{Center for Cosmology and Astro-Particle Physics, The Ohio State University, Columbus, OH 43210, USA}
\affiliation{Department of Astronomy, The Ohio State University, Columbus, OH 43210, USA}
\author{F.~Menanteau}
\affiliation{Department of Astronomy, University of Illinois at Urbana-Champaign, 1002 W. Green Street, Urbana, IL 61801, USA}
\affiliation{National Center for Supercomputing Applications, 1205 West Clark St., Urbana, IL 61801, USA}
\author{C.~J.~Miller}
\affiliation{Department of Astronomy, University of Michigan, Ann Arbor, MI 48109, USA}
\affiliation{Department of Physics, University of Michigan, Ann Arbor, MI 48109, USA}
\author{R.~Miquel}
\affiliation{Instituci\'o Catalana de Recerca i Estudis Avan\c{c}ats, E-08010 Barcelona, Spain}
\affiliation{Institut de F\'{\i}sica d'Altes Energies (IFAE), The Barcelona Institute of Science and Technology, Campus UAB, 08193 Bellaterra (Barcelona) Spain}
\author{R.~C.~Nichol}
\affiliation{Institute of Cosmology \& Gravitation, University of Portsmouth, Portsmouth, PO1 3FX, UK}
\author{A.~A.~Plazas}
\affiliation{Jet Propulsion Laboratory, California Institute of Technology, 4800 Oak Grove Dr., Pasadena, CA 91109, USA}
\author{A.~K.~Romer}
\affiliation{Department of Physics and Astronomy, Pevensey Building, University of Sussex, Brighton, BN1 9QH, UK}
\author{M.~Sako}
\affiliation{Department of Physics and Astronomy, University of Pennsylvania, Philadelphia, PA 19104, USA}
\author{E.~Sanchez}
\affiliation{Centro de Investigaciones Energ\'eticas, Medioambientales y Tecnol\'ogicas (CIEMAT), Madrid, Spain}
\author{V.~Scarpine}
\affiliation{Fermi National Accelerator Laboratory, P. O. Box 500, Batavia, IL 60510, USA}
\author{R.~Schindler}
\affiliation{SLAC National Accelerator Laboratory, Menlo Park, CA 94025, USA}
\author{M.~Schubnell}
\affiliation{Department of Physics, University of Michigan, Ann Arbor, MI 48109, USA}
\author{M.~Smith}
\affiliation{School of Physics and Astronomy, University of Southampton,  Southampton, SO17 1BJ, UK}
\author{R.~C.~Smith}
\affiliation{Cerro Tololo Inter-American Observatory, National Optical Astronomy Observatory, Casilla 603, La Serena, Chile}
\author{F.~Sobreira}
\affiliation{Instituto de F\'isica Gleb Wataghin, Universidade Estadual de Campinas, 13083-859, Campinas, SP, Brazil}
\affiliation{Laborat\'orio Interinstitucional de e-Astronomia - LIneA, Rua Gal. Jos\'e Cristino 77, Rio de Janeiro, RJ - 20921-400, Brazil}
\author{E.~Suchyta}
\affiliation{Computer Science and Mathematics Division, Oak Ridge National Laboratory, Oak Ridge, TN 37831}
\author{M.~E.~C.~Swanson}
\affiliation{National Center for Supercomputing Applications, 1205 West Clark St., Urbana, IL 61801, USA}
\author{G.~Tarle}
\affiliation{Department of Physics, University of Michigan, Ann Arbor, MI 48109, USA}
\author{D.~Thomas}
\affiliation{Institute of Cosmology \& Gravitation, University of Portsmouth, Portsmouth, PO1 3FX, UK}
\author{D.~L.~Tucker}
\affiliation{Fermi National Accelerator Laboratory, P. O. Box 500, Batavia, IL 60510, USA}
\author{A.~R.~Walker}
\affiliation{Cerro Tololo Inter-American Observatory, National Optical Astronomy Observatory, Casilla 603, La Serena, Chile}
\author{R.~H.~Wechsler}
\affiliation{Department of Physics, Stanford University, 382 Via Pueblo Mall, Stanford, CA 94305, USA}
\affiliation{Kavli Institute for Particle Astrophysics \& Cosmology, P. O. Box 2450, Stanford University, Stanford, CA 94305, USA}
\affiliation{SLAC National Accelerator Laboratory, Menlo Park, CA 94025, USA}

\collaboration{(DES Collaboration)}

\email{norashipp@uchicago.edu, kadrlica@fnal.gov}

\begin{abstract}

We perform a search for stellar streams around the Milky Way using the first three years of multi-band optical imaging data from the Dark Energy Survey (DES). 
We use DES data covering $\roughly 5000 \deg^2$ to a depth of $g > 23.5$ with a relative photometric calibration uncertainty of $< 1\%$. 
This data set yields unprecedented sensitivity to the stellar density field in the southern celestial hemisphere, enabling the detection of faint stellar streams to a heliocentric distance of $\roughly 50 \kpc$.
We search for stellar streams using a matched-filter in color-magnitude space derived from a synthetic isochrone of an old, metal-poor stellar population.
Our detection technique recovers \NKNOWN previously known thin stellar streams: Phoenix, ATLAS, Tucana III, and a possible extension of Molonglo. 
In addition, we report the discovery of \NSTREAMS new stellar streams.
In general, the new streams detected by DES are fainter, more distant, and lower surface brightness than streams detected by similar techniques in previous photometric surveys.
As a by-product of our stellar stream search, we find evidence for extra-tidal stellar structure associated with \NGLOB globular clusters: NGC 288, NGC 1261, NGC 1851, and NGC 1904.
The ever-growing sample of stellar streams will provide insight into the formation of the Galactic stellar halo, the Milky Way gravitational potential, as well as the large- and small-scale distribution of dark matter around the Milky Way.
\end{abstract}


\section{Introduction}
\label{sec:intro}

Stellar streams produced by the tidal disruption of globular clusters and dwarf galaxies are a prevalent feature of the Milky Way environs \citep[see][for a recent review]{Newberg:2016}.
Observations of stellar streams can provide important constraints on the formation of the Milky Way stellar halo \citep[e.g.,][]{Johnston:1998,Bullock:2005,Bell:2008}, the shape of the Galactic gravitational field \citep[e.g.,][]{Johnston:2005,Koposov:2010,Law:2010,Bovy:2014,Bonaca:2014,Gibbons:2014,Price-Whelan:2014,Sanders:2014,Bowden:2015,Kupper:2015,Erkal:2016,Bovy:2016}, and the abundance of low-mass dark matter substructure \citep[e.g.,][]{Ibata:2002,Johnston:2002,Carlberg:2009,Yoon:2011,Carlberg:2012,Ngan:2014,Erkal:2015,Carlberg:2016a,Sanderson:2016,Sanders:2016,Bovy:2017,Erkal:2017,Sandford:2017}.
In addition, stellar streams are a direct snapshot of hierarchical structure formation \citep{Peebles:1965,Press:1974,Blumenthal:1984} and support the standard \LCDM cosmological model \citep{Diemand:2008,Springel:2008}.

Wide-area, multi-band digital sky surveys have been essential for finding and characterizing resolved stellar populations in the Galactic halo. 
The Sloan Digital Sky Survey \citep[SDSS;][]{York:2000} revolutionized our understanding of the Milky Way stellar halo, both through improved sensitivity to diffuse components \citep[e.g.,][]{Carollo:2007,Carollo:2010,Dejong:2010,Deason:2011,An:2013,Kafle:2013,Hattori:2013,An:2015,Das:2016} and by vastly increasing the number of known satellite galaxies 
\citep[e.g.,][]{Willman:2005a,Willman:2005b,Zucker:2006,2006ApJ...643L.103Z,2006ApJ...647L.111B,Belokurov:2007}, stellar clouds \citep[e.g.,][]{Newberg:2002,Yanny:2003,Rocha-Pinto:2004}, and stellar streams \citep[e.g.,][]{Odenkirchen:2001,Newberg:2002,Belokurov:2006,Grillmair:2006}. 
Early techniques for detecting stellar streams used simple color and magnitude cuts to select blue main sequence turn-off (MSTO) stars \citep[e.g.,][]{Grillmair:1995,Belokurov:2006}.
More recently, matched-filter techniques have been used to maximize the contrast between distant, metal poor stellar populations and foreground field stars to push the detection limit to lower surface brightnesses \citep[e.g.][]{Rockosi:2002}.
The matched-filter technique has been applied broadly to other digital sky surveys including Pan-STARRs \citep{Bernard:2014,Bernard:2016,Grillmair:2017} and ATLAS \citep{Koposov:2014}.

The Dark Energy Survey \citep[DES;][]{DES:2005,DES:2016} is a deep, wide-area survey with the primary goal of constraining dark energy and the nature of cosmic acceleration \citep[e.g.,][]{DES:2017}. 
While DES was designed to probe the evolution of the universe out to $z \sim 1.2$, it has already had a major impact on ``near-field'' cosmology and Galactic archaeology. 
Specifically, DES has nearly doubled the number of known ultra-faint dwarf galaxies \citep{Bechtol:2015,Koposov:2015,Kim:2015,Drlica-Wagner:2015}, increased the number of known faint outer-halo star clusters \citep{Pieres:2016,Luque:2016,Luque:2017,Luque:2017b}, and identified several diffuse stellar overdensities \citep{Li:2016,Pieres:2017}.
In addition, early data from DES have been used to detect a cold stellar stream in the constellation of Phoenix \citep{Balbinot:2016} and a tidal stream associated with the ultra-faint satellite, Tucana III \citep{Drlica-Wagner:2015}.
Here we extend the search for stellar streams with DES using a deeper, more uniform, and better calibrated data set.

We perform a search for stellar streams using the first three years of DES data.
We search for stellar streams possessing old, metal-poor stellar populations at heliocentric distances between $6\kpc \lesssim \Dsun \lesssim 63\kpc$ ($14 < \mM < 19$).
We recover known streams within the DES footprint, including Sagittarius \citep{Newberg:2002}, ATLAS \citep{Koposov:2014}, Phoenix \citep{Balbinot:2016}, and Tucana III \citep{Drlica-Wagner:2015}. 
In addition, we detect a possible faint extension of the Molonglo stream \citep{Grillmair:2017} in the DES data.
Our search results in the discovery of \NSTREAMS new high-significance curvilinear stellar stream candidates.
These new stream candidates range in distance from \CHECK{$\roughly 13\kpc$} to \CHECK{$\roughly 50\kpc$}.
These streams are low surface brightness, \CHECK{$\mu \gtrsim 32 \magn/\asec^2$}, and push the boundary of detectability using the current generation of wide-area photometric surveys.
As a natural by-product of a global search for extended stellar structures, we find evidence for extended extra-tidal features around the Milky Way globular clusters NGC 288, NGC 1261, NGC 1851, and NGC 1904.

This paper is organized as follows.
In \secref{data} we discuss the DES data and the construction of the stellar sample used in this work. 
In \secref{analysis} we present our matched-filter search algorithm and maximum-likelihood techniques that we implement for characterizing stellar streams.
In \secref{streams} we discuss the properties of previously known and newly detected stellar streams.
We place our results in the larger context of the Milky Way in \secref{discussion} and conclude in \secref{conclusions}.

\section{Data Set}
\label{sec:data}

\begin{figure}
\includegraphics[width=\columnwidth]{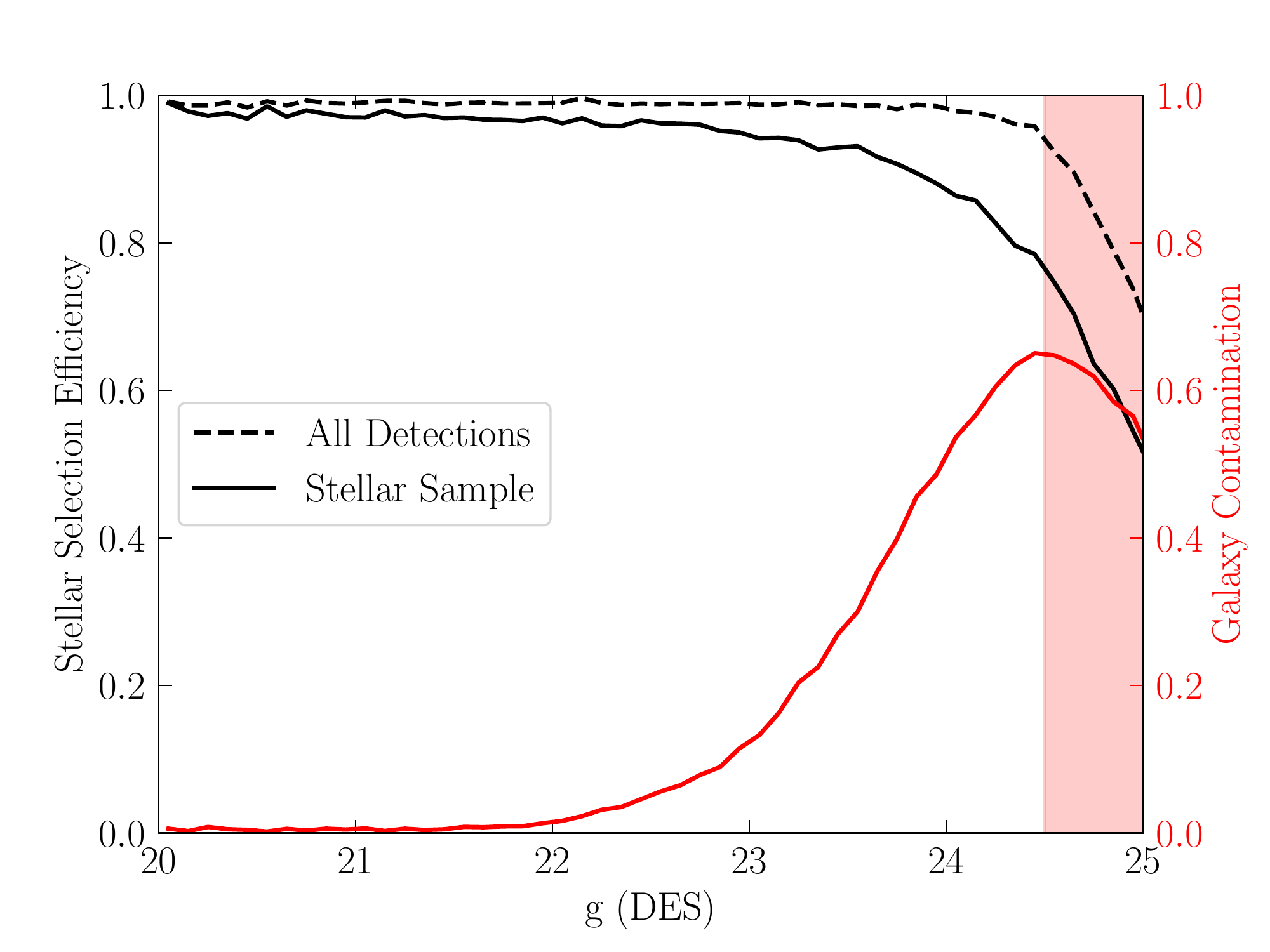}
\caption{Stellar selection efficiency and galaxy contamination for the DES Y3 data evaluated from a comparison to HSC DR1 using $\sim18$~deg$^2$ of overlap in the SDSS Stripe 82 region.
The DES efficiency is evaluated with respect to a stellar sample from HSC.
The red shaded band indicates the faint magnitude range where the HSC data are affected by star-galaxy confusion and may be less reliable as a test sample.
Our stellar classification primarily uses the \ngmix multi-epoch fitting size parameter and error.
\label{fig:class}
}
\end{figure}

DES is a deep, wide-area imaging survey using the Dark Energy Camera \citep[DECam;][]{Flaugher:2015} mounted on the 4-m Blanco Telescope at Cerro Tololo Inter-American Observatory in Chile. 
DES surveys $\roughly 5000 \deg^2$ of the southern Galactic cap in five visible/near-infrared filters, $grizY$.
Here, we use wide-field imaging data from an internal data release of the first three years of DES operations (DES Y3A2).\footnote{DES Y3A2 serves as the basis for the first public DES data release \citep[DES DR1;][]{DES:2018}. However, the internal Y3A2 data release contains improved multi-epoch photometric and morphological measurements as well as other auxiliary data products.}
DES Y3A2 is the first data set to cover the full DES wide-area footprint, and has a median coverage of 5--6 exposures per filter \citep{Diehl:2016}. 
The contiguous, uniform, wide-area imaging of DES allows for the first deep, systematic search for faint features in the Milky Way stellar halo in this region of the southern sky.

The DES Y3A2 images were processed by the DES data management pipeline \citep{Morganson:2018}.
Photometric calibration was performed via the Forward Global Calibration Method \citep[FGCM;][]{Burke:2017}, which utilized ancillary information about atmospheric and environmental conditions at the time of each exposure.
The FGCM photometric calibration is found to have a relative photometric uniformity of $\roughly 7 \mmag$ \citep{Burke:2017} and an absolute calibration accuracy of $\roughly 3 \mmag$ \citep{DES:2018}.
Individual exposures were remapped to a consistent pixel grid and coadded to increase imaging depth \citep{Morganson:2018}.  
Object detection was performed on a combination of the $r + i + z$ coadded images using the \SExtractor toolkit \citep{Bertin:1996,Bertin:2002} with an object detection threshold corresponding to ${\rm S/N} \sim 10$ \citep{Morganson:2018}.

While the coadded images increase our sensitivity to faint sources, depth variations and PSF discontinuities in the coadds can make it difficult to perform precise photometric and morphological measurements \citep[\eg,][]{Drlica-Wagner:2017}.
To circumvent this issue, we used \ngmix\footnote{\url{https://github.com/esheldon/ngmix}} \citep{Sheldon:2014,Jarvis:2015,Drlica-Wagner:2017} to fit the flux and morphology of each source over all individual single-epoch images simultaneously.
When fitting each source we masked nearby neighbors using the \code{uberseg} map, which was derived from the \sextractor coadd segmentation maps \citep[Section 5.2 of][]{Jarvis:2015}.\footnote{Masking nearby sources yields slightly less accurate photometry than the iterative multi-object fitting (MOF) described by \citet{Drlica-Wagner:2017}; however, masking is less computationally intensive and has a lower failure rate than MOF.}
Throughout this paper, quoted magnitudes and errors were derived from fitting a PSF model to each source using \ngmix (\ie, \var{PSF\_MAG} and \var{PSF\_MAG\_ERR}).

To select a high-quality stellar sample, we expanded on the star-galaxy classification procedure outlined in Appendix A of \citet{Rozo:2016}.
We used \ngmix to fit a composite galaxy model (bulge plus disk) to each source  in all bands simultaneously \citep{Drlica-Wagner:2017}. 
We then used the best-fit size, \var{CM\_T}, and associated uncertainty, \var{CM\_T\_ERR}, from this galaxy-model fit to distinguish point-like objects from those that are spatially extended.
Specifically, we defined an extended classification variable, \var{NGMIX\_CLASS}, based on the sum of three selection criteria,

\footnotesize 
\begin{align}\begin{split}
\var{NGMIX\_CLASS} = &~ ((\var{CM\_T} + 5\, \var{CM\_T\_ERR}) > 0.1) \\
 + &~ ((\var{CM\_T} + \var{CM\_T\_ERR}) > 0.05) \\
 + &~ ((\var{CM\_T} - \var{CM\_T\_ERR}) > 0.02). \\
\end{split} \end{align} 
\normalsize

\noindent 
The \ngmix composite galaxy-model fit fails for a small number of bright stars. 
To recover those objects, we defined a second selection based on the weighted-average \sextractor quantity \var{WAVG\_SPREAD\_MODEL} measured in the DES $i$-band \citep{Morganson:2018},

\footnotesize\begin{align}\begin{split}
\var{WA}&\var{VG\_CLASS} = \\
~&((\var{wavg\_spread\_model\_i} + 3\, \var{wavg\_spreaderr\_model\_i}) > 0.005) \\
+&((\var{wavg\_spread\_model\_i} + \var{wavg\_spreaderr\_model\_i}) > 0.003) \\
+&((\var{wavg\_spread\_model\_i} - \var{wavg\_spreaderr\_model\_i}) > 0.001).
 \end{split}\end{align}
\normalsize

\noindent
Both \var{NGMIX\_CLASS} and \var{WAVG\_CLASS} can have values of 0, 1, 2, or 3, with 0 being most star-like and 3 being most galaxy-like.
Our final stellar sample used $\var{NGMIX\_CLASS} \leq 1$ when the composite-model fit succeeded and $\var{WAVG\_CLASS} \leq 1$ otherwise:

\footnotesize \begin{align}
\var{STARS} =
\begin{cases}
\var{NGMIX\_CLASS} \leq 1, & \text{if \ngmix fit succeeds} \\
\var{WAVG\_CLASS} \leq 1, & \text{otherwise}
\label{eqn:stars}
\end{cases}
\end{align}\normalsize

Our stellar selection was designed to yield a compromise between completeness and purity in the resulting stellar sample. 
In \figref{class} we compare our stellar classification with deeper imaging data from Hyper Suprime-Cam DR1 \citep{Aihara:2017}.
We find that our selection is $> 90\%$ complete for $g = 23.5$ with a galaxy contamination rising from $\lesssim 5\%$ at $g \leq 22.5$ to $\roughly 30\%$ by $g \sim 23.5$.
Throughout the paper we refer to the objects passing the selection in \eqnref{stars} as stars.

We constrained our stellar sample to the range $16 < g < 23.5$. 
The bright-end limit was imposed to avoid saturation effects from bright stars, while at the faint end we seek  to avoid spurious density fluctuations resulting from inhomogeneous survey depth and galaxy contamination. 
Since we are primarily interested in Main Sequence (MS) and Red Giant Branch (RGB) stars associated with old, metal-poor stellar populations, we constrain our sample to the color range, $0.0 < g-r < 1.0$.

In contrast to previous DES photometric calibration techniques \citep{Drlica-Wagner:2015,Drlica-Wagner:2017}, no stellar locus regression adjustment was applied to the DES Y3A2 zeropoints derived by the FGCM.
Instead, we followed the procedure described in \citet{DES:2018} to account for interstellar dust extinction.
We started with $E(B-V)$ values from the reddening map of \citet[SFD;][]{Schlegel:1998}.
We computed fiducial interstellar extinction coefficients, $R_b$, for each band so that the corrections to the top-of-the-atmosphere calibrated source magnitudes are $A_b = E(B-V) \times R_b$.
Fiducial coefficients are derived using the \citet{Fitzpatrick:1999} reddening law with $R_V = 3.1$ and the \citet{Schlafly:2011} adjusted reddening normalization parameter, $N = 0.78$. 
We integrated over the DES standard bandpasses considering a fixed source spectrum that is constant in spectral flux density per unit wavelength.
The resulting multiplicative coefficients for the $g$ and $r$ band are $R_g = 3.185$ and $R_r = 2.140$.\footnote{An update to the DECam standard bandpasses changed these coefficients by $<1$ \mmag for the DR1 release \citep{DES:2018}.}
Throughout this paper, all magnitudes refer to extinction corrected PSF magnitudes derived by \ngmix.

We build a high-resolution map of the DES survey coverage to account for missing survey coverage at the boundary of the footprint and gaps associated with saturated stars, bleed trails, and other instrumental signatures.
We follow the procedure described in \citet{Drlica-Wagner:2017} to transform a vectorized representation of the survey coverage calculated by \mangle \citep{Hamilton:2004, Swanson:2008} into a \healpix \citep{Gorski:2005} coverage fraction map.
In each $\nside = 4096$ ($\roughly 0.74 \amin^2$) \healpix pixel, we oversample the \mangle map by a factor of 64 to quantify the simultaneous coverage in the $griz$ bands.
The \healpix nested pixelization scheme makes it trivial to degrade the resolution of this coverage fraction map by summing the coverage fraction of all high resolution pixels nested within a lower resolution pixel.
We restrict our stream search to regions where the $griz$ detection fraction is greater than 50\%, resulting in a total solid angle of $4946 \deg^2$.
Throughout this paper, all coordinates refer to J2000 epoch.

\section{Analysis}
\label{sec:analysis}

\begin{figure*}[th!]
\centering
\includegraphics[width=\textwidth]{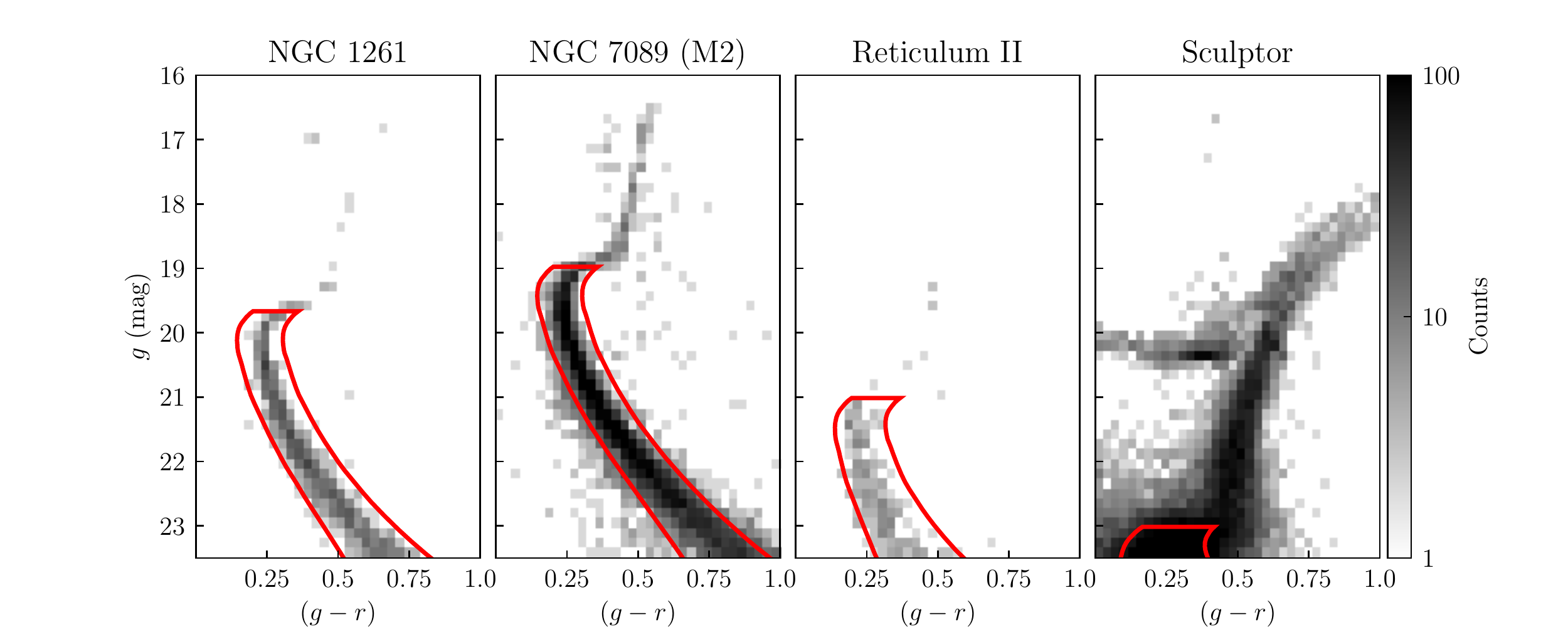}
\caption{
Binned color-magnitude diagram of DES Y3A2 stellar sources selected around the globular clusters NGC 1261 and NGC 7089 (M2), the ultra-faint dwarf galaxy Reticulum II, and the classical dwarf galaxy Sculptor. 
For the two globular clusters, data are selected in an annulus with $0\fdg07 < r < 0\fdg12$. 
Stellar sources are selected within $r < 0\fdg15$ of Reticulum II and $r < 0\fdg20$ of Sculptor.
The matched-filter selection region for our stellar stream search is shown with the red outlines.
}
\label{fig:isosys}
\end{figure*}

\subsection{Matched-Filter Selection}
\label{sec:filter}

We searched for stellar streams using a matched-filter algorithm in color-magnitude space \citep{Rockosi:2002,Grillmair:2006,Bonaca:2012,Jethwa:2017}.
Our matched filter is based on the synthetic isochrone of an old, $\age = 13 \Gyr$, metal-poor, \CHECK{$Z = 0.0002$ ($\feh = -1.9$)}, stellar population as constructed by \citet{Dotter:2008} and implemented in \ugali \citep{Bechtol:2015,Drlica-Wagner:2015}.
We selected stars within a range of colors around the isochrone according to the criteria
\begin{align}\begin{split}
(g - r)_{\rm iso} + &E \times err(g_{\rm iso} + \mu + \Delta \mu / 2) - C_1 \\
&< (g-r) <\\
(g - r)_{\rm iso} + &E \times err(g_{\rm iso} + \mu - \Delta \mu / 2) + C_2.
\label{eqn:select}
\end{split}\end{align}
\noindent where $(g-r)_{\rm iso}$ represents the predicted color from the synthetic isochrone at a distance modulus of $\mu = \mM$. 
We parametrize the magnitude-dependent spread in color due to measurement uncertainties as
\begin{align}\begin{split}
err(g) =  0.001 + e^{(g - 27.09) / 1.09}
\end{split}\end{align}
\noindent where the normalization coefficients were derived from fitting the median photometric error as a function of magnitude in the $g$ band (\var{PSF\_MAG\_G} vs. \var{PSF\_MAG\_ERR\_G}). 
We parameterize the selection region around the isochrone with a symmetric magnitude broadening, $\Delta \mu$, an asymmetric color broadening, $C_{1,2}$, and a multiplicative factor for broadening based on photometric uncertainty, $E$.
We set the values $\Delta \mu = 0.5$, $C_{1,2} = (0.05,0.10)$ and $E = 2$ by comparing to the color-magnitude diagrams (CMDs) of old, metal-poor globular clusters and dwarf galaxies (\figref{isosys}).

In \figref{isosys} we show our matched-filter selection over-plotted on binned CMDs from the globular clusters NGC 1260 and NGC 7089 (M2), the ultra-faint dwarf galaxy Reticulum II, and the classical dwarf galaxy Sculptor. 
Our selection retains $\gtrsim 90\%$ of stellar sources fainter than the MSTO in an annulus of $0\fdg07 < r < 0\fdg12$ around NGC 1260 and NGC 7089. 
This is a conservative estimate of the true efficiency of our selection, since it does not account for contamination from Milky Way foreground stars within this annulus.

\begin{figure*}[ht!]
  \centering
  \includegraphics[width=0.8\textwidth]{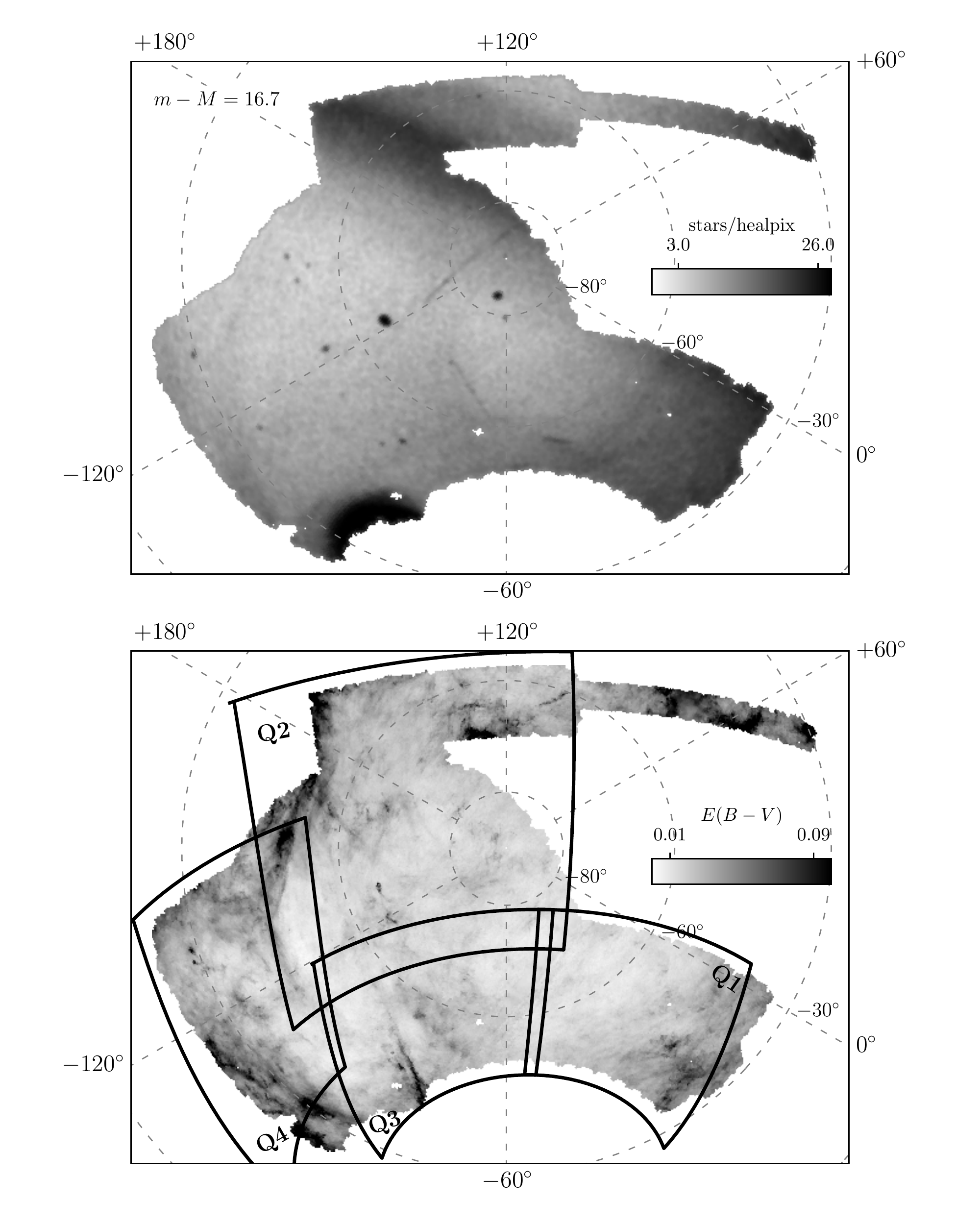}
  \caption{(Top) Density of stars passing the matched-filter isochrone selection at a distance modulus of $\mM = 16.7$. Stars are pixelized into equal-area \healpix pixels with area of $\roughly 0.013 \deg^2$ ($\nside=512$). Contributions from the LMC (lower left), Sagittarius stream (top center), and Galactic thick disk (lower right) can be clearly seen.
    (Bottom) Interstellar extinction, $E(B-V)$, estimated by \citet{Schlegel:1998}.
    Outlines of the four DES ``quadrants'' defined in \secref{streams} are overplotted.
    Both panels are plotted in Galactic coordinates using a polar Lambert equal-area projection.
    An animated version of this figure can be found online \href{http://home.fnal.gov/~kadrlica/movies/density_dust_v17p2.gif}{at this url}.
}
  \label{fig:isosel}
\end{figure*}

In our initial search for stellar streams in Y3A2, we applied our matched-filter isochrone selection over a grid of distance moduli from $14 < m-M < 19$ spaced at intervals of $\Delta(\mM) = 0.3$. 
This spacing between distance modulus steps was chosen so that sequential isochrone selections overlap by $\gtrsim 75\%$ to insure that streams at intermediate distance moduli were detectable \citep{Grillmair:2017}.
For each distance modulus, we binned stars passing our matched-filter selection  into equal-area \healpix pixels with area of $\roughly 0.013 \deg^2$ ($\code{nside} = 512$).
We divided the number of stars selected in each \healpix pixel by a map of the coverage fraction at equivalent resolution to produce a coverage-corrected map of stellar density.
We smoothed the density maps with a 2-dimensional Gaussian symmetric beam with $\sigma = 0\fdg3$.\footnote{\url{http://healpy.readthedocs.io/en/latest/generated/healpy.sphtfunc.smoothing.html}}
We later repeated this procedure at spacings of $\Delta(\mM) = 0.1$ and confirmed that all stream candidates were still detected.
We show the smoothed density map after the isochrone selection in \figref{isosel}.

We note that for our generic search we did not use the Milky Way foreground population to weight stars in our matched filter isochrone selection. 
This was done to ensure that our search was agnostic to changes in the foreground/background stellar population across the footprint (i.e., close to the Galactic disk, LMC, or Sagittarius stream).
As a validation, we repeated our search following the procedure of \citet{Rockosi:2002} to build a matched-filter isochrone from the CMDs of the globular cluster NGC 7089 and the average Milky Way foreground population.
We apply an additive factor of $\Delta(\mM)$ to shift the CMD of NGC 7089 in distance modulus and thus select for stellar populations over a range of distances.
We find that this procedure yields very similar results to our primary un-weighted synthetic isochrone selection technique, and all stream candidates reported here were detected by both analyses.

\begin{figure*}[ht!]
  \centering
  \includegraphics[width=0.8\textwidth]{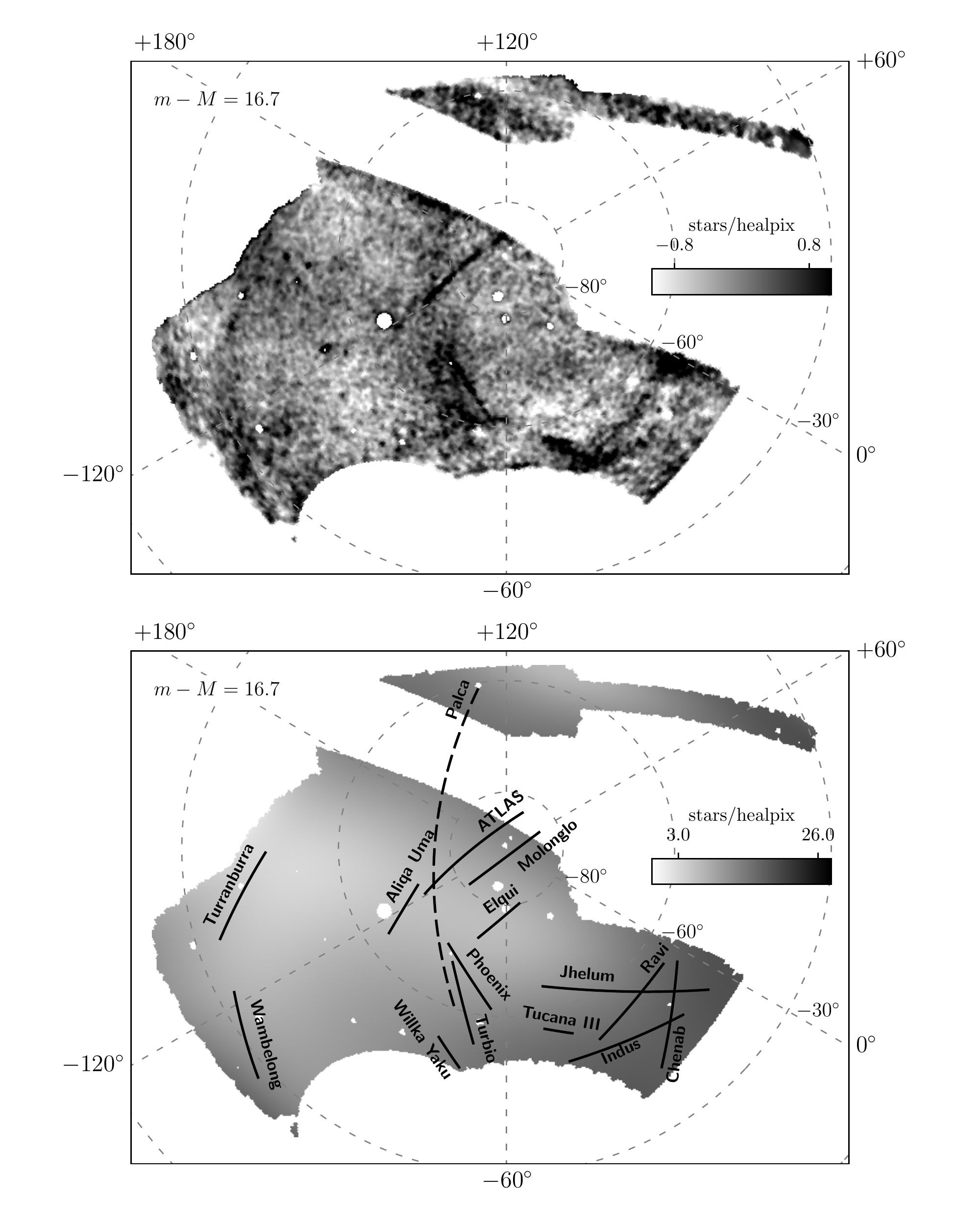}
  \caption{(Top) Residual density of stars passing the matched-filter isochrone selection at a distance modulus of $\mM = 16.7$. 
Regions around the LMC (lower left), Sagittarius stream (top center), and the Galactic disk (lower right) have been masked to improve the quality of the polynomial background fit. 
In addition, small regions ($r < 1\degree$) around bright Milky Way globular clusters, satellite galaxies, and Local Group galaxies have been masked.
    (Bottom) Smooth background fit to the stellar density using a 2-dimensional, 5th-order polynomial (gray-scale) .
Stellar streams labeled and overplotted.
Palca and ATLAS are traced by a second order polynomials (Sections \ref{sec:atlas} and \ref{sec:palca}, respectively), while other streams are traced by great circle arcs.
    Both panels are plotted in Galactic coordinates using a polar Lambert equal-area projection.
    An animated version of this figure can be found online \href{http://home.fnal.gov/~kadrlica/movies/residual_bkg_v17p2.gif}{at this url}.
}
  \label{fig:isosel2}
\end{figure*}

\subsection{Stream Detection}
\label{sec:detect}

The Sagittarius ($l,b \sim 160\degrees, -60\degrees$), ATLAS ($l,b \sim -130\degrees,-80\degrees$), Phoenix ($l,b \sim -70\degrees,-65\degrees$), and Tucana III ($l,b \sim -45\degrees,-55\degrees$) streams are clearly visible in smoothed density maps (\figref{isosel}).
To further increase our sensitivity to faint stellar streams, we created a smooth model for the stellar foreground and mis-classified background galaxies.
We mask the dense stellar regions around the Sagittarius stream, the LMC, and the Galactic plane (\figref{isosel2}) and fit a 2-dimensional, 5th-order polynomial to the distribution of smoothed stellar counts.
We subtracted this model from the stellar density to create smooth maps of the residual stellar density (\figref{isosel2}).
Visual inspection of the foreground-subtracted residual density maps served as the primary technique for identifying new streams.

To facilitate the detection of faint streams, we repeated the procedure described above to generate residual stellar density maps for a range of isochrone selections with distance moduli between $14 \leq \mM \leq 19$ in steps of $\Delta (\mM) = 0.3$. 
We assembled these sequences of residual density maps into animations of the isochrone-selected stellar density as a function of heliocentric distance.
We required that stellar stream candidates appear in the residual density maps for at least two sequential distance moduli.
The animations associated with \figref{isosel} and \figref{isosel2} contain a wealth of information about stellar structure in the Milky Way halo. 
In this paper we specifically focus on the the most prominent stellar streams, leaving other studies of the outer halo to future work.

We perform our visual search by assembling residual density maps of the full DES footprint and in smaller subregions, which we call ``quadrants''.
Candidates identified in the residual density maps were further examined in color-magnitude space for evidence of a stellar population distinct from the Milky Way foreground (\secref{isochrone}).
Only candidates that showed a distinct stellar locus consistent with an old, metal poor isochrone were included in our list of stellar stream candidates.
This search resulted in the detection of the \NKNOWN previously known narrow stellar streams (ATLAS, Phoenix, Tucana III, and Molonglo) and \NSTREAMS new stream candidates.
We report the measured and derived parameters of our stream candidates in Tables \ref{tab:params} and \ref{tab:deriv}.
and discuss each candidate in more detail in \secref{streams}. 

\newcommand{\paramcaption}{Measured parameters of stellar streams \label{tab:params}}
\newcommand{\paramcomments}{
Measured characteristics of stellar streams detected in DES Y3A2 data. 
The first section reports DES measurements of previously known streams, while the second section reports narrow streams discovered by DES. 
The broad stream/stellar overdensity, Palca, is given its own section.
Endpoints and great circle poles are reported in Celestial coordinates and are derived from the residual stellar density analysis described in \secref{detect}. 
Stream lengths are calculated as the angular separation between endpoints, while stream widths, $w$, come from the standard deviation of a Gaussian fit to the transverse stream profile.
Distance moduli, ages, and metallicities were calculated by fitting a \citet{Dotter:2008} isochrone to the Hess diagrams described in \secref{isochrone}. 
The number of stars is calculated by summing MS stars in the background-subtracted Hess diagram.
The significance is calculated as the signal-to-noise ratio between the on-stream and off-stream regions.
}
\begin{deluxetable*}{l cccccccccc }
\tablecolumns{13}
\tablewidth{0pt}
\tabletypesize{\scriptsize}
\tablecaption{ \paramcaption }
\tablehead{
Name & Celestial End Points & Celestial Pole & Length & Width & $\mM$ & Age & Z & $N_*$ & Significance \\
 & (deg) & (deg) & (deg) & (deg) &  & (Gyr) &  &  & 
}
\startdata
Tucana III      & $( -6.3,-59.7),(  3.2,-59.4)$ & $(354.2, 30.3)$ &  4.8 & 0.18 & 17.0 & 13.5 & 0.0001 &   700 & 17.0 \\
ATLAS           & $(  9.3,-20.9),( 30.7,-33.2)$ & $( 74.3, 47.9)$ & 22.6 & 0.24 & 16.8 & 11.0 & 0.0007 &  1600 & 13.9 \\
Molonglo        & $(  6.4,-24.4),( 13.6,-28.1)$ & $( 62.3, 51.0)$ &  7.4 & 0.32 & 16.8 & 13.5 & 0.0010 &   700 &  5.2 \\
Phoenix         & $( 20.1,-55.3),( 27.9,-42.7)$ & $(311.2, 14.0)$ & 13.6 & 0.16 & 16.4 & 13.0 & 0.0004 &   700 & 11.1 \\
[+0.5em]\tableline\\[-1em]
Indus           & $(-36.3,-50.7),( -8.0,-64.8)$ & $( 24.8, 21.6)$ & 20.3 & 0.83 & 16.1 & 13.0 & 0.0007 &  9700 & 21.4 \\
Jhelum          & $(-38.8,-45.1),(  4.7,-51.7)$ & $(359.1, 38.2)$ & 29.2 & 1.16 & 15.6 & 12.0 & 0.0009 &  4600 & 18.6 \\
Ravi            & $(-25.2,-44.1),(-16.0,-59.7)$ & $( 53.2, 11.7)$ & 16.6 & 0.72 & 16.8 & 13.5 & 0.0003 &  2300 & 10.3 \\
Chenab          & $(-40.7,-59.9),(-28.3,-43.0)$ & $(255.5, 14.4)$ & 18.5 & 0.71 & 18.0 & 13.0 & 0.0004 &  1700 & 15.1 \\
Elqui           & $( 10.7,-36.9),( 20.6,-42.4)$ & $( 64.0, 38.5)$ &  9.4 & 0.54 & 18.5 & 12.0 & 0.0004 &   700 & 18.4 \\
Aliqa Uma       & $( 31.7,-31.5),( 40.6,-38.3)$ & $( 94.5, 36.7)$ & 10.0 & 0.26 & 17.3 & 13.0 & 0.0004 &   400 &  9.1 \\
Turbio          & $( 28.0,-61.0),( 27.9,-46.0)$ & $(297.8, -0.1)$ & 15.0 & 0.25 & 16.1 & 13.0 & 0.0004 &  1000 &  7.9 \\
Willka Yaku     & $( 36.1,-64.6),( 38.4,-58.3)$ & $(316.0,  4.7)$ &  6.4 & 0.21 & 17.7 & 11.0 & 0.0006 &   600 &  7.1 \\
Turranburra     & $( 59.3,-18.0),( 75.2,-26.4)$ & $(123.5, 53.3)$ & 16.9 & 0.60 & 17.2 & 13.5 & 0.0003 &  1300 & 14.4 \\
Wambelong       & $( 90.5,-45.6),( 79.3,-34.3)$ & $(328.7,-27.3)$ & 14.2 & 0.40 & 15.9 & 11.0 & 0.0001 &   500 &  5.9 \\
[+0.5em]\tableline\\[-1em]
Palca           & $( 30.3,-53.7),( 16.2,  2.4)$ & $(286.6, -9.9)$ & 57.3 & \ldots & 17.8 & 13.0 & 0.0004 & \ldots & \ldots \\
\enddata
{\footnotesize \tablecomments{ \paramcomments }}
\end{deluxetable*}

\newcommand{\derivcaption}{Derived parameters of stellar streams \label{tab:deriv}}
\newcommand{\derivcomments}{
Derived physical parameters of stellar streams detected in DES Y3A2 data. 
Heliocentric distances are calculated by fitting a \citet{Dotter:2008} isochrone to the Hess diagrams described in \secref{isochrone} and have an error of $\roughly 20\%$. 
Other physical parameters are derived from the measured parameters in \tabref{params} assuming this distance. 
Stream lengths are calculated from the angular distance between the stream endpoints, while the widths represent the standard deviation of the best-fit Gaussian.
Stellar masses, absolute magnitudes, and surface brightnesses are derived from the richness of the best-fit isochrone model assuming a \citet{Chabrier:2001} IMF.
The absolute magnitude is derived from the DES $g$ and $r$ bands following the prescription of \citet{Bechtol:2015}.
The surface brightness is derived assuming that 68\% of the luminosity is contained within the stream width of $\pm w$. 
The progenitor masses are estimated using the results of \citet{Erkal:2016}. 
}
\begin{deluxetable*}{l cccccccc }
\tablecolumns{13}
\tablewidth{0pt}
\tabletypesize{\scriptsize}
\tablecaption{ \derivcaption }
\tablehead{
Name & Distance & Length & Width & Stellar Mass & $M_{V}$ & $\mu_{V}$ & Progenitor Mass \\
 & (kpc) & (kpc) & (pc) & ($10^3 \Msun$) & (mag) & ($\magn \asec^{-2}$) & ($10^4 \Msun$)
}
\startdata
Tucana III      & 25.1 &  2.1 &   79 &  3.8 & -3.8 & 32.0 &    8 \\
ATLAS           & 22.9 &  9.0 &   96 &  7.4 & -4.5 & 33.0 &   12 \\
Molonglo        & 22.9 &  3.0 &  128 &  3.5 & -3.7 & 33.0 &   30 \\
Phoenix         & 19.1 &  4.5 &   53 &  2.8 & -3.6 & 32.6 &    3 \\
[+0.5em]\tableline\\[-1em]
Indus           & 16.6 &  5.9 &  240 & 34.0 & -6.2 & 31.9 &  650 \\
Jhelum          & 13.2 &  6.7 &  267 & 13.0 & -5.1 & 33.3 & 1300 \\
Ravi            & 22.9 &  6.6 &  288 & 10.4 & -5.0 & 33.4 &  520 \\
Chenab          & 39.8 & 12.9 &  493 & 18.3 & -5.7 & 34.1 &  780 \\
Elqui           & 50.1 &  8.2 &  472 & 10.4 & -4.9 & 34.3 &  320 \\
Aliqa Uma       & 28.8 &  5.0 &  131 &  2.3 & -3.4 & 33.8 &   18 \\
Turbio          & 16.6 &  4.3 &   72 &  3.5 & -3.9 & 32.6 &   10 \\
Willka Yaku     & 34.7 &  3.9 &  127 &  4.6 & -4.1 & 32.9 &   14 \\
Turranburra     & 27.5 &  8.1 &  288 &  7.6 & -4.7 & 34.0 &  180 \\
Wambelong       & 15.1 &  3.7 &  106 &  1.6 & -3.0 & 33.7 &   26 \\
[+0.5em]\tableline\\[-1em]
Palca           & 36.3 & 36.3 & \ldots & \ldots & \ldots & \ldots & \ldots \\
\enddata
{\footnotesize \tablecomments{ \derivcomments }}
\end{deluxetable*}

\newcommand{\galactocaption}{Galactocentric parameters of stellar streams \label{tab:galacto}}
\newcommand{\galactocomments}{
Galactocentric parameters of stellar streams detected in DES Y3A2 data. 
Transformation into Galactocentric coordinates performed assuming the Earth resides at $(8.3\kpc,0,0)$.
The Galactocentric azimuthal and polar angles, ($\phi,\psi$), are defined as in Fig.\ 1 of \citet{Erkal:2016}.
Both endpoints are assumed to be at the heliocentric distance quoted in \tabref{params}.
}
\begin{deluxetable*}{l ccccc }
\tablecolumns{13}
\tablewidth{0pt}
\tabletypesize{\footnotesize}
\tablecaption{ \galactocaption }
\tablehead{
Name & $x_1,y_1,z_1$ & $x_2,y_2,z_2$ & $\overline{\rm R_{GC}}$ & $(\phi,\psi)$ \\
 & (kpc) & (kpc) & (kpc) & (deg)
}
\startdata
Tucana III      & $(  2.7, -9.4,-20.5)$ & $(  0.8,-10.2,-21.1)$ & $  23$ & $(285.9, 64.6)$ \\
ATLAS           & $( -8.5,  2.8,-22.7)$ & $(-11.7, -5.6,-22.0)$ & $  25$ & $(157.3, 68.5)$ \\
Molonglo        & $( -6.9,  2.1,-22.8)$ & $( -8.3, -0.5,-22.9)$ & $  24$ & $(152.3, 72.7)$ \\
Phoenix         & $( -4.5, -8.3,-16.7)$ & $( -8.4, -6.4,-17.9)$ & $  20$ & $(235.4, 60.6)$ \\
[+0.5em]\tableline\\[-1em]
Indus           & $(  2.9, -2.6,-12.0)$ & $( -0.6, -7.3,-12.7)$ & $  14$ & $(321.3, 72.0)$ \\
Jhelum          & $(  0.9, -0.8, -9.4)$ & $( -4.3, -4.0,-11.9)$ & $  11$ & $(298.6, 83.1)$ \\
Ravi            & $(  4.7, -1.4,-18.8)$ & $(  3.4, -7.9,-18.0)$ & $  20$ & $(350.6, 75.4)$ \\
Chenab          & $( 18.9,-12.5,-26.3)$ & $( 15.5, -1.5,-31.9)$ & $  35$ & $( 28.7, 68.0)$ \\
Elqui           & $( -2.4, -6.3,-49.4)$ & $( -5.2,-13.9,-48.0)$ & $  50$ & $(159.6, 89.9)$ \\
Aliqa Uma       & $(-13.4, -6.7,-27.6)$ & $(-13.5,-11.4,-26.0)$ & $  31$ & $(171.2, 66.0)$ \\
Turbio          & $( -5.0, -9.0,-13.5)$ & $( -7.8, -6.3,-15.4)$ & $  18$ & $(208.7, 57.3)$ \\
Willka Yaku     & $( -1.5,-21.4,-26.4)$ & $( -4.7,-20.0,-28.1)$ & $  34$ & $(229.7, 56.9)$ \\
Turranburra     & $(-24.6, -9.8,-19.9)$ & $(-23.6,-16.6,-15.8)$ & $  33$ & $(155.3, 47.5)$ \\
Wambelong       & $(-12.3,-12.8, -6.9)$ & $(-15.0,-10.7, -8.3)$ & $  20$ & $(183.1, 28.1)$ \\
[+0.5em]\tableline\\[-1em]
Palca           & $( -4.6,-17.5,-31.6)$ & $(-19.8, 13.8,-31.5)$ & $  38$ & $(205.8, 69.5)$ \\
\enddata
{\footnotesize \tablecomments{ \galactocomments }}
\end{deluxetable*}

\subsection{Stream Characterization}
\label{sec:spatial}
\label{sec:isochrone}

After candidates are identified in the residual density images, we perform an iterative process to fit the characteristics of each stream:

\setlist{nolistsep}
\begin{enumerate}[noitemsep]
\item Define stream endpoints from the residual density maps.
\item Fit the transverse stream width to define an ``on-stream'' region.
\item Fit isochrone parameters to the CMD of on-stream minus off-stream stars.
\item Re-fit the transverse stream width using the best-fit isochrone selection.
\end{enumerate}
\noindent We describe each of these steps in more detail below.

We defined the endpoints of each stream from the residual stellar density map.
The residual stellar density map has a pixel scale of $\roughly 0\fdg1$ ($\code{nside} = 512$) and is smoothed by a Gaussian kernel with standard deviation $\sigma = 0\fdg3$.
For narrow, prominent streams the endpoints can be measured with an accuracy of better than $0\fdg1$; however, for fainter and/or more diffuse structures, measuring endpoints becomes more uncertain.
The stream length reported in \tabref{params} was calculated as the angular separation between the endpoints assuming that the streams follow a great circle on the sky.

For each stream, we calculate the pole of a great circle passing through the endpoints and rotate into a coordinate system where the fundamental plane is aligned with the long axis of the stream.
Following the convention of \citet{Majewski:2003}, we define (heliocentric) longitudinal and latitudinal coordinates ($\Lambda,B$) for the rotated coordinate system associated with each stream. 

As an initial estimate for the width of each stream, we rotate the \healpix pixels of our raw and residual density maps into the frame of each stream.
We then sum the content of \healpix pixels along the transverse stream dimension to provide the transverse stream profile.
We fit the transverse stream profile with a linear foreground component and a Gaussian stream model with free normalization and standard deviation.
The standard deviation of the best-fit Gaussian was taken as an estimate of the stream width, and was used to define signal and background regions for the color-magnitude analysis in the following section.
We repeated this procedure after selecting stars consistent with the best-fit isochrone to derive the final stream width.

Stream candidates were examined in color-magnitude space to confirm the presence of a distinct stellar population matched to an old, metal poor isochrone.
An ``on-stream'' region was selected along the great circle connecting the stream endpoints. For most streams, this region had a width of $\pm 2w$, where $w$ is the stream width derived from the standard deviation of the best-fit Gaussian.
Two ``off-stream'' regions were selected with the same shape as the on-stream region, but offset perpendicular to the stream axis by $\pm 4w$. In some cases this on-stream and off-stream geometry was impossible due to the boundary of the survey or the presence of large resolved stellar populations (e.g., other streams, large dwarf galaxies, etc.). The precise on- and off-stream regions selected for each stream are described in \tabref{fits}.
When building on- and off-stream regions, we excise regions around known globular clusters and dwarf galaxies to avoid contaminating the CMD analysis.
We calculate the effective solid angle of each region accounting for the excised regions and incomplete survey coverage using the maps described in \secref{filter}.
We binned the stars in the on-stream region in color-magnitude space with bin size $\Delta g = 0.167$ mag and $\Delta (g-r) = 0.04$ mag.
We calculated the effective foreground contribution in each bin of the CMD using the off-stream regions and correcting the difference in effective solid angle.
Hess diagrams were smoothed by a 0.75 pixel Gaussian kernel, and the resulting smooth residual color-magnitude diagram was examined for the presence of a distinct stellar population. 

We performed a binned maximum-likelihood fit of the smoothed two-dimensional background-subtracted Hess diagrams using a synthetic isochrone from \citet{Dotter:2008} weighted by a \citet{Chabrier:2001} initial mass function (IMF).
We built a binned Poisson likelihood function for the observed number of stars in each CMD bin given the number of stars predicted by our isochrone model convolved by the empirically determined photometric measurement uncertainties (\secref{data}). 
We simultaneously fit the richness, distance modulus, age, and metallicity of the isochrone model to the observed excess counts in the Hess diagram.
The richness is a normalization parameter representing the total number of stream member stars with mass ${>}0.1 \Msolar$ \citep{Bechtol:2015}.
For roughly half the stream candidates, the data were unable to reliably constrain all four parameters simultaneously, and we fixed the age ($\age = 13 \Gyr$) and metallicity ($Z = 0.0004$) while fitting richness and distance modulus.
Furthermore, we note that there is a significant degeneracy between the age, metallicity, and distance modulus. 
We estimate a systematic uncertainty on the distance modulus of $\sigma(\mM) \sim 0.4 \magn$, while spectroscopic observations are essential to break the degeneracy between age and metallicity.

Several streams, specifically those closer to the Galactic plane, suffer from over- or under-subtraction due to gradients in the surrounding stellar density.
Mis-subtraction will bias estimates of the richness and total luminosity.
To mitigate these issues, we estimate the stellar content of each stream based on the number of MS stars within a region around the best-fit isochrone.
We apply a narrow isochrone selection based on \eqnref{select} using the best-fit age, metallicity, and distance modulus (\tabref{params}) and the selection parameters $\Delta \mu = 0.5$, $C_{1,2} = (0.05, 0.05)$, and $E=1$.\footnote{We find that this selection is $67 \%$ efficient for  stars in the globular cluster NGC 7089 (M2). This efficiency is taken into account when calculating the richness and stellar mass.}
We sum the content of the background-subtracted Hess diagram within this selection region.
We then correct the number of stars for the fraction of the stream width contained in the spatial selection region, to estimate the total number of MS stream stars within the spatial and magnitude range of DES.
We record this value as $N_*$ in \tabref{params}. 
We then use the isochrone model along with a \citet{Chabrier:2001} IMF to estimate to total stellar mass, luminosity, and absolute magnitude in \tabref{deriv}. 

We re-fit the stream width after applying an isochrone selection consistent with the best-fit isochrone.
We also calculate the statistical significance of each stream from the on- and area-corrected off-stream regions, $S \equiv {\rm (on - off)/\sqrt{off}}$. For consistency, all on-stream regions had a width of $\pm w$ for this calculation.
We report the measured parameters of each stream in \tabref{params} and derived physical parameters in \tabref{deriv}. 
In addition, we use this absolute magnitude to calculate an average surface brightness, which we estimate assuming 68\% of the luminosity is contained within $\pm w$ of the stream axis.

In \tabref{deriv} we also provide estimates of the progenitor masses. We use the relation between the stream width and the progenitor mass derived in \citet{Erkal:2016}. More precisely, we use Equation~(27) from \citet{Erkal:2016b}, where the progenitor mass is given in terms of the stream width as viewed from the galactic center, and the enclosed mass of the Milky Way at the stream's location. For the mass of the Milky Way, we use the best fit model in \citet{McMillan:2017} who used a range of data to constrain the Milky Way potential. We then use \textsc{galpot} \citep{Dehnen:1998} to evaluate the circular velocity (and hence the enclosed mass), as a function of Galactocentric radius. We note that this method assumes that the streams are on a circular orbit and only works on average for streams on eccentric orbits \citep{Erkal:2016}. Furthermore, this method also assumes that the streams have not fanned out significantly due to being in a non-spherical potential \citep{Erkal:2016}. As such, this method should be seen as giving a rough estimate of the progenitor mass.

In \tabref{galacto} we present stream parameters in Galactocentric coordinates, assuming that the Sun is located $8.3\kpc$ from the Galactic center \citep{deGrijs:2016, Gillessen:2009}.
Galactocentric Cartesian coordinates are provided for endpoints of each stream assuming the heliocentric distance derived in \tabref{deriv}.
We provide longitude and co-latitude ($\phi,\psi$) for the pole of a Galactocentric orbit passing through the endpoints of each stream.\footnote{Our definition of $\phi$ and $\psi$ conforms to Figure 1 of \citet[][]{Erkal:2016}.}
Assuming a single heliocentric distance for each stream naturally introduces a gradient in the Galactocentric radius. 
We use the average Galactocentric radius when calculating Galactocentric great-circle orbits.
These Galactocentric parameters are primarily used to identify potential associations in \secref{assocs}.

\section{Stellar Stream Candidates}
\label{sec:streams}

As part of our search for stellar streams, we divided the DES footprint into four ``quadrants'' (Q1 -- Q4).
These quadrants were designed to be large enough to fully contain streams spanning $>20\degrees$, while providing a more detailed view than maps of the full footprint could offer.
These quadrants offer a useful unit to subdivide the DES stellar stream candidates and we discuss each quadrant in turn.
We choose to name our stellar stream candidates after aquatic terms used by the geographically distinct cultures of India (Q1), Chile (Q2 and Q3), and Australia (Q4).

\begin{figure}[]
  \centering
  \includegraphics[width=\columnwidth]{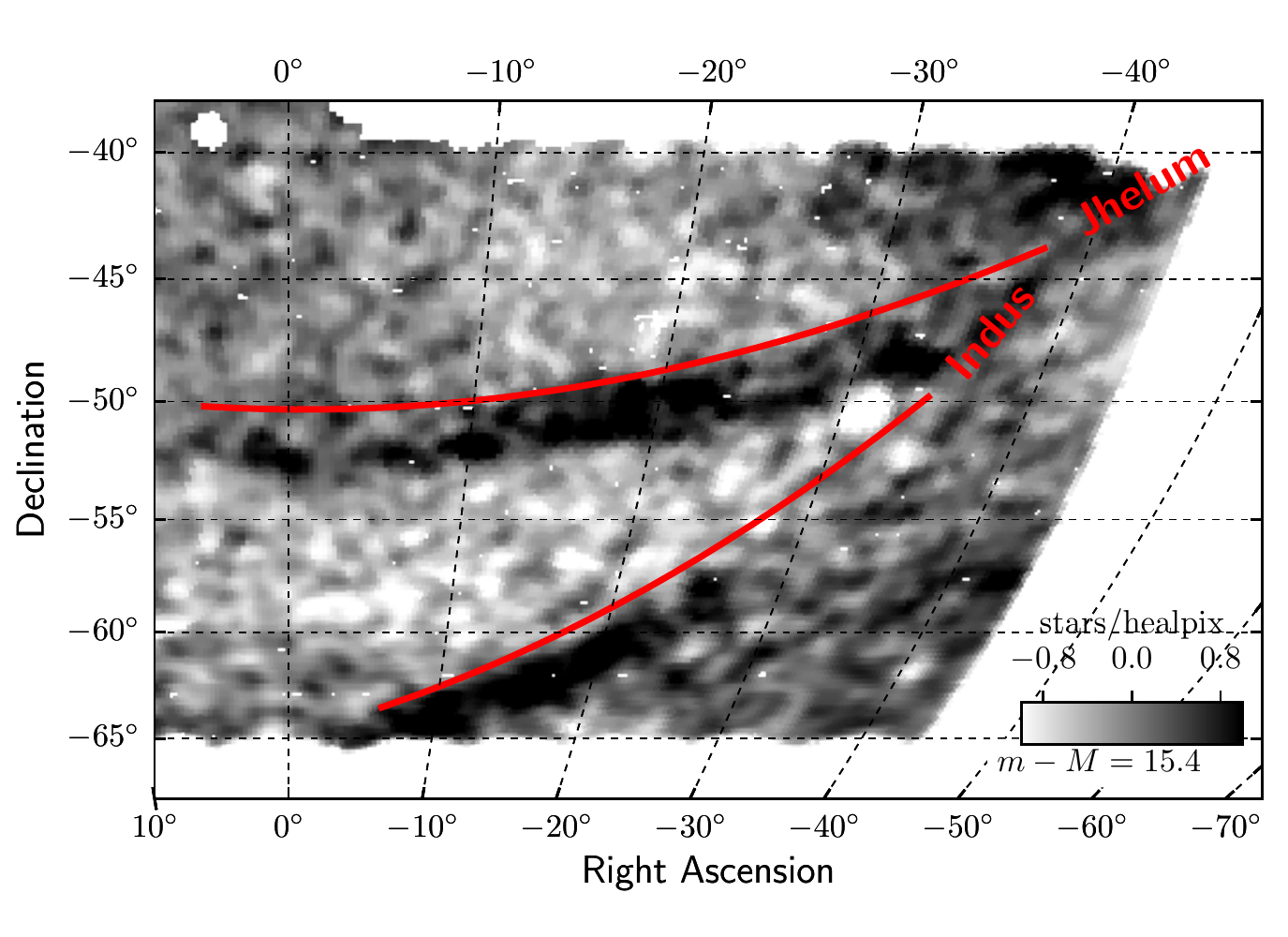}
  \includegraphics[width=\columnwidth]{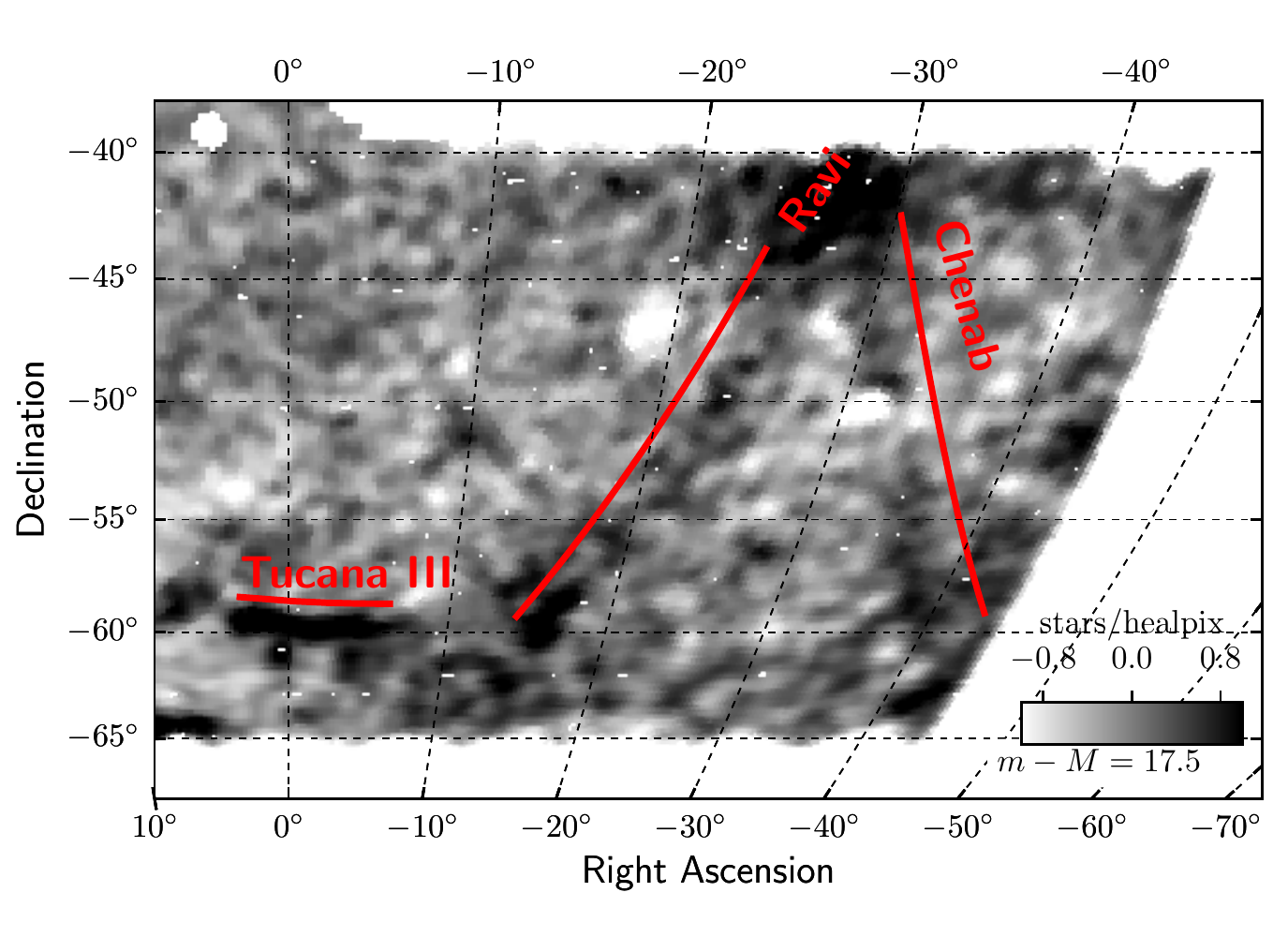}
  \includegraphics[width=\columnwidth]{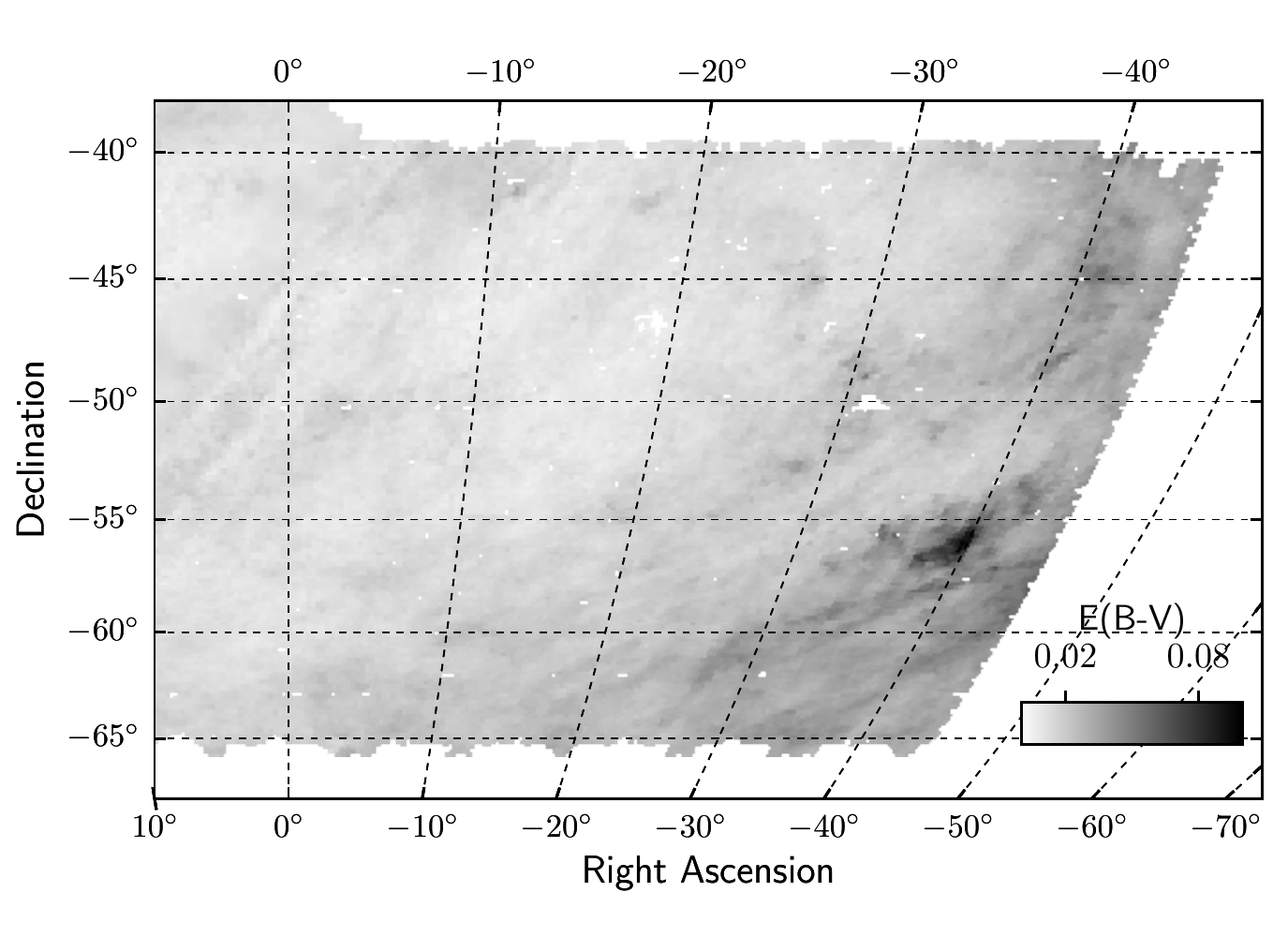}
  \caption{Residual stellar density map in Q1 after subtracting a smooth background model from the distribution of isochrone filtered stars (equal area McBryde-Thomas flat-polar quartic projection). 
    Streams are marked with great circles that are aligned with the major axis of the stream and offset perpendicularly by 1\fdg5.
(Top) Isochrone selection with $\mM = 16.4$. 
(Middle) Isochrone selection with $\mM = 17.5$. 
(Bottom) Interstellar reddening, $E(B-V)$, from \citet{Schlegel:1998}.
    An animated version of this figure can be found online \href{http://home.fnal.gov/~kadrlica/movies/residual_q1_v17p2_label.gif}{at this url}.
}
  \label{fig:q1}
\end{figure}

\subsection{First Quadrant\,\footnote{
Due to overlap with the constellation Indus, which shares a name with an Indian river, and to honor the tradition of astronomy in India, new stellar stream candidates in this quadrant are named after rivers in India (including the Indus River itself).
}}
\label{sec:q1}

The first quadrant (Q1) covers the western portion of the DES footprint from $-45\degrees \lesssim \ra \lesssim 10\degrees$ and $-65\degrees \lesssim \dec \lesssim -40\degrees$.
While the DES data extend to $\ra \gtrsim -55\degrees$, the low-order polynomial background fit has difficulty modeling the rapidly varying stellar density at these lower Galactic latitudes.
Q1 includes the Tucana III satellite and stream, the two most prominent new stellar streams, Indus and Jhelum, and two lower significance streams, Chenab and Ravi.
In addition, diffuse stellar overdensities are found at the northern ($\ra,\dec \sim -28\degrees,-42\degrees$) and southern ($\ra,\dec \sim -10\degrees, -63\degrees$) edges of Q1; however, the footprint boundary makes it difficult to perform a quantitative evaluation of these structures.

\subsubsection{Tucana III Stream}
\label{sec:tuc}

The Tucana III stellar stream is located at a distance of $\roughly 25 \kpc$, extending at least $\pm 2\degrees$ from the ultra-faint satellite Tucana III \citep{Drlica-Wagner:2015}. 
We find that Tucana III appears prominently in the Q1 residual density maps with a projected length of 5\degrees extending from \CHECK{$(-6\fdg3, -59\fdg7)$ to $(3\fdg2, -59\fdg4)$} (\figref{q1}).

\begin{figure}
\centering
\includegraphics[width=0.7\columnwidth]{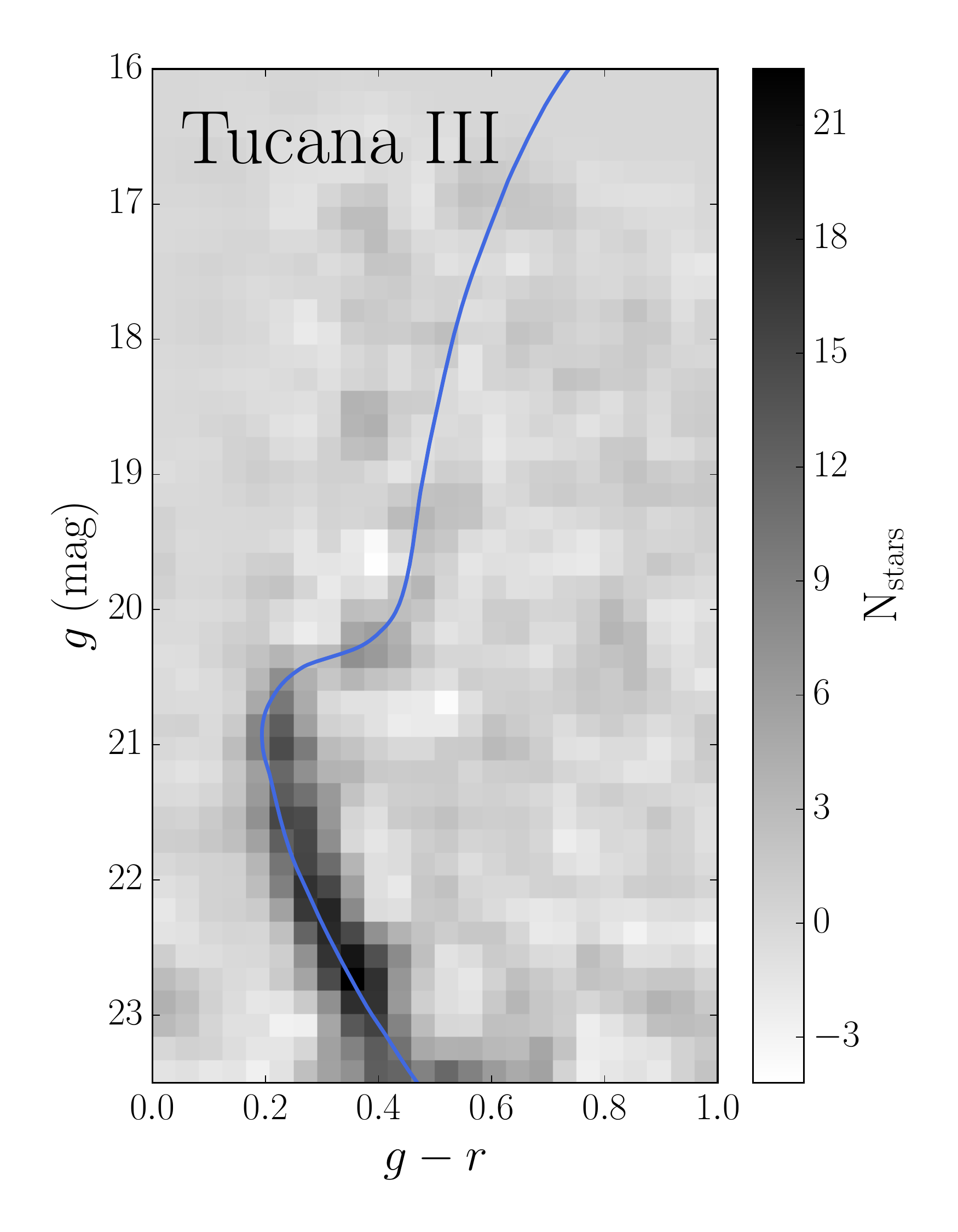}
\caption{Background-subtracted binned color-magnitude Hess diagram for stars associated with the Tucana III stellar stream. 
The background is estimated from an off-stream region parallel to the stream and is area corrected and subtracted from the onstream region.
The Hess diagram is smoothed with a 2D-Gaussian kernel with a standard deviation of 0.75 pixels.
Darkly colored pixels correspond to higher residual stellar density while lighter pixels represent underdense regions in color-magnitude space.
A synthetic isochrone from \citet{Dotter:2008} is over-plotted with best-fit parameters described in \tabref{params}.
The isochrone fitting procedures are described in \secref{isochrone} and \secref{streams}.
}
\label{fig:tucana}
\end{figure}

In \figref{tucana}, we show a Hess diagram calculated by subtracting a local background estimate derived from off-stream regions on either side of the Tucana III stream.
Despite a well-defined MS and visible RGB, our likelihood analysis has trouble simultaneously fitting the richness, distance modulus, age, and metallicity of the Tucana III stream.
Assuming a distance modulus of $\mM = 17.0$ \citep{Drlica-Wagner:2015} and a spectroscopically determined metallicity  $Z = 0.0001$ ($\feh = -2.24$) \citep{Simon:2017}, we find that the MSTO of Tucana III is well-described by an age of $\age = 13.5\Gyr$. 
This is older than the $\tau = 10.9\Gyr$ reported by \citet{Drlica-Wagner:2015}, which is due in part to a correction to the synthetic isochrones using an updated version of the DECam filter throughput \citep{Li:2018}.
In addition, the change in photometric calibration between the DES Y2Q1 and Y3A2 data sets is found to introduce a small color shift.

Spectroscopic observations have been unable to conclusively classify Tucana III as an ultra-faint galaxy or star cluster \citep{Simon:2017}. 
The unresolved velocity dispersion ($\sigma_v < 1.5 \kms$ at 95.5\% confidence) and metallicity spread ($\sigma_{\feh} < 0.19$ at 95\% confidence) are both low for an ultra-faint dwarf galaxy. 
However, the mean metallicity ($\feh = -2.42^{+0.07}_{-0.08}$) and large physical size ($\mathrm{r_{1/2} = 44 \pm 6\ pc}$) are both unusual for a globular cluster. 
In addition, \citet{Simon:2017} argue that the mass-to-light ratio of the core of Tucana III is larger than that of a globular cluster, $M/L >20 \Msun/\Lsun$, based on its proximity to the Galactic center and the non-detection of a velocity gradient out to 90\pc.
The core of Tucana III lies slightly offset from the luminosity-metallicity relationship for ultra-faint galaxies \citep{Kirby:2013}.
\citet{Simon:2017} note that if Tucana III has been stripped of $\roughly 70\%$ of its stellar mass, then it would lie directly on the metallicity-luminosity relation of ultra-faint dwarfs.
We find that the total stellar mass of the Tucana III stream is \CHECK{$2.6\times 10^3 \Msun$}, which is \CHECK{$3.25$} times the stellar mass of the Tucana III core \citep{Drlica-Wagner:2015}.
This corresponds to a mass loss of \CHECK{$69\%$}, \CHECK{which moves the Tucana III progenitor system onto the luminosity-metallicity relationship for dwarf galaxies.}

We search for indications of a distance gradient following a similar procedure to that applied to the Sagittarius stream by \citet{Koposov:2012}. 
We transform to a coordinate system oriented along the stream axis and divide the stellar counts into 8 longitudinal bins.
Within each longitudinal bin, we examine the mean magnitude of MSTO stars satisfying the criteria $0.20 < (g-r) < 0.24$.
We find that the mean magnitude of the MSTO changes from $g \sim 16.75$ at the western end of the stream to $g \sim 17.19$ on the eastern end. 
Fitting a linear gradient model to these data yields a distance gradient of $0.16 \pm 0.06 \magn \deg^{-1}$. 
This measurement implies that the Tucana III stream spans \CHECK{$\roughly 4\kpc$} in distance with a total physical extent of \CHECK{$\roughly 4.5 \kpc$} and that it is on a radial orbit.

\subsubsection{Indus Stream}

\begin{figure*}[]
  \centering
  \includegraphics[width=0.24\textwidth]{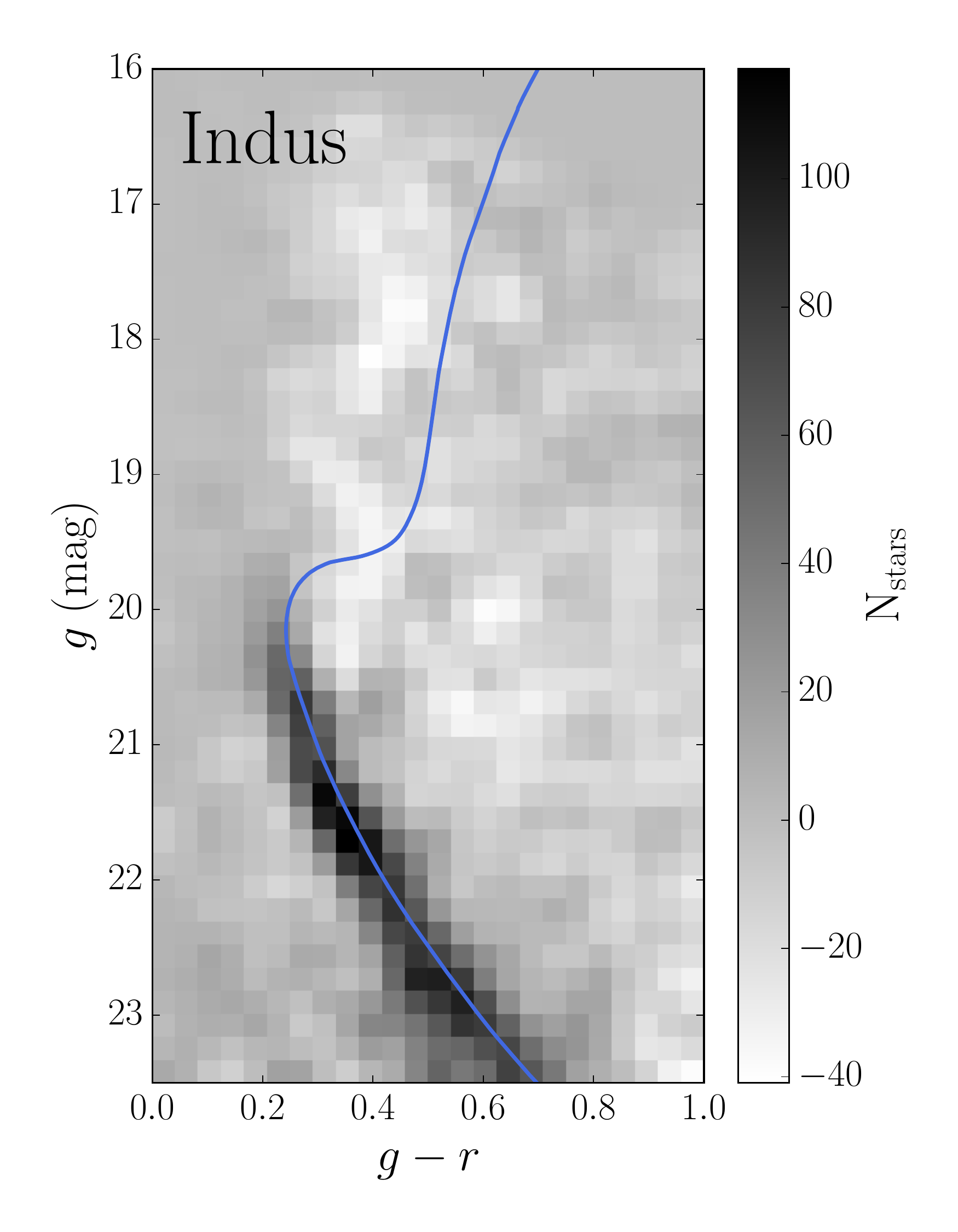}
  \includegraphics[width=0.24\textwidth]{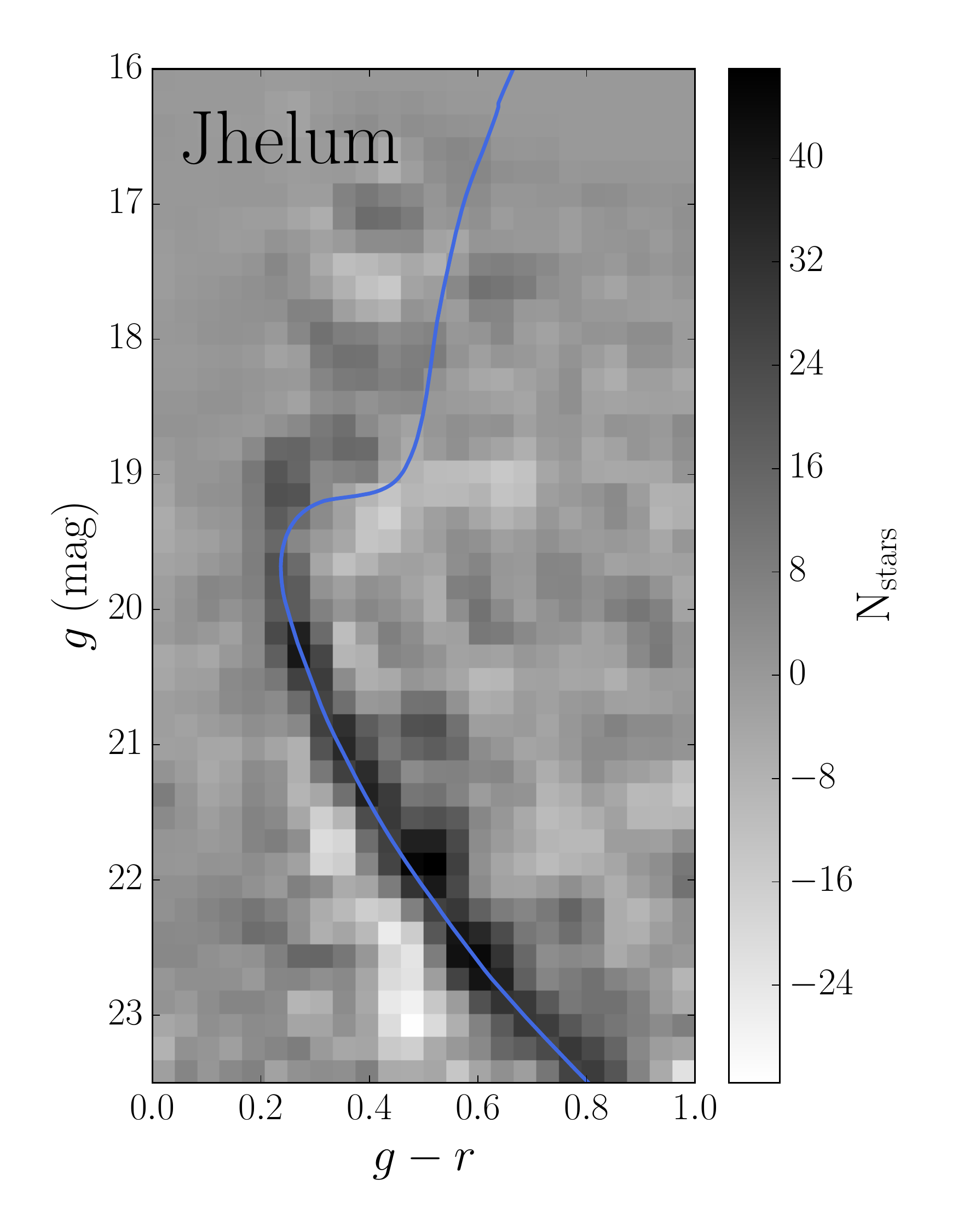}
  \includegraphics[width=0.24\textwidth]{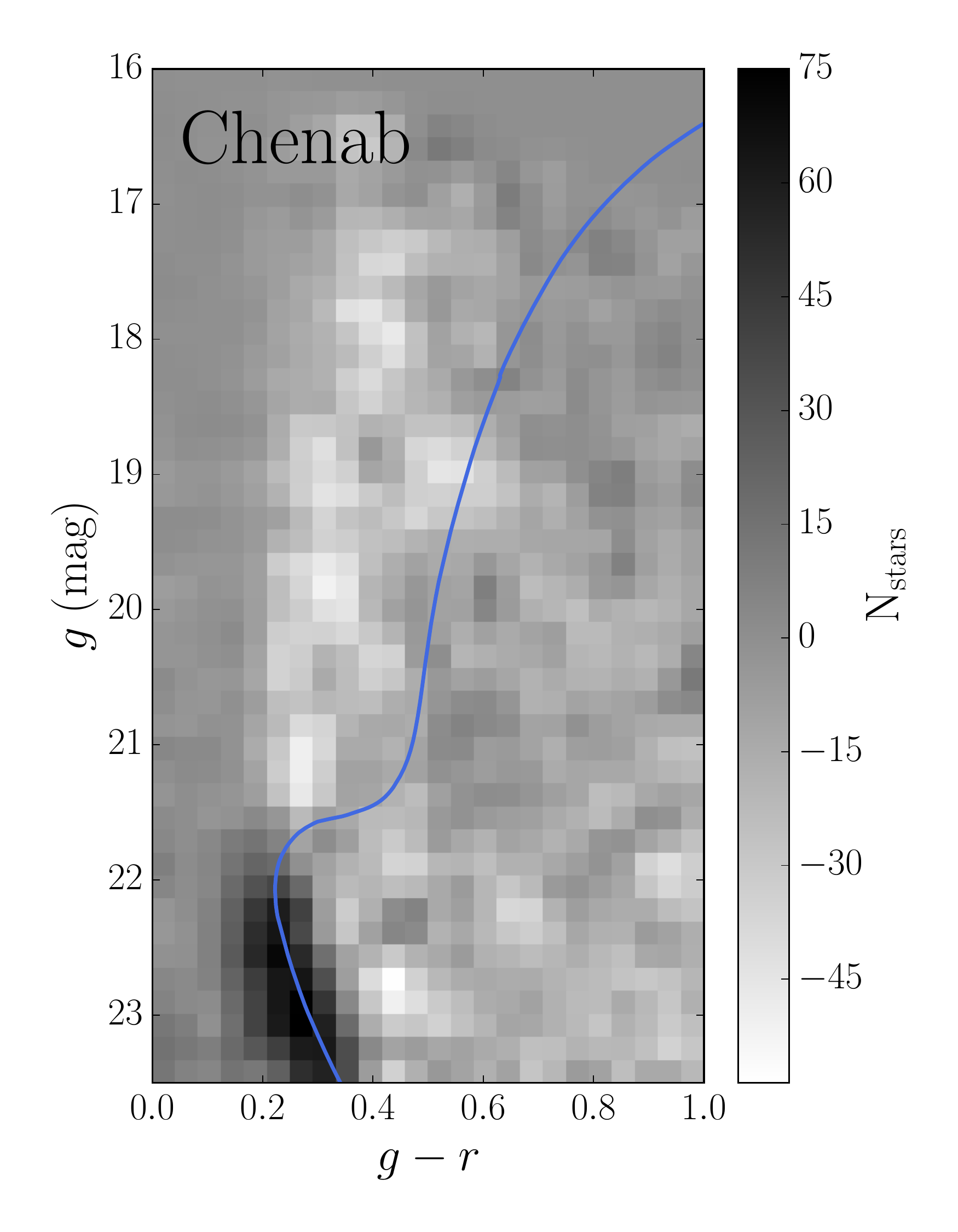}
  \includegraphics[width=0.24\textwidth]{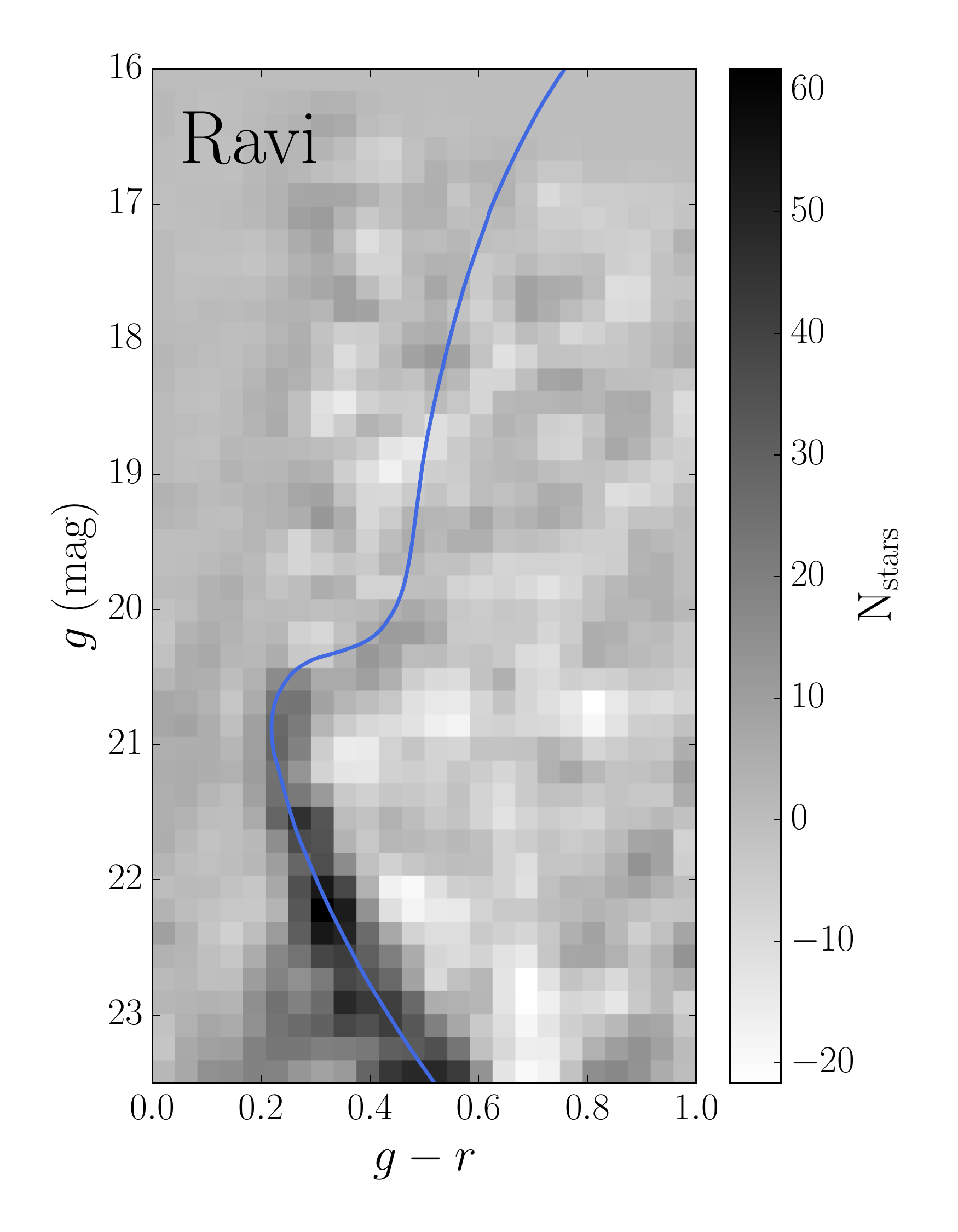}
  \caption{Hess diagrams for Indus (far left), Jhelum (center left), Chenab (center right), and Ravi (far right) stellar stream candidates in DES quadrant one.
Over-plotted in each panel is a \citet{Dotter:2008} synthetic stellar isochrone with parameters determined from the best-fit in \tabref{params}.
These panels are similar to \figref{tucana}.
}
  \label{fig:q1_hess}
\end{figure*}

Indus is the first of \CHECK{4} new stellar stream candidates detected in Q1.
Indus has an angular extent of \CHECK{$20\fdg3$} with a projected width of \CHECK{$0\fdg83$}. 
It may extend beyond the southern edge of the DES footprint in the direction of the SMC; however, at a best-fit distance of \CHECK{16.6\kpc ($\mM = 16.1$)}, it is unlikely that there is a physical association between the Indus stream and the Magellanic Clouds (located $\roughly 3$ times farther away).
The physical width of the Indus stream, \CHECK{$\sigma = 240 \pc$ (${\rm FWHM} = 565 \pc$)}, is comparable to that of the Orphan stream \citep[${\rm FWHM} = 688\pc$;][]{Belokurov:2007b}, and considerably larger than known globular cluster streams.
There is no obvious progenitor for Indus (\secref{assocs}); however, its width may indicate that the Indus stream is the disrupted remains of a faint dwarf galaxy.

The MS of the Indus stream is seen prominently in color-magnitude space (\figref{q1_hess}), with an estimated absolute magnitude of \CHECK{$M_V = -6.2$} (\tabref{deriv}).
The measured metallicity of the Indus stream, \CHECK{$Z = 0.0007$ ($\feh = -1.4$)}, is considerably higher than would be expected for a dwarf galaxy with similar luminosity \citep{Kirby:2013}.
The proximity of the Galactic bulge makes it difficult to model the stellar foreground in the vicinity of Indus, and it is possible that foreground contamination in the RGB of Indus may be artificially inflating the measured metallicity (an even more pronounced example can be seen in Jhelum).

The southern portion of the Indus stream becomes confused with a more distant diffuse stellar structure ($\mM \geq 16.5$) that extends towards the Tucana III stream. 
Due to the incomplete southern coverage of DES, we cannot determine whether this is the signature of another stream or a diffuse stellar cloud. 
Other DECam imaging in the regions of the Magellanic Clouds -- e.g. the Survey of the Magellanic Stellar History \citep[SMASH;][]{Nidever:2017} and the Magellanic Satellites Survey \citep[MagLiteS;]{Drlica-Wagner:2016, Pieres:2017} -- may be able to clarify this question in the near future.
However, kinematic information will be necessary to test for any physical connection between Indus/Tucana III and this putative diffuse structure.

\subsubsection{Jhelum Stream}

The Jhelum stream is comparable to Indus in width, \CHECK{$w = 1\fdg16$}, and due to its orientation on the sky, a longer portion of the stream is contained within the DES footprint (\CHECK{$L \sim 29\fdg2$}).
At a distance of 13.2\kpc ($\mM = 15.6$), Jhelum is closer than Indus; however, both streams can be detected by our isochrone selection simultaneously for distance moduli $15.0 \lesssim \mM \lesssim 16.2$.
The average physical width of Jhelum is \CHECK{267 pc}, though narrowing is seen at the eastern end of the stream.
While Jhelum appears curved in \figref{q1}, the observed curvature is well-matched by a celestial great circle on the sky.

The Hess diagram for the Jhelum stream shows a prominent MS, but also shows some foreground contamination above the MSTO as well as some evidence of over-subtraction. In order to reduce contamination, we selected a narrower on-stream region of width $\pm w$.
Additionally, to reduce the impact of Galactic foreground stars, we first fit the richness, distance modulus, age, and metallicity using just the eastern portion of the stream (higher Galactic latitude).
We fix the age and metallicity at the best-fit values from this initial fit and then refit the richness and distance modulus using the full extent of the stream.
Similar to Indus, we find a high metallicity, \CHECK{$Z = 0.0009$}, which is likely influenced by foreground contamination.

The physical similarity and proximity of the Indus and Jhelum streams is suggestive of a possible physical connection between the two streams. 
To investigate the possibility that Indus and Jhelum may be different orbital arms of the same progenitor, we transformed both into Galactocentric coordinates (\tabref{galacto}). 
We find that the two streams are at a similar Galactocentric radius, \CHECK{$R_{\rm GC} \sim 11 - 13 \kpc$}, but that the Galactocentric great-circle orbits have poles that differ by \CHECK{$\roughly22\degrees$}.
For a flattening $q = 0.9$ and an initial polar angle \CHECK{$\psi = 75\degrees$}, the expected precession after one orbit is \CHECK{$\Delta \phi \sim -10\degrees$} \cite{Erkal:2016}. This suggests that if Indus and Jhelum are associated with the same progenitor, that progenitor has experienced a more highly asymmetric gravitational potential.
Such precession is possible if the progenitor is on an eccentric orbit that takes it close to the Galactic plane, or if the Milky Way halo is more heavily flattened than previously expected.
Additional kinematic information is necessary to confirm or refute this hypothesis.

\subsubsection{Ravi Stream}
\label{sec:ravi}

The Ravi stream candidate is a tenuous feature detected in Q1.
It extends from the lower region of the DES footprint up to the Q1 northern overdensity described in \secref{chenab}.
The Ravi stream is close to the Tucana II dwarf galaxy in projection; however, a fit to the Hess diagram in \figref{q1_hess} puts it at less than half the distance.
Due to its higher Galactic latitude and orientation nearly parallel to the Galactic plane, the CMD of Ravi appears considerably cleaner than some of the other streams in this quadrant, with a pronounced main sequence, and less indication of foreground contamination at bright magnitudes.
We report the best-fit the age, metallicity, and distance modulus in \tabref{params}.

\subsubsection{Chenab Stream}
\label{sec:chenab}

The Chenab stream candidate runs nearly perpendicular to Indus and Jhelum, but at a significantly larger distance of \CHECK{$39.8 \kpc$ ($\mM = 18.0$)}.
Like the other new streams in this quadrant, Chenab has a large angular size, \CHECK{$0\fdg71$}, and a physical width of \CHECK{$\sigma = 493 \pc$}.
The measured extent of Chenab is \CHECK{$18\fdg5$ ($12.9 \kpc$)}. 
The analysis of the Chenab stream is complicated by contamination from Milky Way foreground.

Chenab intersects a diffuse stellar overdensity near the northern edge of the DES footprint.
This overdensity spans from $-32\degrees \lesssim \ra \lesssim -22\degrees$ and $-45\degrees \lesssim \dec \lesssim -40\degrees$ and is apparent in the residual density maps selected for isochrones between $17 < \mM < 19$.
It is possible that this overdensity could be an extended spur of the Sagittarius stream, the ridge line of which passes $\roughly 11\degrees$ from the northwest corner of the DES footprint at a heliocentric distance of $\roughly 25 \kpc$ ($\mM \roughly 17$) \citep{Majewski:2004, Law:2005}.

\subsection{Second Quadrant\,\footnote{To honor the long astronomical tradition in Chile (home of the Blanco telescope), we name stellar streams in Q2 and Q3 after Chilean rivers and aquatic terms in native Chilean tongues. Aliqa Uma is the Aymara term for "peaceful water" and Willka Yaku is the Quechua term for "sacred water."}}

\begin{figure*}[]
  \centering
  \includegraphics[width=0.32\textwidth]{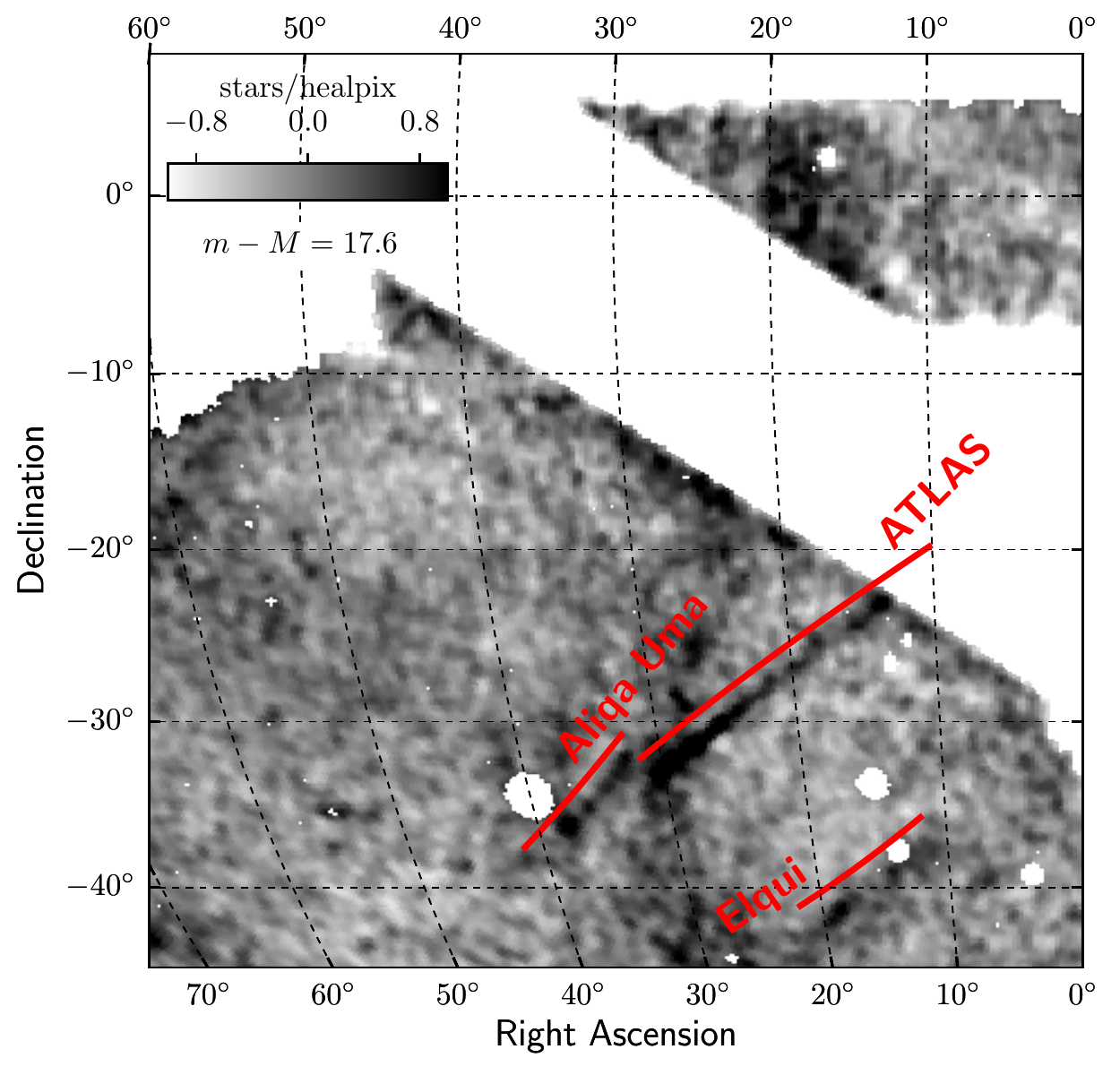}
  \includegraphics[width=0.32\textwidth]{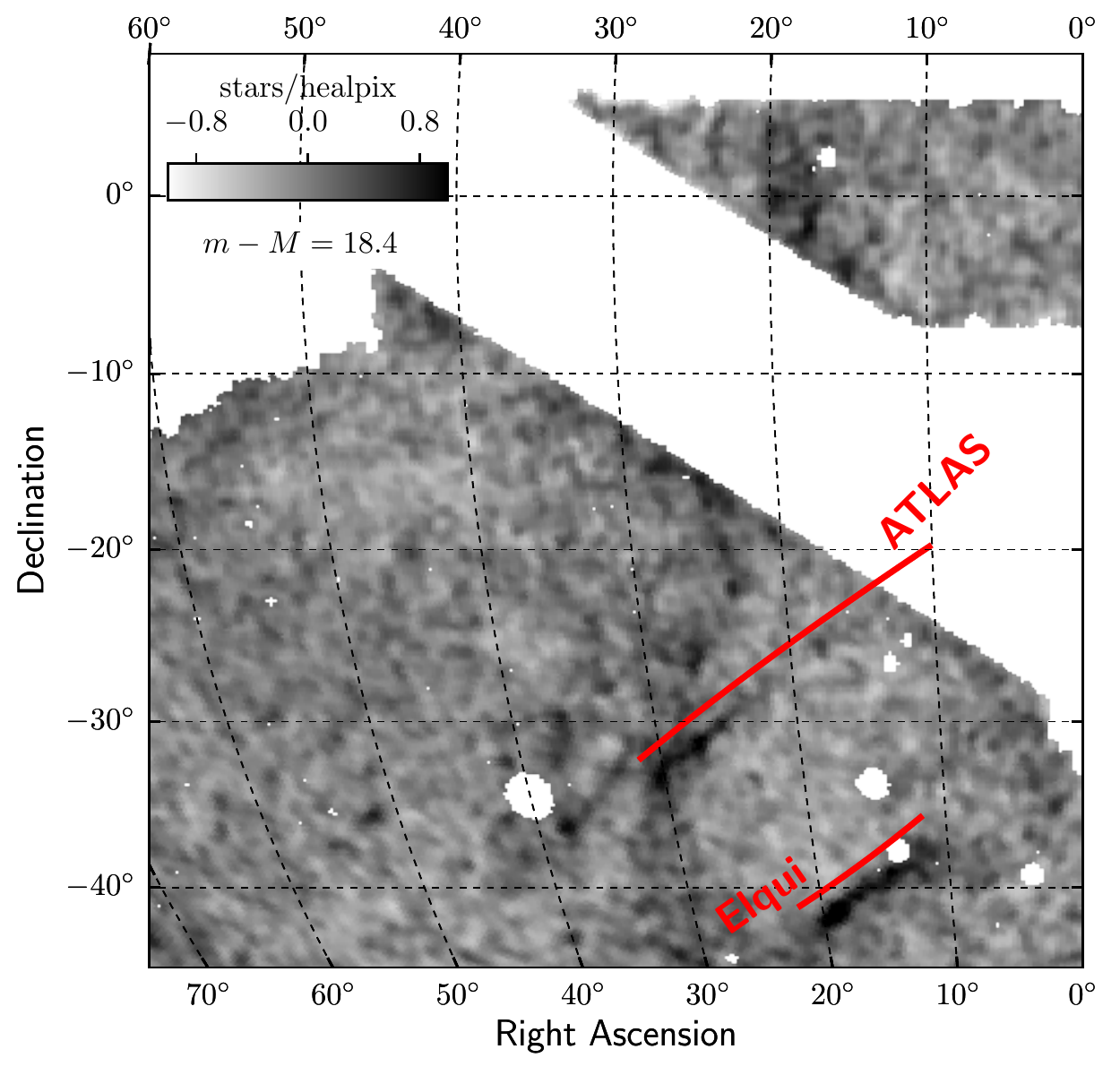}
  \includegraphics[width=0.32\textwidth]{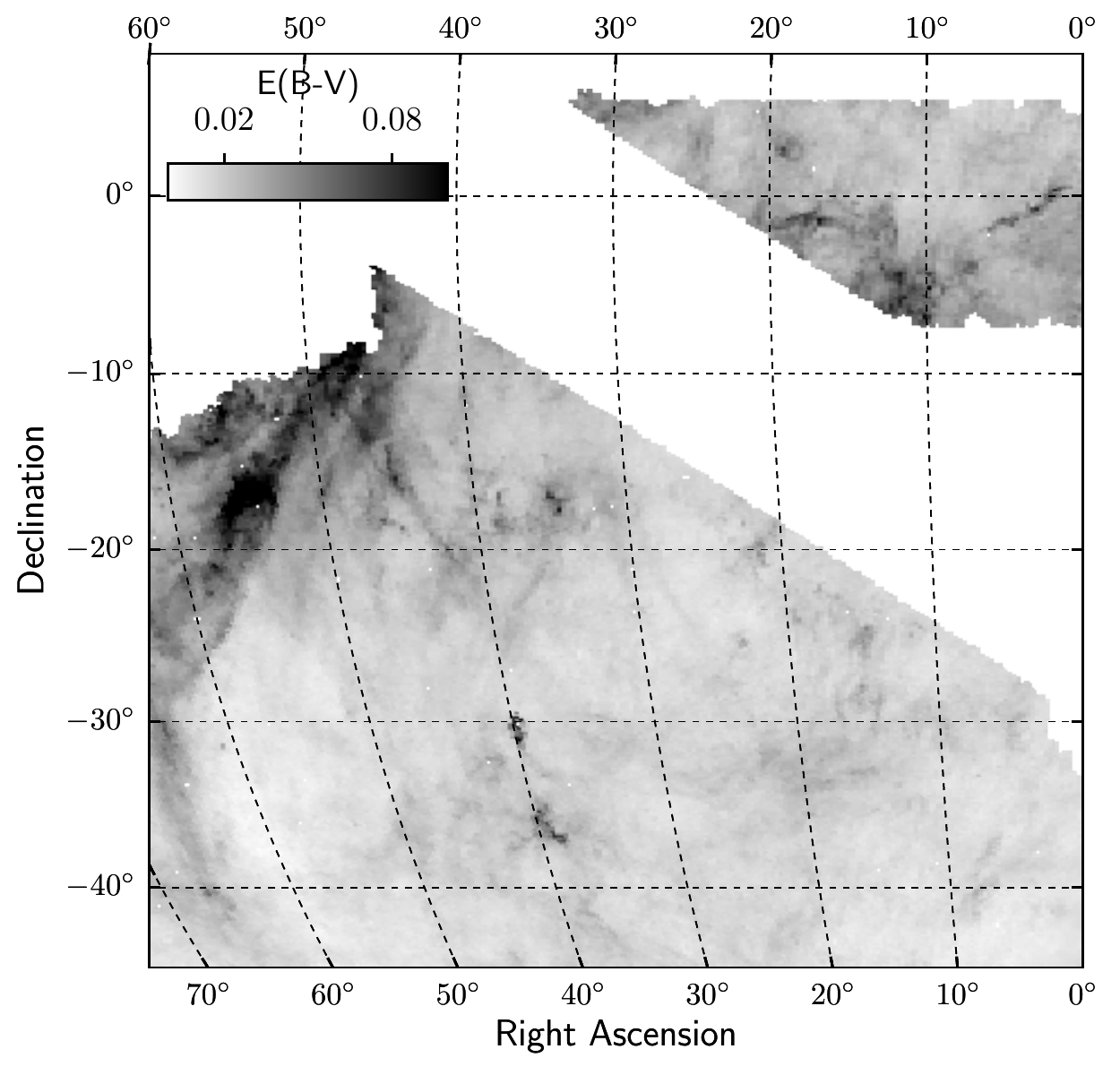}
  \caption{Residual density maps in Q2 (similar to \figref{q1}).
    The Sagittarius stream has been masked to optimize the search for faint stellar streams.
    The Fornax ($\ra,\dec = 40\fdg0,-34\fdg4$) and Sculptor ($\ra,\dec = 15\fdg0,-33\fdg7$) dwarf galaxies have been masked along with several globular clusters and Local Group galaxies.
    (Left) The residual density map for $m-M = 16.4$ showing the ATLAS and Aliqa Uma streams, with a hint of the Elqui stream.
    (Middle) The residual density map for $m-M = 18.4$. 
    The Elqui stream appears prominently, while the southeastern portion of ATLAS is still visible.
    Molonglo has a relatively high metallicity (Z=0.001) and is therefore not visible with this selection.
    (Right) Interstellar reddening,  $E(B-V)$, from \citet{Schlegel:1998}. 
    An animated version can be found online \href{http://home.fnal.gov/~kadrlica/movies/residual_q2_v17p2_label.gif}{at this url}.
}
  \label{fig:q2}
\end{figure*}

The second quadrant (Q2) spans from $0\degrees < \ra \lesssim 60\degrees$ and $-42\degrees < \dec < 7\degrees$ (\figref{q2}). 
This quadrant contains the Sagittarius stream, which we have masked from our analysis in order to increase our sensitivity to fainter new streams. 
We detect \CHECK{4} narrow streams in this region, including the previously known ATLAS \citep{Koposov:2014} stream and a possible extension of the Molonglo \citep{Grillmair:2017} stream, and two newly detected streams that we name Elqui and Aliqa Uma. 
In addition, we find a long, diffuse structure that extends along the height of Q2 intersecting the Eridanus-Phoenix stellar overdensity lower in the DES footprint \citep{Li:2016}. 
We name this structure the Palca stream and discuss it in more detail in \secref{palca}.

\begin{figure*}
  \centering
  \includegraphics[width=0.24\textwidth]{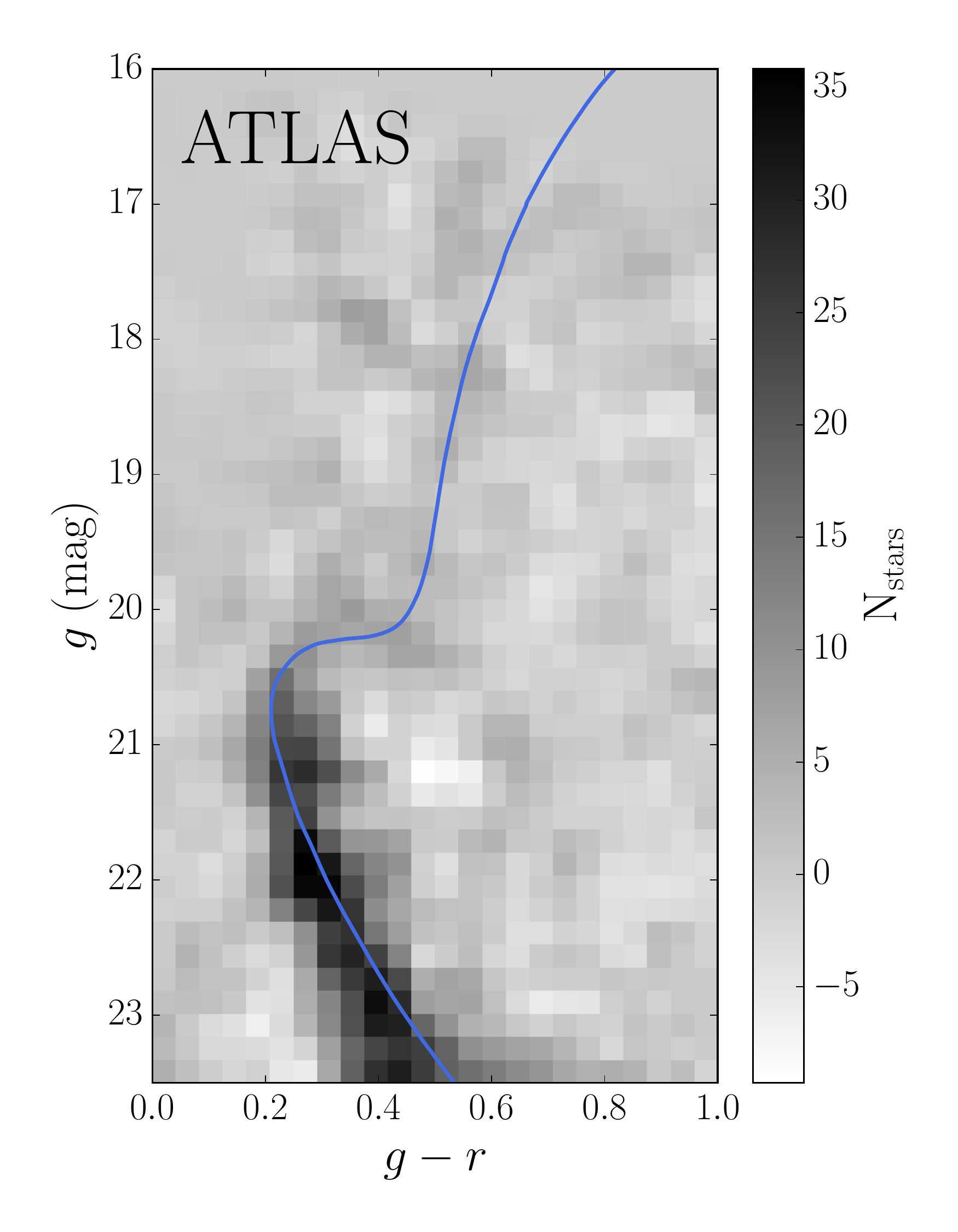}
  \includegraphics[width=0.24\textwidth]{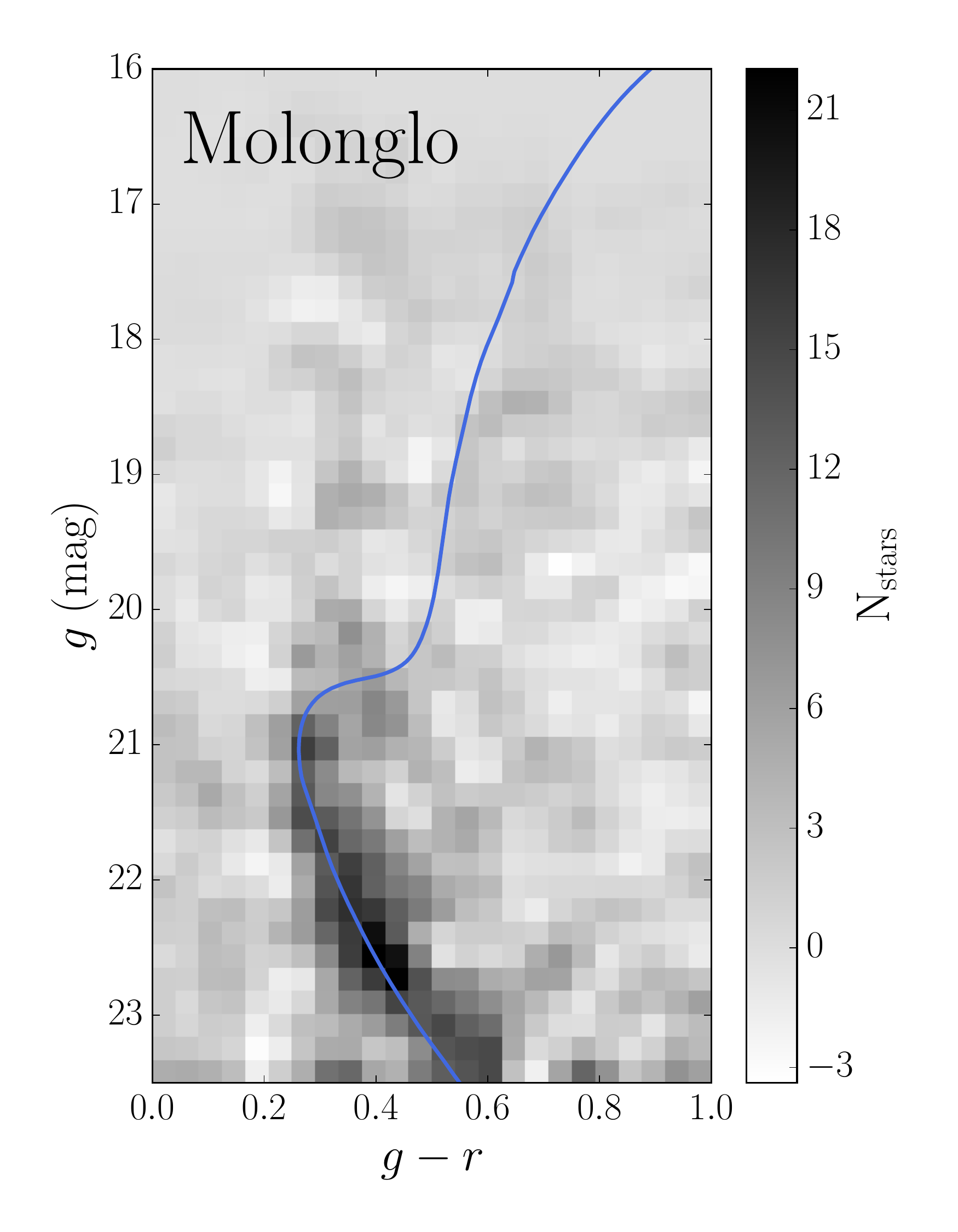}
  \includegraphics[width=0.24\textwidth]{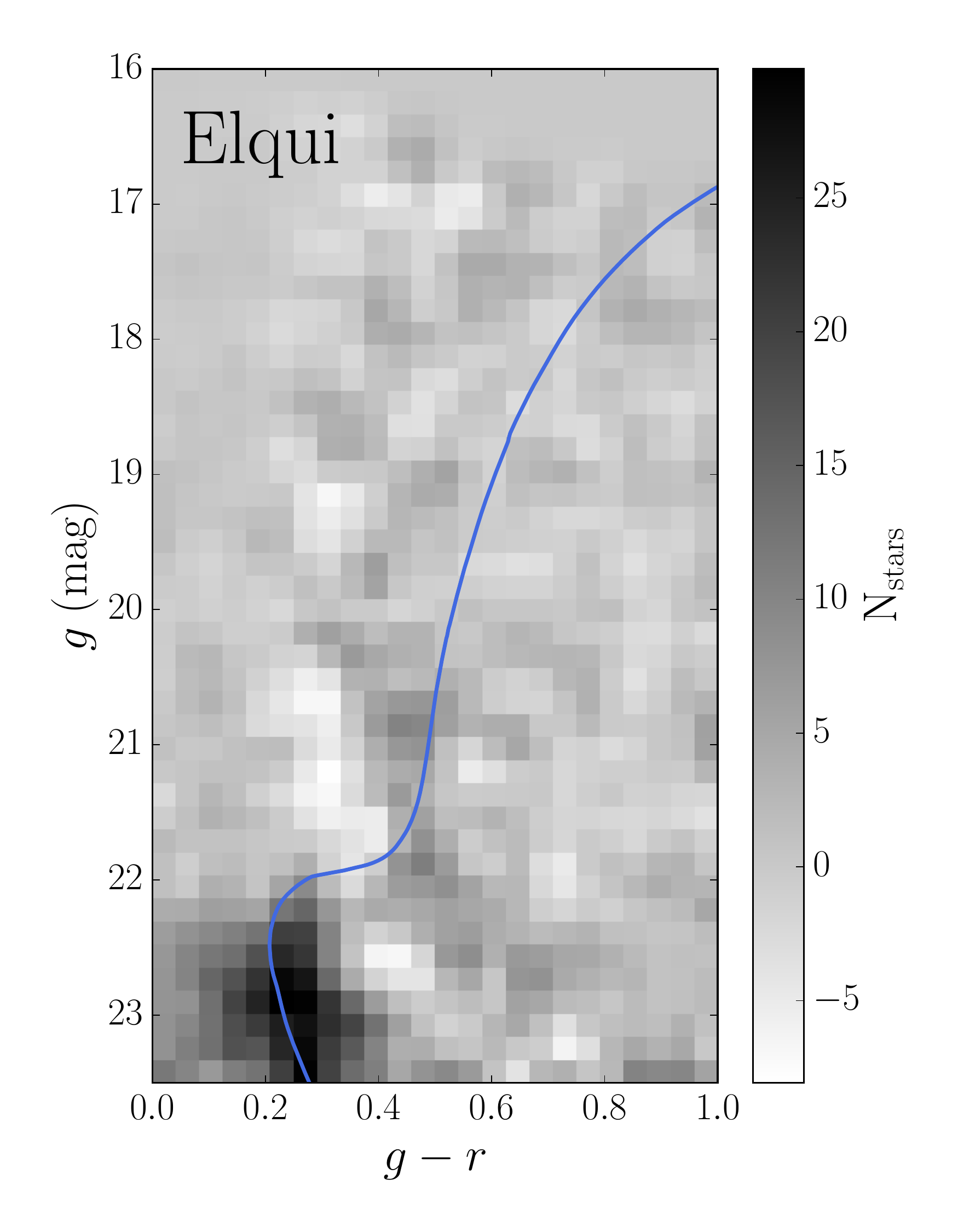}
  \includegraphics[width=0.24\textwidth]{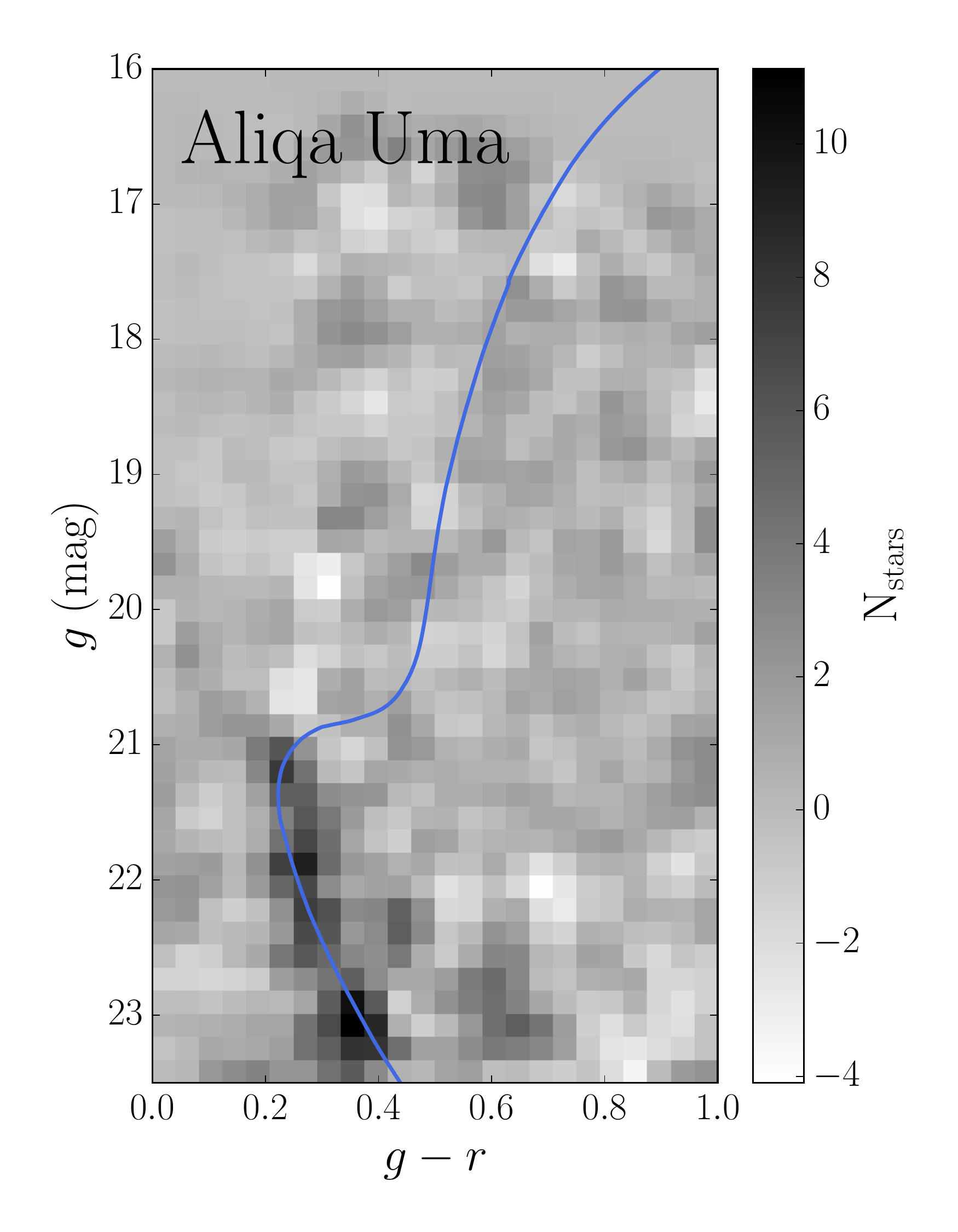}
  \caption{Hess diagrams for the ATLAS (far left), Molonglo (center left), Elqui (center right), and Aliqa Uma (far right) stream candidates in DES quadrant two (Q2). 
Over-plotted in each panel is a \citet{Dotter:2008} synthetic stellar isochrone with parameters determined from the best-fit in \tabref{params}.
These panels are similar to \figref{tucana}.
  \label{fig:q2_hess}}
\end{figure*}

\subsubsection{ATLAS Stream}
\label{sec:atlas}

The ATLAS stream is a narrow stellar stream discovered in the first data release of the VST ATLAS survey, which covered a declination slice around the stream from $-37\degrees \lesssim \dec \lesssim -25\degrees$ to a limiting magnitude of $r \sim 22$ \citep{Koposov:2014}.
The ATLAS stream was later studied by \citet{Bernard:2016} using the larger sky coverage provided by Pan-STARRS.
The DES analysis is deeper than both VST ATLAS and Pan-STARRS, $g = 23.5$, and extends the sky coverage around ATLAS to lower declinations.
The ATLAS stream does not appear to extend significantly beyond the length described by \citet{Koposov:2014}, ending at \CHECK{$(\ra,\dec) \sim (30\fdg7, -33\fdg2)$}.
At higher declination it becomes difficult to disentangle the ATLAS stream from the much more luminous Sagittarius stream before hitting the boundary of the DES footprint at $(\ra,\dec) \sim (9\fdg3, -20\fdg9)$.
Using Pan-STARRS data, \citet{Bernard:2016} have extended ATLAS to $\dec \sim -15\degrees$ leading to a total length of $\roughly 28\degrees$, of which $22\fdg6$ is contained within DES.
 
We follow the procedure described in \secref{detect} to characterize the physical properties of the ATLAS stream. 
The deeper DES data prefer a slightly larger distance of \CHECK{22.9\kpc ($\mM \sim 16.8$)}, which is marginally consistent with the previously measured distance, $20 \pm 2 \kpc$, derived using the VST ATLAS data \citep{Koposov:2014}.
At a distance of $22.9\kpc$, the visible portion of the ATLAS stream extends \CHECK{$9.0 \kpc$}.
We independently fit the distance modulus to each half of the ATLAS stream and find evidence that the southwestern portion of ATLAS has a distance modulus that is $\roughly 0.3 \magn$ larger than the northeastern portion.  
The southwestern portion of ATLAS is detectable in the residual density maps at distance modulus of $\mM > 18.0$
The stellar density is not uniform along the length of ATLAS and we note a roughly spherical overdensity in the southwestern portion at $(\ra,\dec) = (25\fdg37,-30\fdg13)$. 
This overdensity is visible at lower significance in Figure 1 of \citealt{Koposov:2014}.
The DES data suggest a fainter absolute magnitude for the ATLAS stream, \CHECK{$M_V = -4.5$}, compared to that estimated in the VST ATLAS data, $M_V \sim -6$ \citep{Koposov:2014}.

\begin{figure*}[htb!]
\centering
\includegraphics[width=0.8\textwidth]{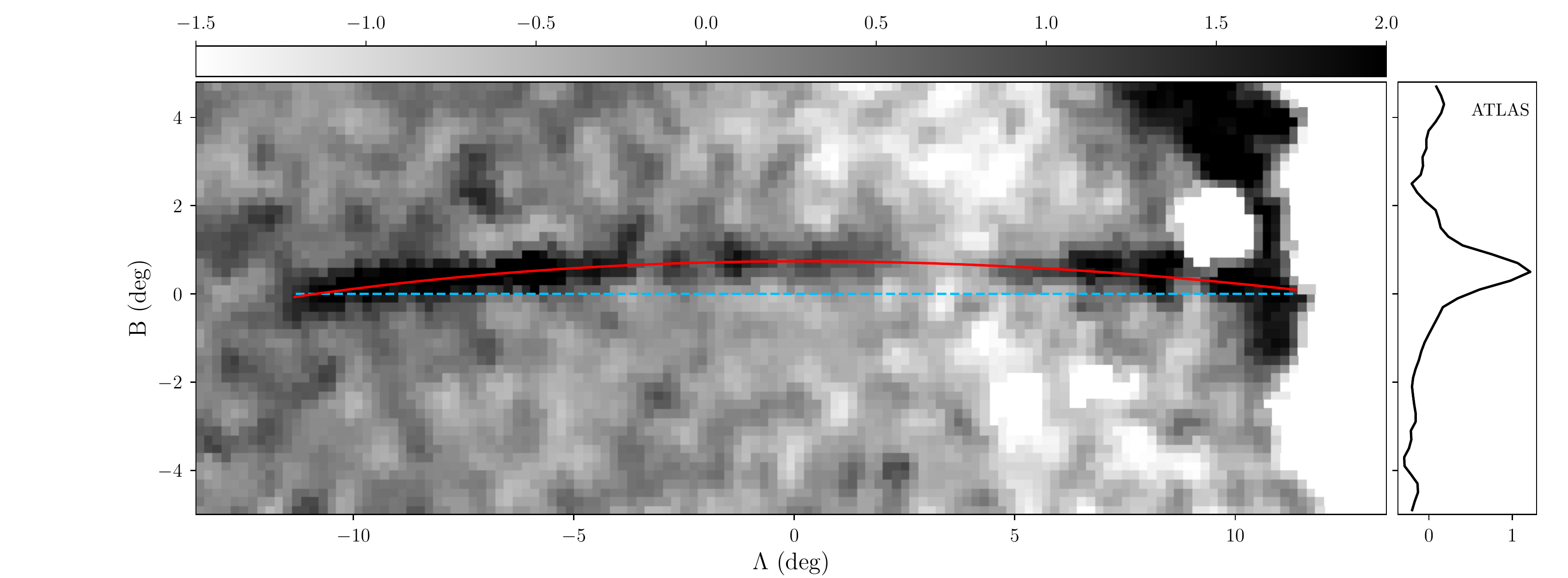}
\caption{Residual stellar density along the ATLAS stream for a distance modulus selection of $\mM = 16.8$. Both the stellar counts and background model have been smoothed by a Gaussian kernel with width 0\fdg3. The ATLAS stream has noticeable curvature, which can be seen by comparing the best-fit second-order polynomial from \eqnref{atlas} (red solid line) to the plane of a great circle on the sky (blue dashed line).}
\label{fig:atlas}
\end{figure*}

\figref{atlas} shows the spatial distribution of stars in a coordinate system aligned with the endpoints of the ATLAS stream.  
Following the procedure described in \secref{spatial}, we fit the width of the ATLAS stream with a Gaussian model on top of a linear background. 
Our measured width of the ATLAS stream, $w = 0\fdg24$, is consistent with $w = 0\fdg25$ reported by \citet{Koposov:2014}. 
However, it is also clear in \figref{atlas} that the ATLAS stream deviates appreciably from a great circle on the sky, which would lie along the equator.
We find that the ridgeline of the ATLAS stream is well-described over the range $9\fdg3 < \ra < 30\fdg7$ by a second-order polynomial of the form
\begin{equation}
\label{eqn:atlas}
\dec = -15.637 - 0.545 (\ra) - 0.001(\ra)^2.
\end{equation}

Interestingly, \figref{atlas} also appears to show an underdensity in the stream at $\Lambda \sim 4\degrees$ which is approximately $2\fdg5$ in size. 
We caution that this underdensity occurs in a region where the polynomial background fit is complicated by the proximity of the Sagittarius stream. 
Over-subtraction of the background could manifest as an underdensity in the residual map.
However, if this underdensity is real it could be due to perturbations by subhaloes around the Milky Way \citep[e.g.][]{Ibata:2002,Johnston:2002}. \citet{Erkal:2016b} estimated the typical size and number of gaps in the ATLAS stream due to subhaloes and found a characteristic gap size of $\sim 4 \degrees$ with 0.1 gaps expected. However, this prediction depends on the length of the ATLAS stream and therefore given the increased length of the ATLAS stream detected in this work and in Pan-STARRs \citep{Bernard:2016}, the predicted number of gaps is an underestimate. If the underdensity is confirmed then the gap can be used to infer the properties of the subhalo which created the gap \citep[][]{Erkal:2015b} and the statistical properties of the stream density can be used to place constraints on the number of subhaloes in the Milky Way \citep[][]{Bovy:2017}.

\subsubsection{Molonglo Stream}
\label{sec:molonglo}

The Molonglo stream was identified as a faint, narrow feature in data from Pan-STARRS \citep{Grillmair:2017}.
We extrapolate Equation~(2) from \citet{Grillmair:2017} to detect a narrow, $\roughly 8\degree$ extension of the Molonglo stream in the DES data offset by $\lesssim 2 \degrees$.
Molonglo is the only stream that is detected based on prior information from another survey and there is some risk of confirmation bias.
In fact, Molonglo is the least apparent feature in the residual density maps, due in part to its proximity to the Sagittarius stream.
However, the Hess diagram for Molonglo shows a clear MS and MSTO (\figref{q2_hess}), and the stream has a detection significance of \CHECK{$5.2\sigma$}.
After fixing the age ($\age = 13.5\Gyr$) and metallicity ($Z=0.0010$) to match the Hess diagram, we fit the distance modulus and richness of Molonglo.
We determine a distance of \CHECK{22.9\kpc ($\mM = 16.8$)}, which agrees with the estimated distance of 20\kpc from \citet{Grillmair:2017}.
However, the foreground subtraction in this region is difficult, and the remaining contamination may have artificially inflated the metallicity.
Our best-fit width of \CHECK{$0\fdg32$} is narrower than the $\roughly0\fdg5$ reported by \citet{Grillmair:2017}; however, it is unclear whether his width was measured after convolving with a $0\fdg4$ Gaussian kernel.
We do not find evidence of the other three streams identified by \citet{Grillmair:2017}.

\subsubsection{Elqui Stream}
\label{sec:elqui}

At a heliocentric distance of \CHECK{$D_{\sun} = 50.1 \kpc$}, the Elqui stream is the most distant stream discovered in the DES Y3A2 data.
Fitting the southwestern and northeastern halves of Elqui independently shows a shift in distance modulus from \CHECK{18.2 to 18.5}, corresponding to a physical change in distance of \CHECK{6.5 \kpc} over a length of \CHECK{8.2 \kpc}.
The Elqui stream is broad and may show slight curvature on the sky, which is unexpected for a stream at large Galactocentric radius. 
The background galaxy NGC 300 resides at $(\ra,\dec) \sim (13\fdg7,-37\fdg7)$, but is unrelated to the much closer Elqui stream.

Elqui resides at similar Galactocentric distance to the LMC and overlaps (in projection) the gaseous component of the Magellanic Stream \citep{Nidever:2008}.
Transforming to the Magellanic Stream coordinates defined by \citet{Nidever:2008}, the endpoints of Elqui are located at \CHECK{$L_{\rm MS},B_{\rm MS} \approx (-49\fdg1, 1\fdg8), (-40\fdg7, 6\fdg1)$}.
The proximity between Elqui and the Magellanic Stream suggests that Elqui may in fact be stellar ejecta from a past collision between the LMC and SMC \citep{Besla:2010,Besla:2012}.
\citet{Besla:2012} suggests that a recent collision between the LMC and SMC could explain many of the observed features of the Magellanic system.
If Elqui indeed formed as the result of such a collision, its existence could be used to further constrain the infall history of the Magellanic Clouds.

Evidence of stellar ejecta from the Magellanic Clouds is observed elsewhere in the DES footprint.
The so-called ``SMC northern overdensity'' \citep{Pieres:2016} is located $\roughly 8\degrees$ from the SMC on the southern edge of the DES footprint $(\ra,\dec) \sim (15\degrees, -65\degrees)$ and is visible in isochrone selections with $\mM \gtrsim 18$.
Establishing complete DECam coverage around the Magellanic Clouds promises additional insight into how the Magellanic Clouds have shaped the Milky Way halo.

\subsubsection{Aliqa Uma Stream}
\label{sec:fornax}

The Aliqa Uma stream resides at the southern end of the ATLAS stream, extending from \CHECK{$\ra,\dec \sim (31\fdg7, -31\fdg5)$ to $\ra,\dec \sim (40\fdg6, -38\fdg3)$}. 
While the northern end of this stream is in close proximity to the southern end of the ATLAS stream, the difference in orientation and distance modulus, \CHECK{$\mM = 17.3$}, leads us to classify it as a distinct system rather than an extension of ATLAS.
While this stream crosses close to the Fornax dwarf galaxy in projection, it is substantially closer and the two systems are very unlikely to be physically associated.
The presence of Fornax and ATLAS bracketing the much fainter Aliqa Uma stream makes it difficult to establish a good background selection region.

\subsection{Third Quadrant}
\label{sec:q3}

\begin{figure}[]
  \centering
  \includegraphics[width=\columnwidth]{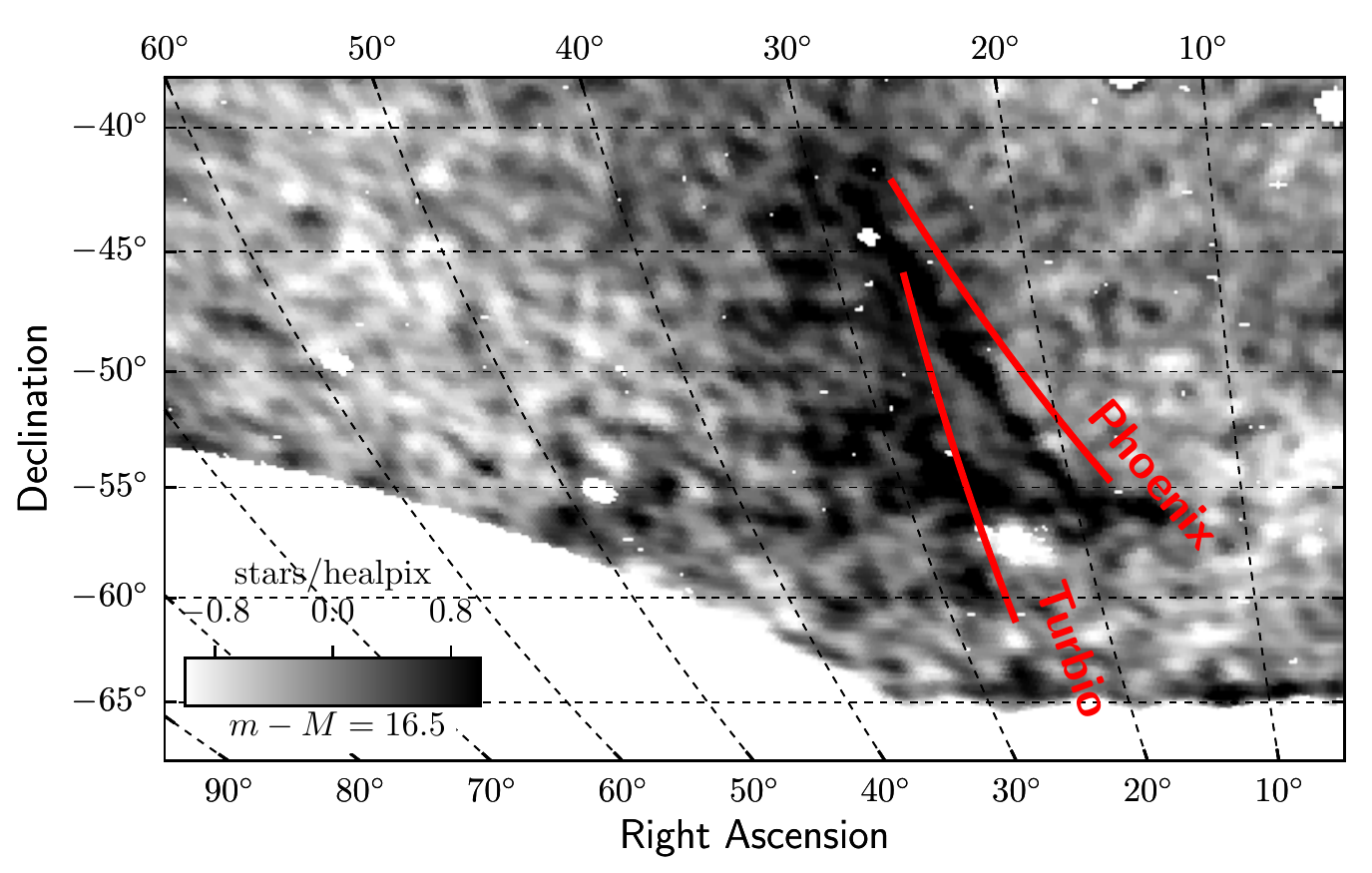}  
  \includegraphics[width=\columnwidth]{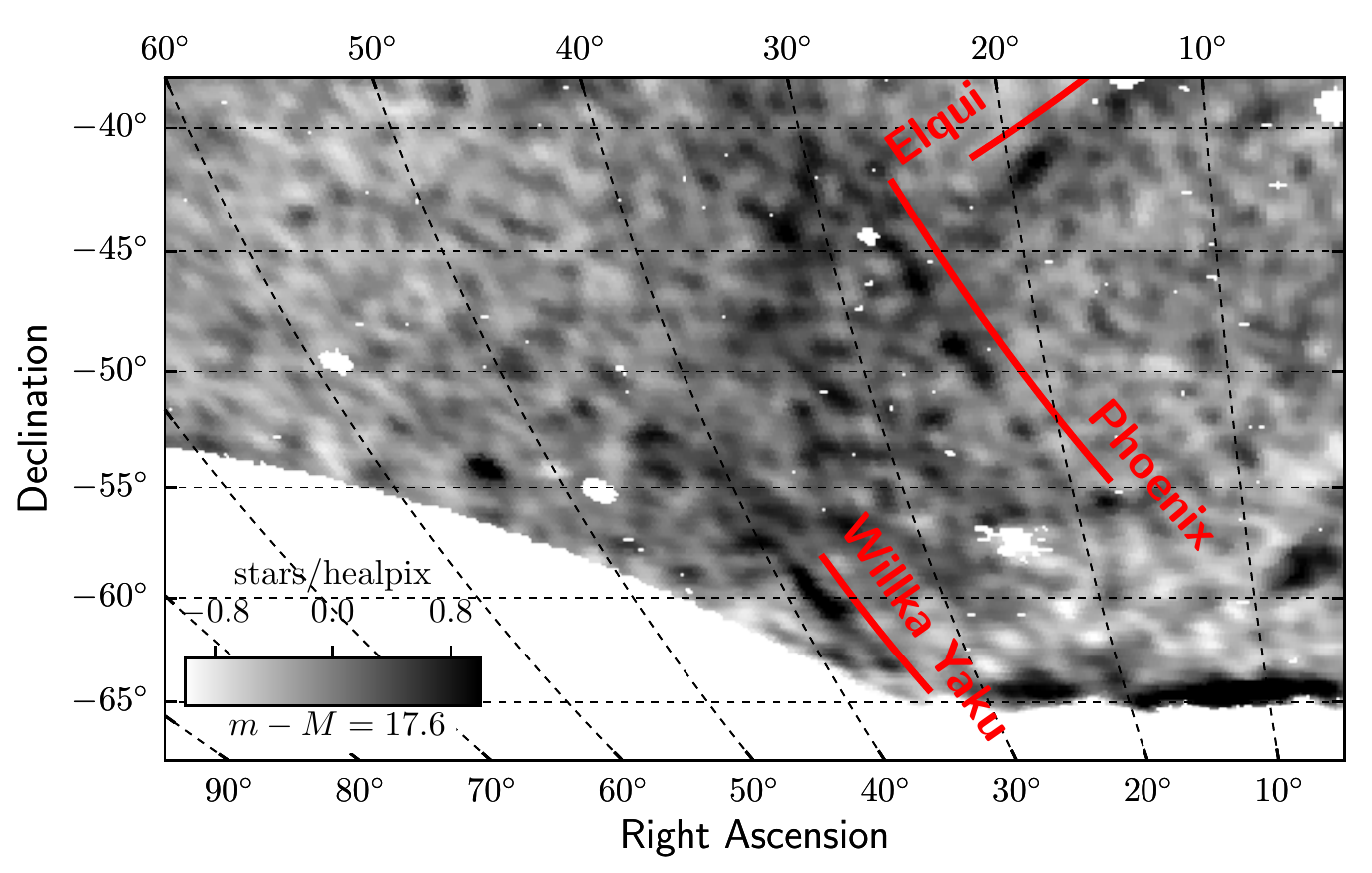}
  \includegraphics[width=\columnwidth]{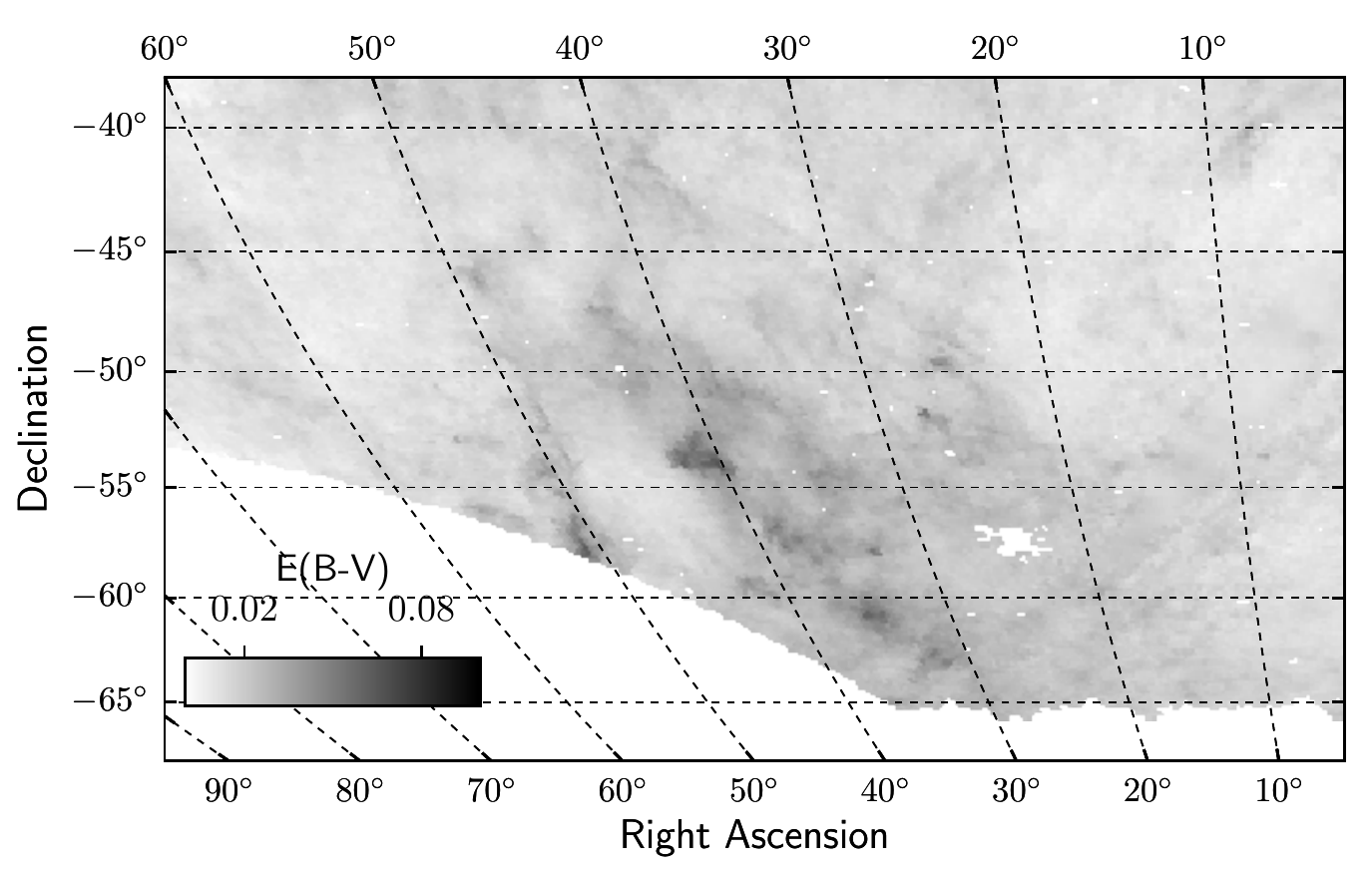}
  \caption{Residual density maps in Q3 (similar to \figref{q1}). 
    (Top) An isochrone selection with $\mM = 16.5$ shows the Phoenix and Tubio streams superposed on the EriPhe overdensity.
    (Middle) The Willka Yaku stream can be seen in a selection for $\mM = 17.6$.
    (Bottom) Interstellar reddening,  $E(B-V)$, from \citet{Schlegel:1998}.
    An animated version of this figure can be found online \href{http://home.fnal.gov/~kadrlica/movies/residual_q3_v17p2_label.gif}{at this url}.
}
  \label{fig:q3}
\end{figure}

\begin{figure*}[t]
  \centering
  \includegraphics[width=0.25\textwidth]{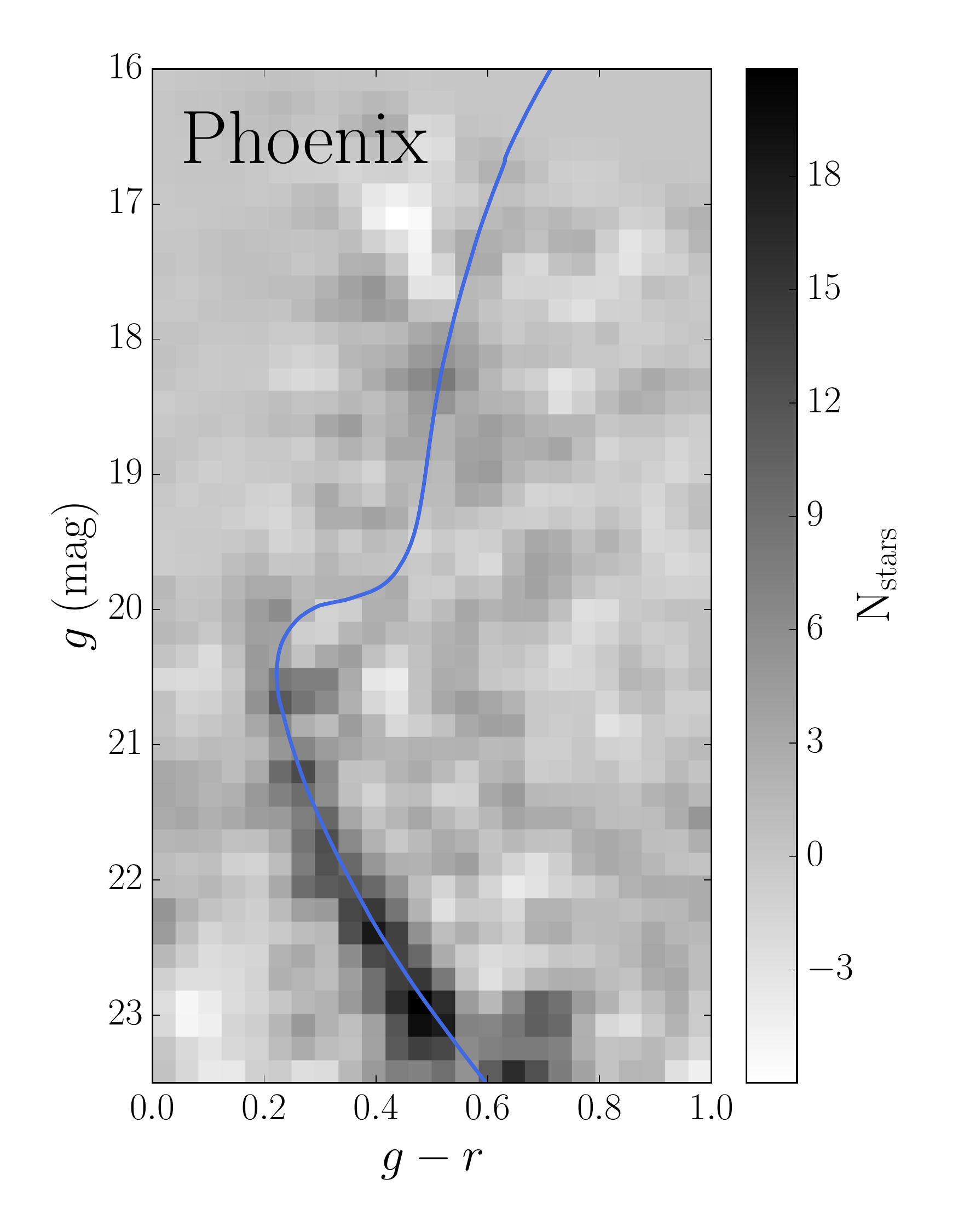}
  \includegraphics[width=0.25\textwidth]{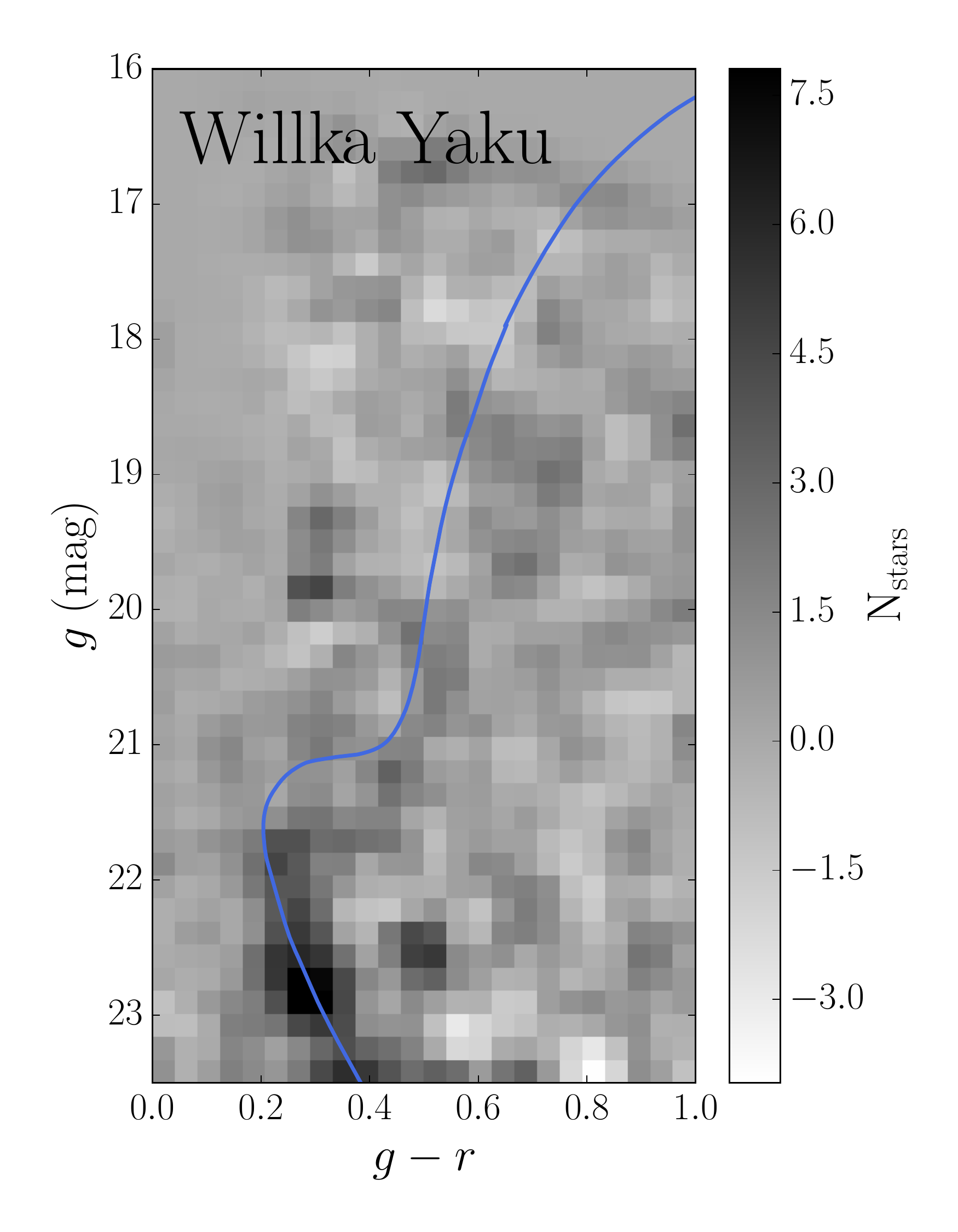}
  \includegraphics[width=0.25\textwidth]{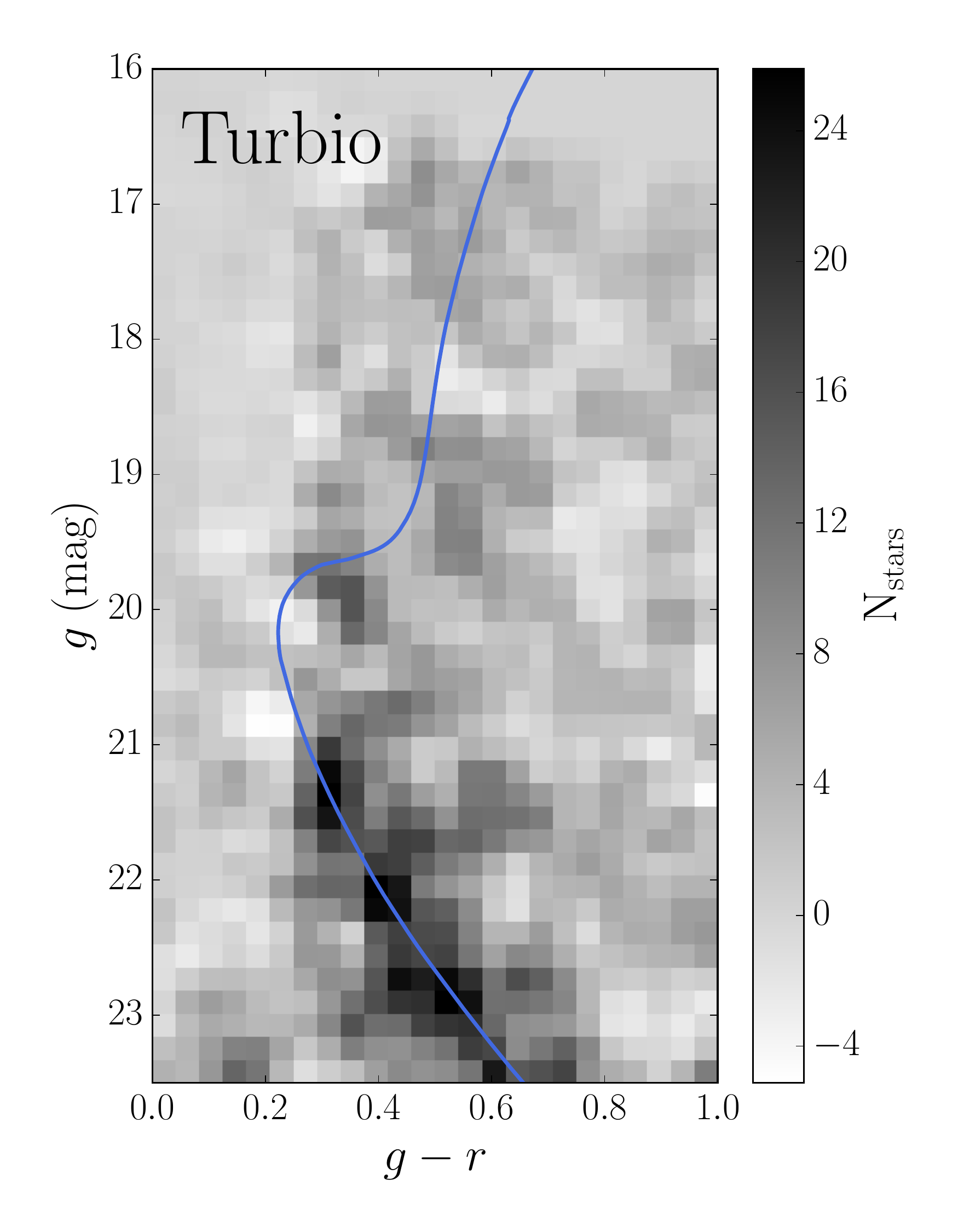}
  \caption{Hess diagrams for the Phoenix (left),  Willka Yaku (center), and Turbio (right) streams found in DES Q3. 
Over-plotted in each panel is a \citet{Dotter:2008} synthetic stellar isochrone with parameters determined from the best-fit in \tabref{params}.
These panels are similar to \figref{tucana}.
  \label{fig:q3_hess}}
\end{figure*}

The third DES quadrant (Q3) covers the region from $5\degrees \lesssim \ra \lesssim 60\degrees$ and $-65\degrees \lesssim \dec \lesssim -42\degrees$. 
This quadrant resides above the LMC and SMC, and overlaps heavily with the EriPhe stellar overdensity \citep{Li:2016}.
Several linear structures are detected in this region; however, it is difficult to conclusively differentiate them from a diffuse component of EriPhe.
In addition to the Phoenix stream \citep{Balbinot:2016}, we identify two new stream candidates, Turbio and Willka Yaku. 

\subsubsection{Phoenix Stream}
\label{sec:phoenix}

The Phoenix stream is a narrow stellar stream discovered in the DES Y1A1 data \citep{Balbinot:2016}.
Since the DES analysis of the Phoenix stream is discussed in great detail in \citet{Balbinot:2016}, we offer only a brief discussion here.
The Y3A2 data provide a deeper and more complete catalog with improved photometric accuracy; however, the qualitative characteristics of the Phoenix stream are predominantly unchanged.
It remains a clumpy and knotted stream, consisting of a more-or-less symmetric distribution of overdensities.
The parameters that we derive for the Phoenix stream largely agree with those of \citet{Balbinot:2016}.
 We measure a distance of \CHECK{$19.1 \kpc$ ($\mM=16.4$)}, which is slightly larger than the value measured previously ($17.5 \pm 9 \kpc$).
Similar to \citet{Balbinot:2016} we find no indication of a distance gradient.

\subsubsection{Turbio Stream}
\label{sec:turbio}

We find a linear feature near the center of the EriPhe stellar overdensity, which constitutes a candidate stream named Turbio.
While Turbio is detected above the background of EriPhe with a significance of \CHECK{$7.9\sigma$}, it would be very surprising if the two were not physically associated.
It is likely that Turbio stands out more prominently in our analysis compared to that of \citet{Li:2016} due to improved photometric calibration ($\roughly 0.7\%$ vs.\ $\roughly 2\%$), increased depth ($g < 23.5$ vs.\ $g < 22.5$), and improved spatial resolution (smoothing kernel of $0\fdg3$ vs.\ $0\fdg5$).

The MS of Turbio is detected in a Hess diagram that subtracts a neighboring region of EriPhe to the east of the structure (\figref{q3_hess}).
We fit an isochrone with fixed age and metallicity ($\age = 13.0\Gyr$, $Z=0.0004$) and find a best-fit distance modulus of $\mM = 16.1$, which is indistinguishable from that of EriPhe and the nearby globular cluster NGC 1261.
\citet{Li:2016} suggest that EriPhe may have a common origin with the Virgo Overdensity \citep{Juric:2008} and the Hercules-Aquila cloud \citep{Belokurov:2007c}.
However, the orientation of Turbio is nearly perpendicular with the orbit necessary to connect these three diffuse structures.
In addition, Turbio may constitute a dense portion of the larger Palca structure that is seen to extend northward from the EriPhe cloud.

\subsubsection{Willka Yaku Stream}
\label{sec:reticulum}

Willka Yaku is a short and relatively narrow stream that extends $\roughly 6\degrees$ from the southern edge of the DES footprint.
Willka Yaku sits on the south eastern boundary of the EriPhe cloud and we find a significant gradient in the stellar density transverse to the stream.
To reduce the effects of foreground contamination, we fit only the more prominent northern half of the stream (\figref{q3_hess}).
A simultaneous fit of age, metallicity, and distance modulus yields a best-fit distance of \CHECK{$34.7\kpc$ ($\mM = 17.7$)}.
This places Willka Yaku at a significantly larger distance than EriPhe, and makes a  physical association less likely.

\subsection{Fourth Quadrant\,\footnote{To honor the long tradition of astronomy in Australia (stretching back tens of thousands of years), stellar stream candidates in this quadrant are named after Aboriginal terms for rivers in Australia.}}
\label{sec:q4}

The fourth quadrant of DES (Q4) spans from $60\degrees \lesssim \ra \lesssim 95\degrees$ and $-65\degrees < \dec < -15\degrees$.
It contains the globular clusters NGC 1851 and NGC 1904, and structure from the Monoceros Ring can be seen in the direction of the Galactic anti-center. 
The proximity of the Monoceros and the Galactic plane make background subtraction difficult near the eastern edge of this region.
Inspection of the residual density maps yields two stellar stream candidates, Turranburra and Wambelong, which are reasonably wide ($w \sim 0\fdg5$) and detected at moderate significance.

\begin{figure*}[t]
  \centering
  \includegraphics[width=0.32\textwidth]{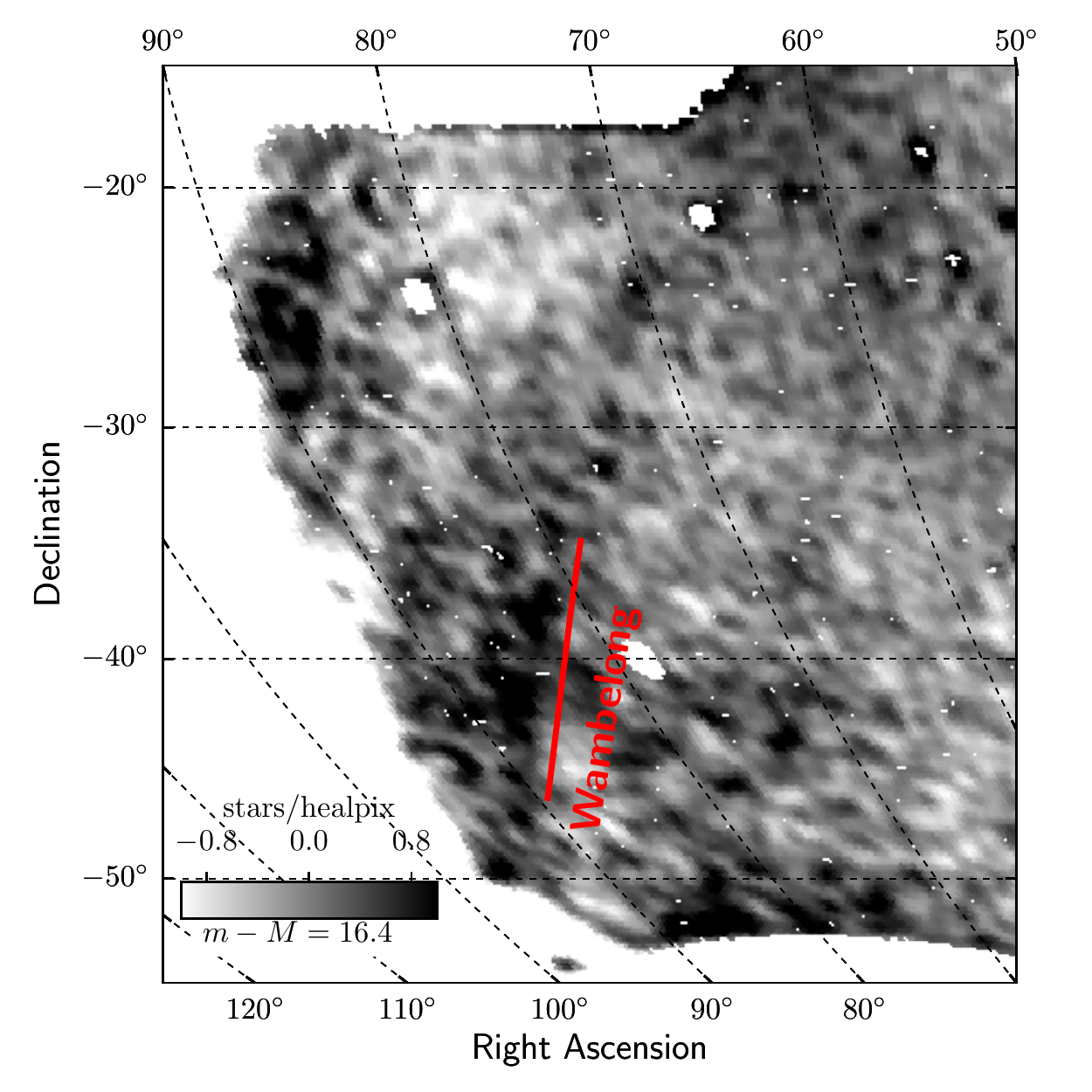}
  \includegraphics[width=0.32\textwidth]{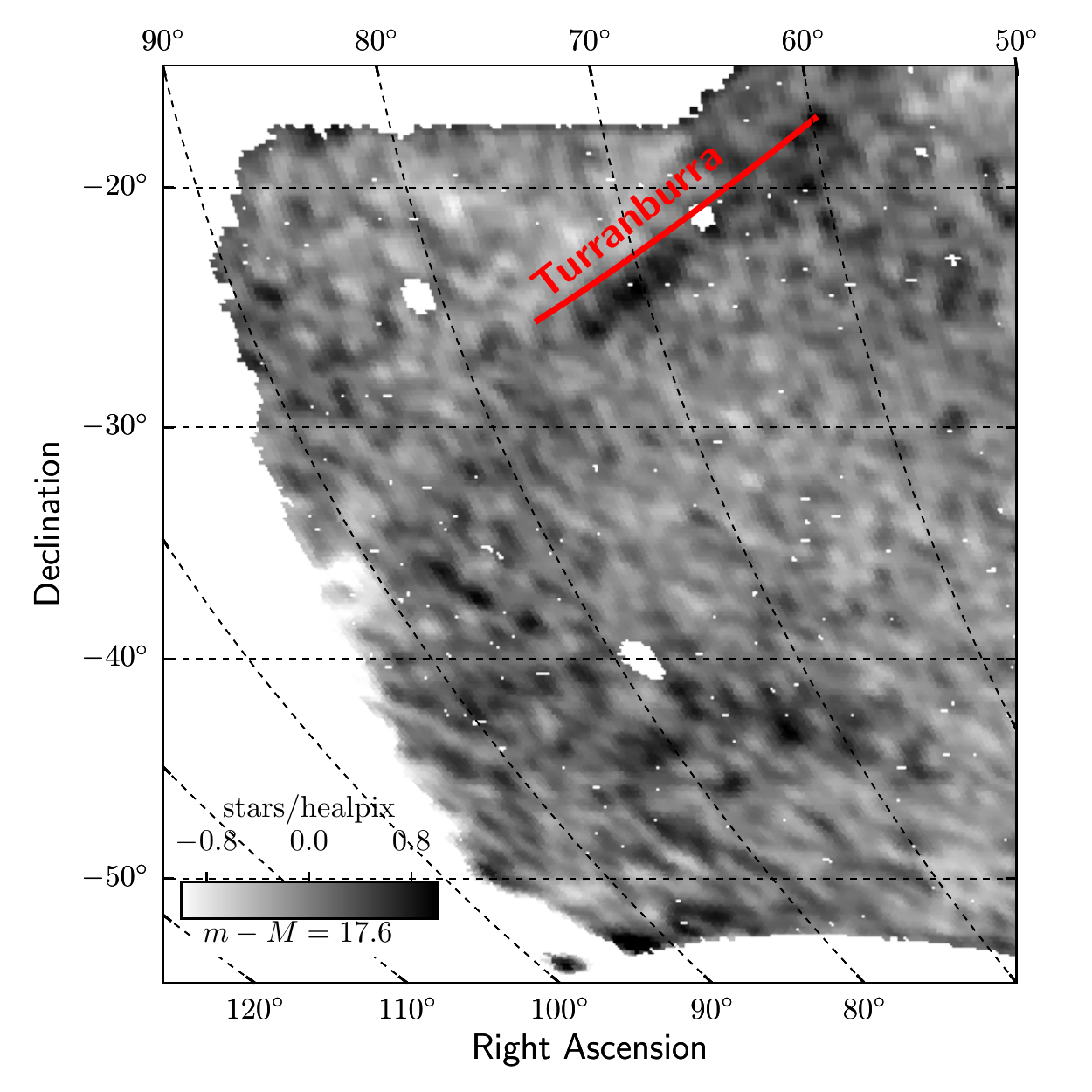}
  \includegraphics[width=0.32\textwidth]{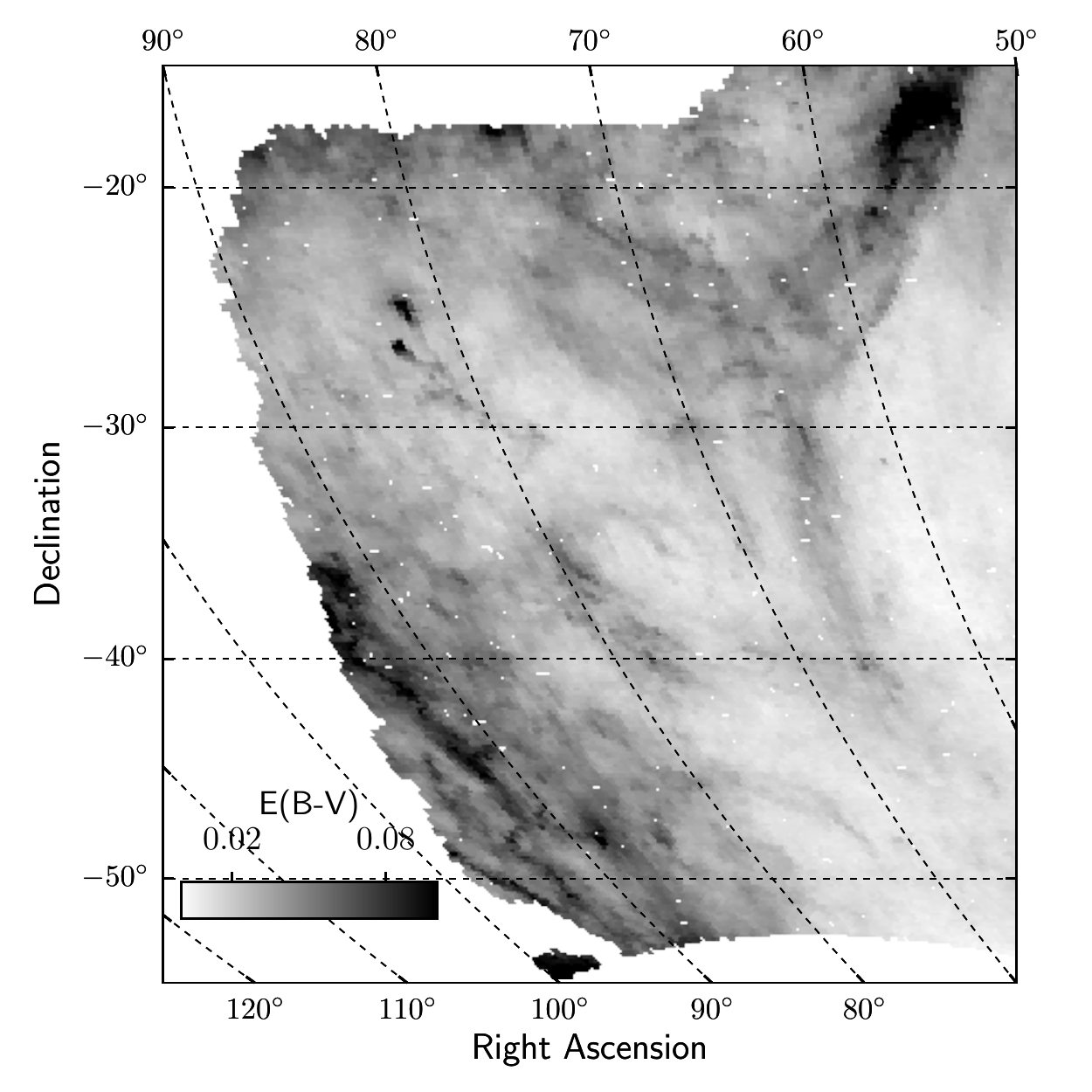}
  \caption{Residual density maps in Q4 (similar to \figref{q1}). 
    (Left) Residual stellar density at $\mM = 16.4$ showing the Wambelong stream. 
    (Middle) The Turranburra stream can be seen in a selection with $\mM = 17.6$.
    (Right) Interstellar reddening,  $E(B-V)$, from \citet{Schlegel:1998}.
    An animated version of this figure can be found online \href{http://home.fnal.gov/~kadrlica/movies/residual_q4_v17p2_label.gif}{at this url}.
}
  \label{fig:q4}
\end{figure*}

\begin{figure}
  \centering
  \includegraphics[width=0.49\columnwidth]{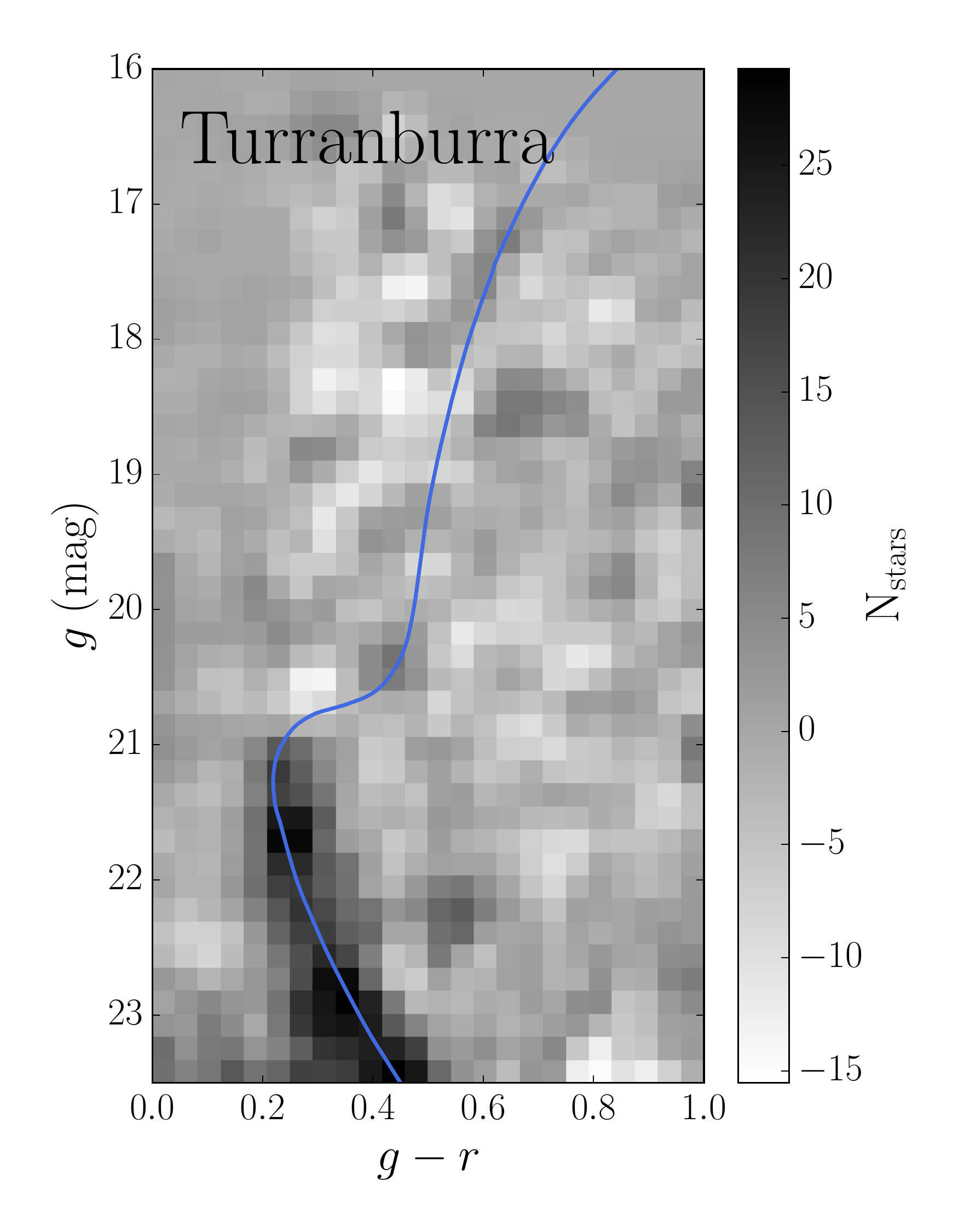}
  \includegraphics[width=0.49\columnwidth]{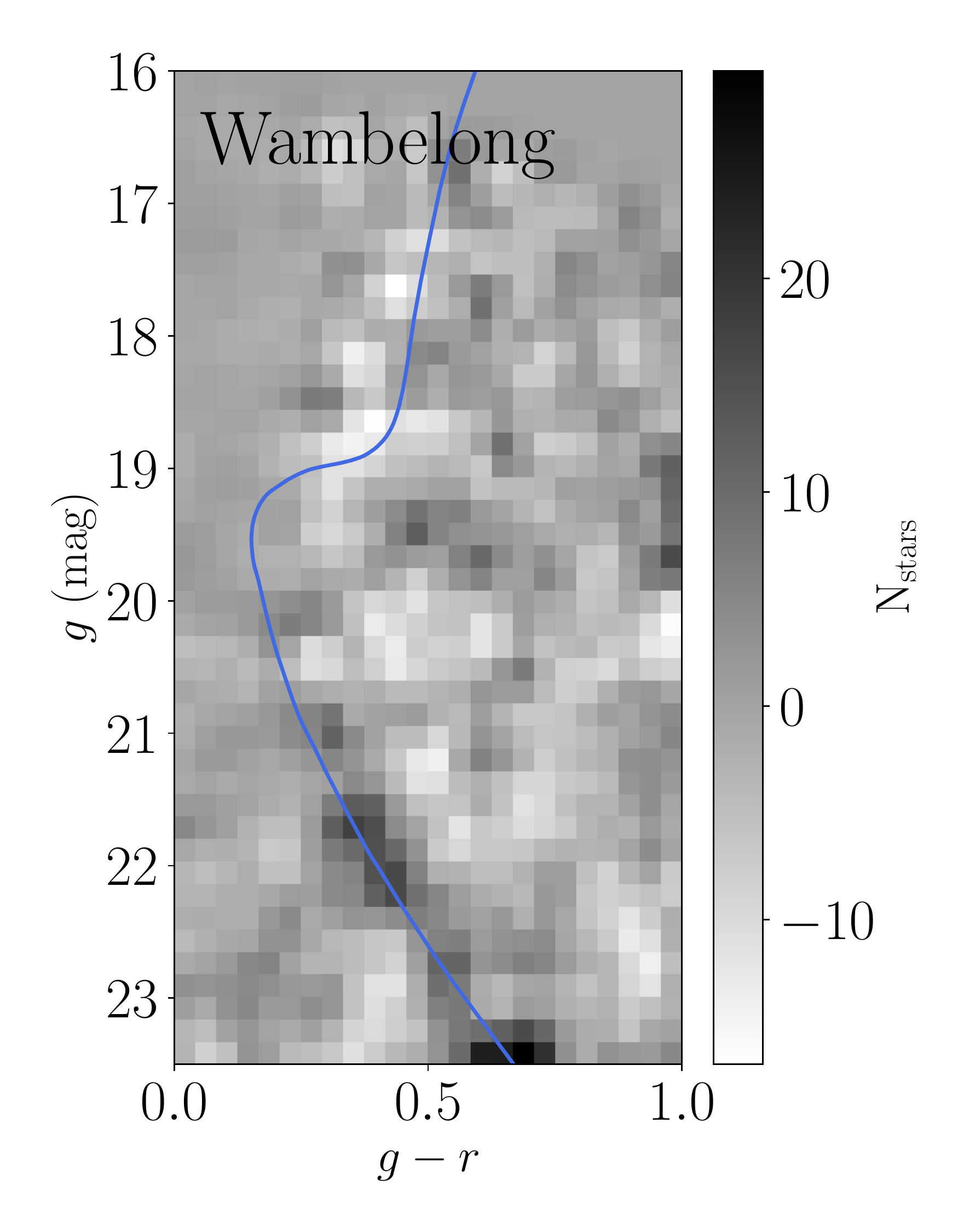}
  \caption{Hess diagrams for the Turranburra (left) and Wambelong (right) stream candidates found in DES quadrant four (Q4). 
Over-plotted in each panel is a \citet{Dotter:2008} synthetic stellar isochrone with parameters determined from the best-fit in \tabref{params}.
These panels are similar to \figref{tucana}.
  \label{fig:q4_hess}}
\end{figure}

\subsubsection{Turranburra Stream}
\label{sec:eridanus}

The Turranburra stream stretches across the northern portion of Q4, extending from \CHECK{$(\ra,\dec) = (59\fdg3, -18\fdg0)$ to $(\ra,\dec) = (75\fdg2, -26\fdg4)$}.
The Hess diagram of this stream shows a prominent MS and a hint of an RGB (\figref{q4_hess}).
We estimate that this structure resides at a distance of \CHECK{$27.5\kpc$} and has a width of \CHECK{$288 \pc$}.
The physical width of Turranburra is more consistent with a dwarf galaxy progenitor, and the photometric estimate of metallicity of $Z = 0.0003$ is comparable to photometric metallicities determined for ultra-faint galaxies \citep[\eg,][]{Bechtol:2015}. 
The extent of Turranburra is somewhat uncertain due to its proximity to the edge of the DES footprint and its diffuse nature, especially at its north western end.
The orbit of Turranburra appears to be distinct from other known streams and globular clusters, without any obvious association or potential progenitor.

\subsubsection{Wambelong Stream}
\label{sec:wambelong}

The Wambelong stream stretches northward from the eastern edge of the DES footprint, spanning \CHECK{$\roughly 14\degrees$} from \CHECK{$(\ra, \dec) \sim (90\fdg5, -45\fdg6)$ to $(\ra, \dec) \sim (79\fdg3, -34\fdg3)$}.
The eastern extent of Wambelong is difficult to determine due to confusion with foreground stars associated with the Galactic anticenter and the Monoceros Ring \citep{Newberg:2002,Yanny:2003}.
The signature of Wambelong peaks at a heliocentric distance of \CHECK{$\roughly 15.1\kpc$ ($\mM \sim 15.9$)} with no strong indication of a distance gradient.
A residual overdensity at $(\ra,\dec) \sim (70\fdg4,-23\fdg9)$ is aligned with the Wambelong stream and may suggest that this stream is nearly twice as long as our conservative estimate.

\subsection{Diffuse Overdensities}
\label{sec:palca}

While the search described here is optimized for $\lesssim 1\degree$-wide stellar features, we note that we are sensitive to more diffuse stellar systems.
Without the directionality of a narrow stream, interpreting the origin and physical parameters of these diffuse structures becomes much more difficult.
However, we do note that the DES Y3A2 data unambiguously confirms the existence of the EriPhe overdensity \citep{Li:2016}, and shows strong indications of additional substructure around or within this system.
Apart from the aforementioned Phoenix, Turbio, and Willka Yaku streams, there are several lower significance linear structures that overlap EriPhe at a heliocentric distances of \CHECK{$\roughly 16.6\kpc$ ($\mM = 16.1$)}.
One pronounced feature runs nearly east-west with $\delta \approx -56\degrees$ (\figref{q3}). 
Unfortunately, image-level masking around the super-saturated star Achernar ($\ra,\dec = 24\fdg43,-57\fdg24$) complicates the interpretation of the southern portion of EriPhe.

\begin{figure*}
  \centering
  \includegraphics[width=0.75\textwidth]{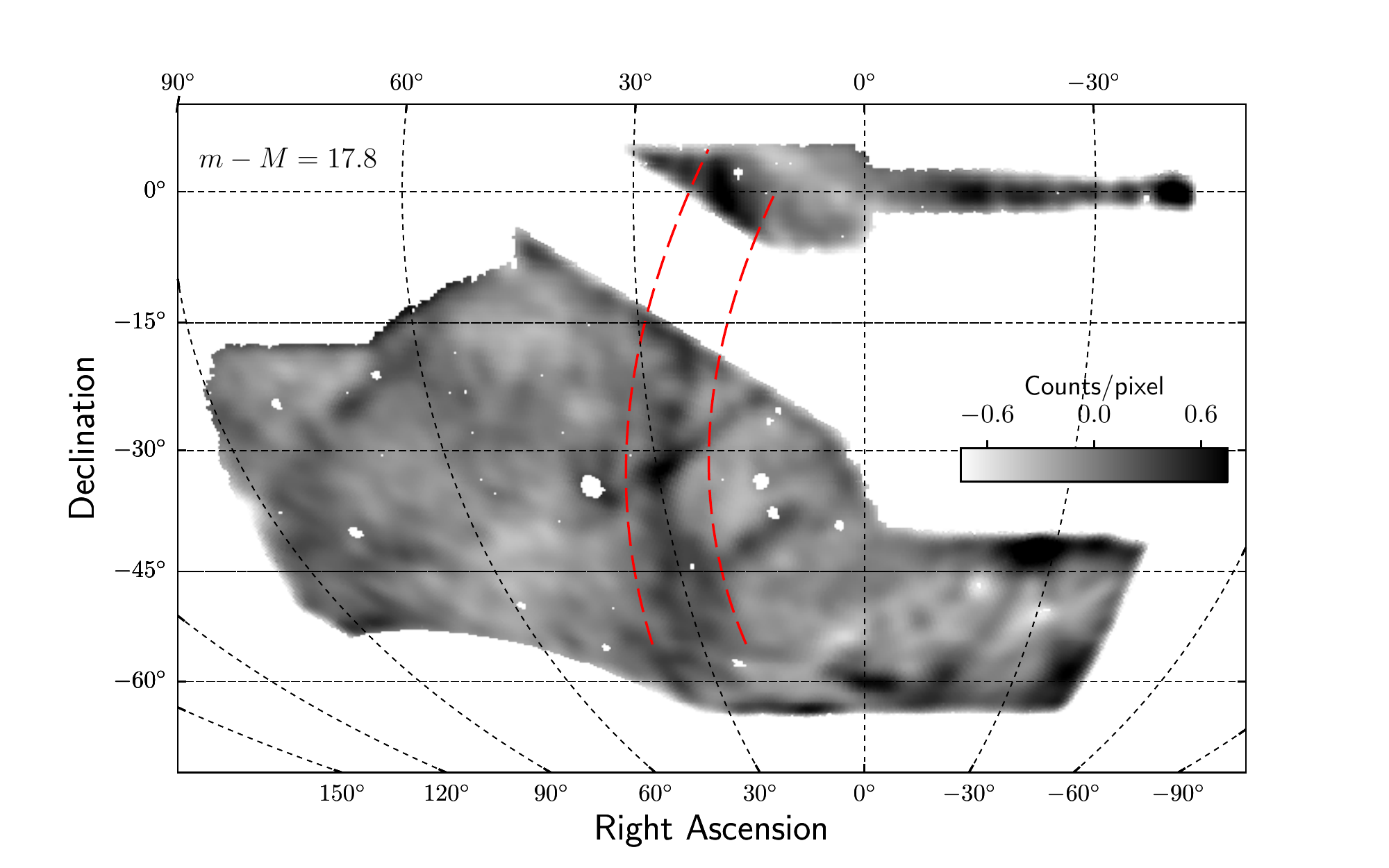}
  \caption{Residual density map for an isochrone selection at $\mM = 17.8$ smoothed by a 1\degree Gaussian kernel. 
The Palca overdensity (bracketed in red) runs along the north-south direction at $\ra \sim 30\degrees$. 
At low declination Palca overlaps the EriPhe stellar overdensity with roughly the same orientation as the Turbio stream.
The ATLAS, Elqui, Tucana III, Chenab, Willka Yaku, and Turranburra streams are also visible in this map.
 \label{fig:palca}
  }
\end{figure*}

The Palca \footnote{Palca is the Quechua word for "cross of rivers."} stream is a broad curvilinear overdensity in the Y3A2 data that extends northward from EriPhe at $\ra \sim 30\degrees$.
Palca extends from the Turbio stream in the south to the northern boundary of the DES footprint, crossing the Sagittarius stream at $\ra \sim 20\degrees$.
Palca is more diffuse than the other streams discovered in DES Y3A2 (FWHM $\roughly 2\degrees$).
To increase our sensitivity to this broad feature, we convolve the residual density maps by a 1\degrees kernel and plot in celestial coordinates (\figref{palca}).
We find that the northern part of Palca is at a significantly larger distance, $\roughly 36\kpc$ ($\mM \sim 17.8$), than EriPhe.
It is unclear whether this is a signature of a distance gradient from EriPhe to Palca, or whether these are two distinct systems.
Assuming that Palca spans the DES footprint (i.e., overlapping with EriPhe), the ridgeline of this structure can be approximated with a second-order polynomial of the form:
\begin{equation}
\ra = 17.277 - 0.495 \dec -0.0046 \dec^2.
\end{equation}
The above equation is found to be valid for $-55\degrees < \dec < 2\fdg5$.
Palca shows appreciable curvature on the sky, suggesting that it may be in a modestly elliptical orbit.
The orbital trajectory of Palca is strongly misaligned with the proposed polar orbit connecting EriPhe with the Virgo Overdensity and the Hercules-Aquila cloud \citep{Li:2016}. 
A physical association between Palca and EriPhe would disfavor this hypothesis.
However, the orbit of Palca may be broadly consistent with another scenario proposed by \citet{Li:2016} suggesting that EriPhe may be the remnants of a disrupted dwarf galaxy originally associated with NGC 1261 and the Phoenix stream.

\subsection{Globular Clusters}

\begin{figure*}[t]
\centering
  \includegraphics[width=0.24\textwidth]{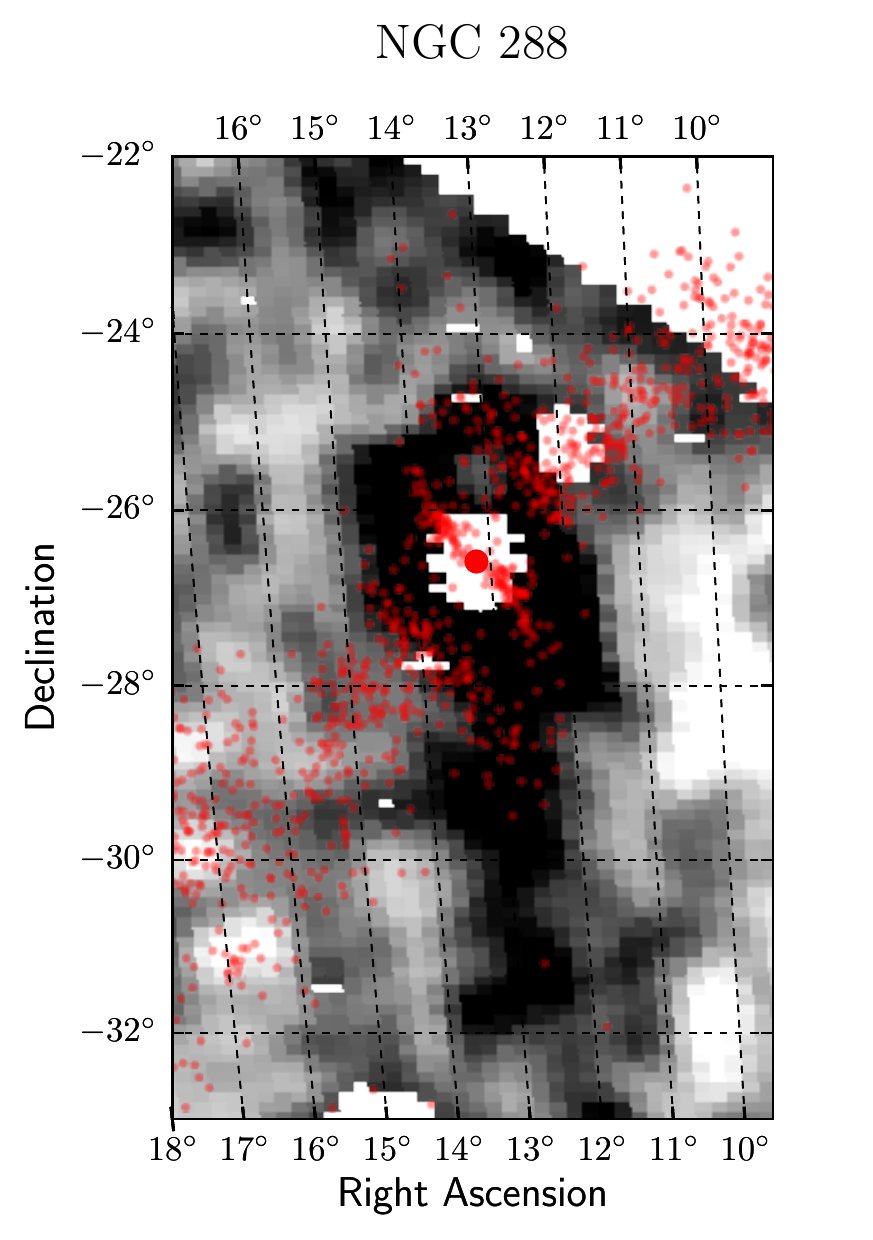}
  \includegraphics[width=0.24\textwidth]{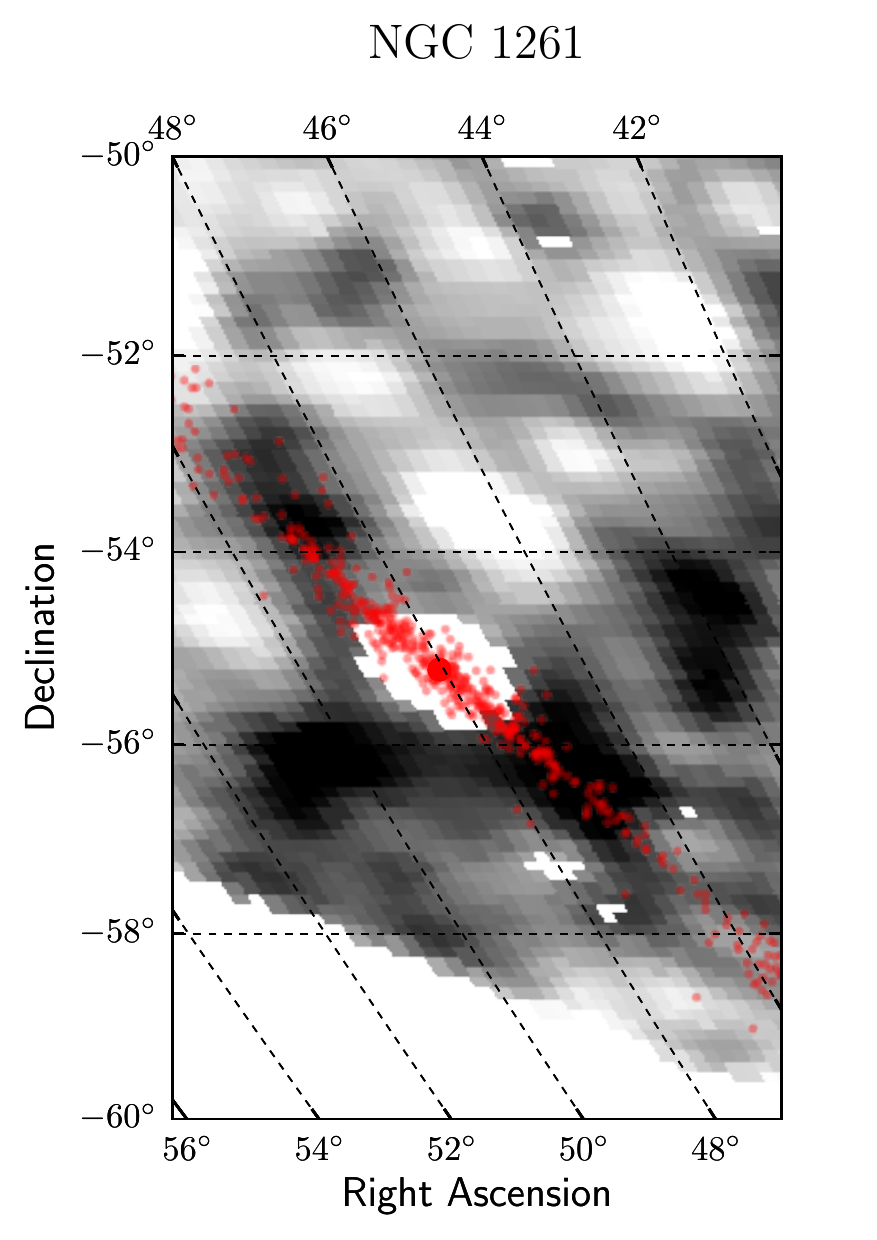}
  \includegraphics[width=0.24\textwidth]{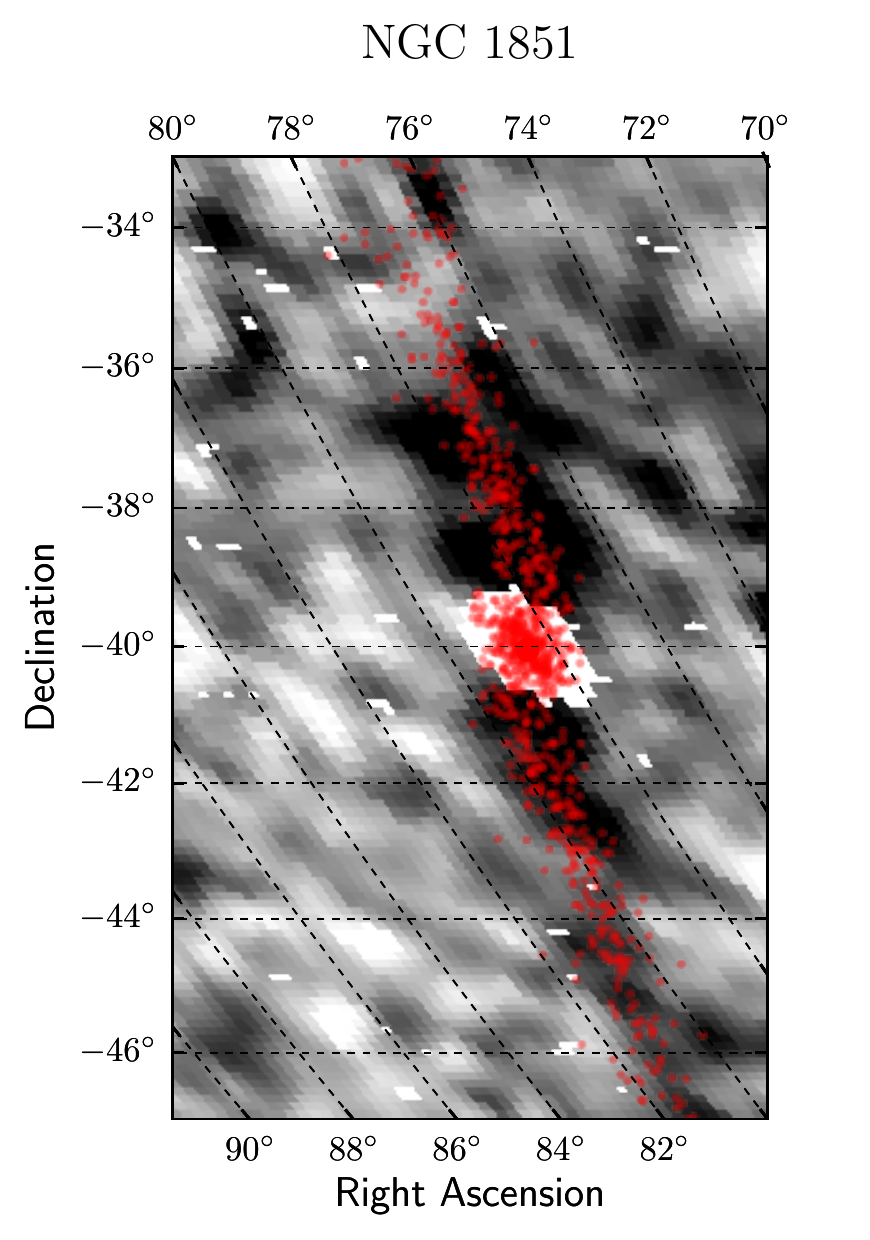}
  \includegraphics[width=0.24\textwidth]{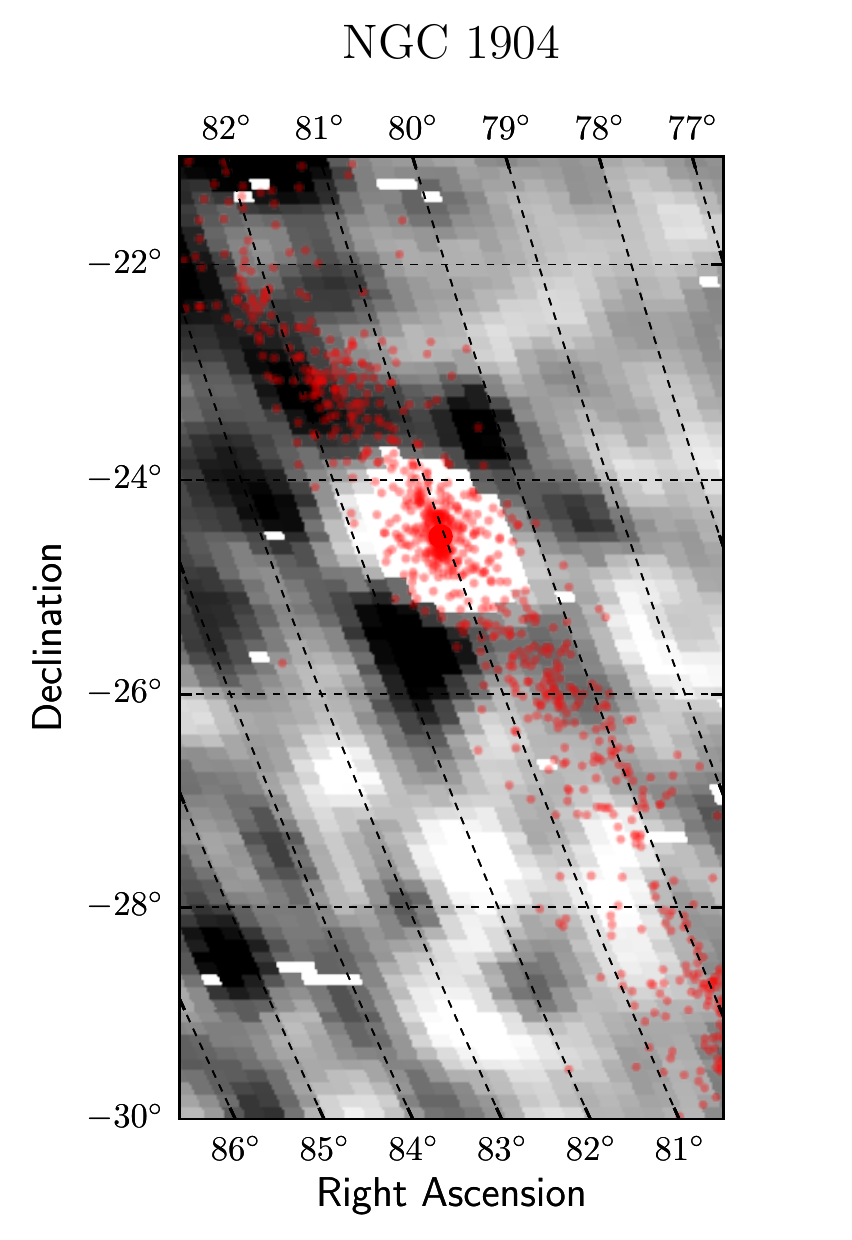}
  \caption{
Distribution of matched-filtered stars selected around 4 globular clusters (gray-scale image). 
Stellar counts within the Jacobi radii of each cluster have been masked to show extra-tidal features.
Star particles simulated using the spray-particle implementation of \citet{Kuepper:2012} are shown in red (see text for details).
Extra-tidal features in NGC 1261 and NGC 1851 appear aligned with the simulated orbit; however, this is not the case for the other two systems.
}
\label{fig:globulars}
\end{figure*}

The DES Y3A2 footprint contains five classical globular clusters (NGC 288, NGC 1261, NGC 1851, NGC 1904, and NGC 7089) and four more distant clusters (Whiting 1, AM-1, Eridanus, and Reticulum).
While a full investigation of globular clusters is outside the scope of the current paper, we note that our analysis is sensitive to stellar features around these clusters.
Four of the classical globular classical clusters (NGC 288, NGC 1261, NGC 1851, and NGC 1904) show hints of extended stellar structure.\footnote{NGC 7089 (M2) is located in the narrow Stripe 82 region of the DES footprint. The narrow width of this region and the large density of foreground stars make a search for extended structure challenging.}
These features are detectable with the generic isochrone selection described in \secref{analysis} and can be seen in the animations associated with \figref{isosel2}.
However, to optimize our sensitivity to faint features, we built individual matched-filter selections for each cluster using the CMD of stars within an annulus of $4\farcm2 < r < 7\farcm2$ around each cluster.
We create an optimal weighting by taking the ratio between the density (in color-magnitude space) of cluster member stars compared to the Milky Way foreground population averaged over the DES footprint, 
\begin{equation}
w_{i,j} = f_{\rm gc}(i,j)/f_{\rm mw}(i,j)
\end{equation}
where $i,j$ index the color and magnitude bins, $f_{\rm gc}(i,j)$ is the normalized density of cluster stars per bin, $f_{\rm mw}(i,j)$ is the normalized density of Milky Way stars, and $w_{i,j}$ is the weighting \citep{Rockosi:2002}.
We mask circular regions comparable to the Jacobi radii of each cluster (\tabref{globulars}) and convolve the selected stellar density with a Gaussian kernel with $\sigma = 0\fdg25$.
We follow the same procedure to derive a global polynomial fit to the smoothed density of selected stars and create residual density maps from the difference between the data and the polynomial fit (\figref{globulars}).

We compare the observed stellar features to predictions about orbital motion and tidal tail formation in each globular cluster.
We simulated the orbits of the globular clusters using the spray-particle implementation by \citet{Kuepper:2012}, where the  escape velocity was modified to match that observed in $N$-body simulations with realistic tidal fields \citep{Claydon:2016}.
We assume a Milky Way potential similar to the best-fit Palomar 5 model \citep{Kuepper:2015}, but with a Jaffe bulge. 
The cluster initial mass is obtained using the method outlined in \citet{Balbinot:2018}.  
The tidal tail formation was simulated for the last 6\Gyr of the cluster history and particles were released every 1\Myr.

The heliocentric distance, sky position, line-of-sight velocity, and integrated magnitude for each cluster was taken from \citet[updated 2010][]{Harris:1996}.
Proper motions for NGC 288 and NGC 1851 are taken from \citet{Dinescu:1997}, for NGC 1904 we used values from \citet{Dinescu:1999}, and for NGC 1261 we used values from \citet{Dambis:2006}.
These parameters are summarized in \tabref{globulars}.
Our simulations assumed a Galactocentric solar position of $(8.3\kpc, 0, 0)$, a local reflex motion of $U,V,W = (11.1, 12.24, 7.25) \kms$ \citep{Schonrich:2010}, and a circular solar velocity of $U,V,W = (0, 233, 0) \kms$ \citep{Kuepper:2015}.

\begin{deluxetable}{l ccccccc }
\tablecolumns{13}
\tablewidth{0pt}
\tabletypesize{\scriptsize}
\tablecaption{ Globular Cluster Parameters \label{tab:globulars}}
\tablehead{
Name & \ra & \dec & $\Dsun$ & $r_{\rm J}$ & $\mu_{\alpha} \cos(\delta)$ & $\mu_{\delta}$ \\
 & (deg) & (deg) & (kpc)    & (pc)     & ($\mas/\yr$) & ($\mas/\yr$) }
\startdata
NGC 288  & 13.189 & -26.583 & 76.4  & 8.9  & 4.48 & -6.04 \\
NGC 1261 & 48.068 & -55.216 & 146.4 & 16.3 & 1.33 & -3.06 \\
NGC 1851 & 78.528 & -40.047 & 166.5 & 12.1 & 1.29 &  2.38 \\
NGC 1904 & 81.046 & -24.525 & 153.8 & 8.9  & 2.12 & -0.02 \\
\enddata
{\footnotesize \tablecomments{Centroids, heliocentric distances, Jacobi radii, and proper motions for four classical globular clusters in the DES footprint. Values taken from: \citet[][2010 edition]{Harris:1996}, \citet{Balbinot:2018}, \citet{Dinescu:1997}, \citet{Dambis:2006}, \citet{Dinescu:1999} }}
\end{deluxetable}


\subsubsection{NGC 288}
\label{sec:ngc288}

NGC 288 is a globular cluster with a dynamical evolution that is strongly driven by tidal shocks \citep{Gnedin:1997}.
\citet{Grillmair:1995} showed initial evidence for extra-tidal features using photographic photometry. 
The analysis of NGC 288 was extended to a larger field by \citet{Leon:2000}, who reported evidence of two sets of tidal tails, extended along the direction of motion and in the direction of the Galactic center.
Subsequently, \citet{Grillmair:2004} used 2MASS data to suggest a $\roughly 17\degrees$ tidal tail; however, this claim was later refuted by \citet{Piatti:2018} using deeper data from Pan-STARRS PS1.
In contrast, \citet{Piatti:2018} found evidence for clumpy extra-tidal structure extending $120\pc$ ($0\fdg8$) from the cluster center.
The DES Y3A2 data is deeper than the Pan-STARRS data analzyed by \citet{Piatti:2018} and supports previous reports of clumpy extra-tidal structure extending $\roughly 1\fdg5$ from the core of NGC 288.
In addition, there is evidence that these extra tidal features may extend $\roughly 5\fdg5$ southward of NGC 288 (\figref{globulars}).
This structure is misaligned with the orbital motion of NGC 288 and the vector connecting NGC 288 to the Galactic center.

\subsubsection{NGC 1261}
\label{sec:ngc1261}

NGC 1261 resides in the southern portion of the DES footprint and it has been suggested that it may be associated with the Pheonix stream \citep{Balbinot:2016} and/or the EriPhe stellar overdensity \citep{Li:2016}.
The analysis of \citet{Leon:2000} suggests the existence of a tidal tail oriented in the direction of the Galactic center.
Recent observations with DECam have similarly detected evidence that the stellar halo of NGC 1261 extends beyond its nominal Wilson tidal radius \citep{Kuzma:2018,Carballo-Bello:2018}, but do not see any evidence of tail-like structure.
Our observations support the existence of extra-tidal structure around NGC 1261. 
While some of this structure appears to be aligned with the orbital motion of the cluster, it is difficult to draw any firm conclusion without more detailed analysis.

\subsubsection{NGC 1851}
\label{sec:ngc1851}

Tidal tails were reported around NGC 1851 by \citet{Leon:2000}, who claimed to detect low-surface-brightness features oriented with the direction of motion of the cluster.
However, \citet{Olszewski:2009} found no evidence of tidal tails in a more recent analysis of deeper data.
Rather, \citet{Olszewski:2009} reported a low-surface-brightness extended stellar halo extending to \roughly 1\fdg25.
Our residual density maps shows evidence of both an extended stellar halo around NGC 1851, and a set of faint linear features aligned with the predicted orbit (\figref{globulars}).
The residual stellar density extends prominently to a radius $\gtr 1 \degree$ from the cluster core, agreeing with measurements of an extended stellar halo by \citet{Olszewski:2009}.
The linear feature extends at least 5\degree to the north and south of NGC 1851.
The orientation of these features are well-aligned with the orbital motion of NGC 1851, suggesting that these features may be tidal tails. 
In fact, the annimation associated with \figref{isosel2} suggests that these putative tidal tails may extend $\pm 15\degrees$ or more from NGC 1851.
The detection of extra-tidal structure associated with NGC 1851 agrees with recent work by \citet{Kuzma:2018} and \citet{Carballo-Bello:2018}.
The DES data greatly extend the coverage around NGC 1851, and make a strong case for a ``vast stellar structure'' \citep{Carballo-Bello:2018} extending both northward and southward of this cluster.

\subsubsection{NGC 1904}
\label{sec:ngc1904}

An extended halo of extra-tidal stars around NGC 1904 was first recognized by \citet{Grillmair:1995} and later by \citet{Leon:2000}. 
We confirm the existence of extra-tidal structure extending $\roughly 1\fdg5$ from the cluster center.
\citet{Leon:2000} suggest that the short relaxation time of NGC 1904 would cause mass segregation in the tidal tails.
This may explain why these features were not seen with shallower observations.
Interestingly, these structures appear to be symmetric, but are misaligned with the orbital motion of the cluster.
Observations of NGC 1904 by \citet{Carballo-Bello:2018} reach a similar conclusion that the stellar distribution of this cluster deviates from the conventional King and Wilson models to fill, and slightly overflow, the Jacobi radius of the cluster.

\section{Discussion}
\label{sec:discussion}

\subsection{Potential Associations with Known Systems}
\label{sec:assocs}
\begin{figure*}[t]
\centering
\includegraphics[width=0.49\textwidth]{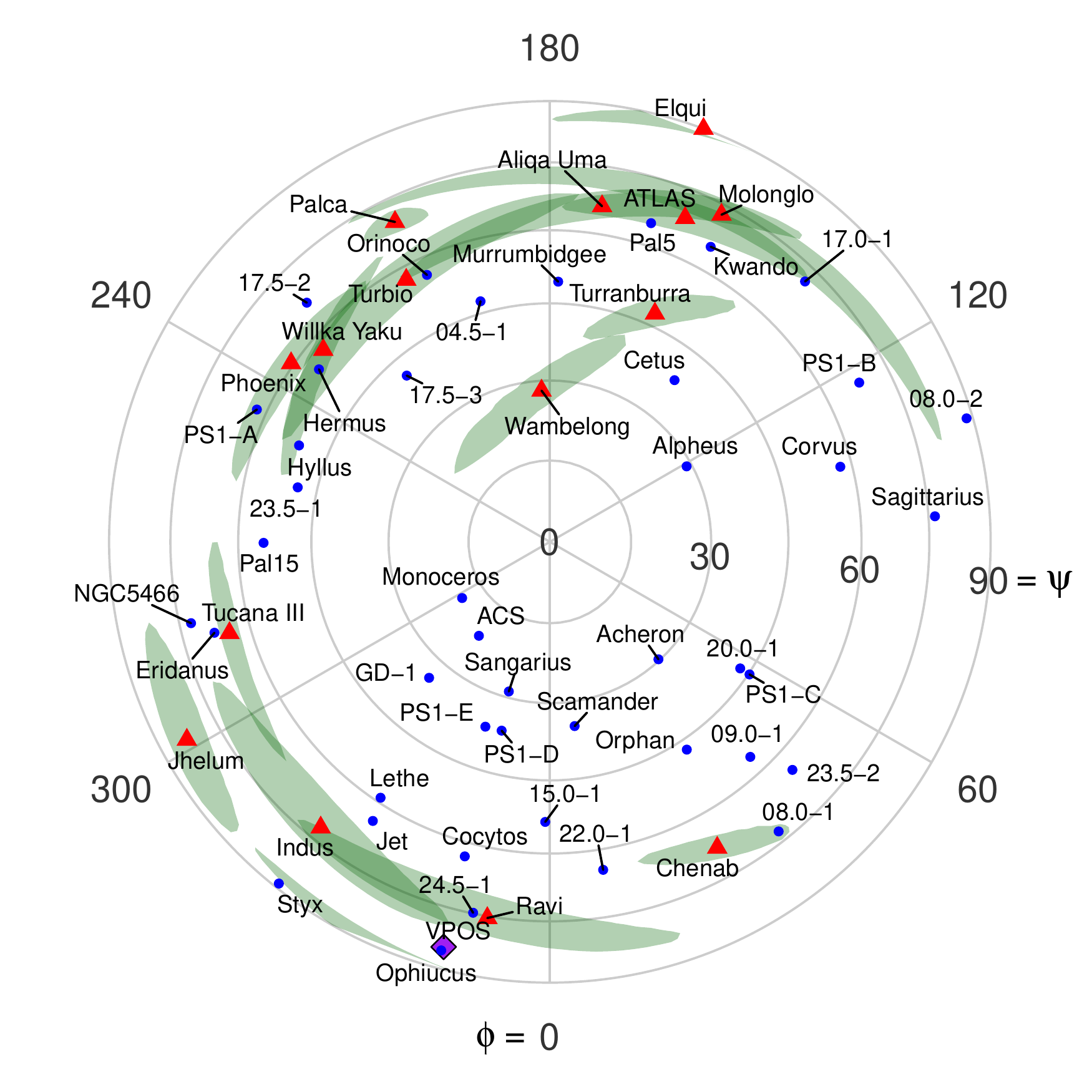}
\includegraphics[width=0.49\textwidth]{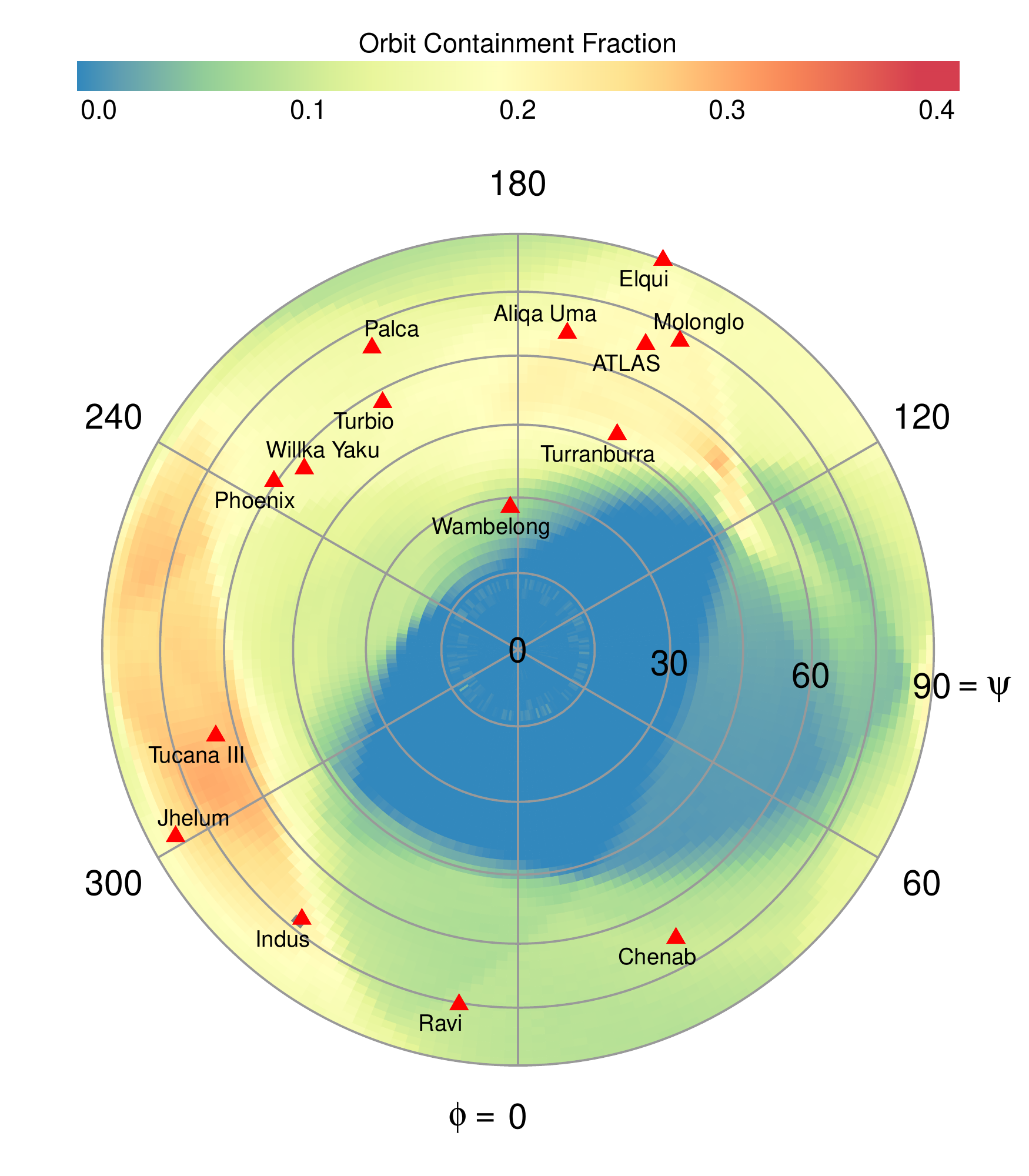}
\caption{
(Left) Galactocentric orbital poles for DES stream candidates (red triangles) and previously known streams \citep[blue circles;][]{Mateu:2017,Jethwa:2017}.
Green shaded regions represent the $1\sigma$ uncertainty ellipses for the poles of the DES streams assuming a distance uncertainty of 20\%.
The purple diamond shows the mean orbital pole of the proposed vast polar structure \citep{Pawlowski:2015}.
(Right) The fraction of uniformly distributed Galactocentric great-circle orbits with $R_{GC}= 25 \kpc$ that are contained within the DES footprint.
The DES footprint imposes a geometric bias against the detection of streams with $-60\degrees < \phi_{\rm pole} < 120\degrees$ and $\psi_{\rm pole} \lesssim 30\degrees$.
}
\label{fig:stream_poles}
\end{figure*}

We use the recent catalog of stellar streams compiled by \citet{Mateu:2017}, augmented with the recently discovered Jet stream \citep{Jethwa:2017} to assess whether any of our stream candidates may be associated with previously detected streams located in other regions of the sky. 
We begin by transforming the endpoints of each stream into Galactocentric Cartesian coordinates. 
The ellipticity of these streams is poorly constrained so we assume only that the streams orbit in a plane around the Galactic Center. 
We find the pole of each stream, $(\phi,\psi)$, defined as the positive normal vector of a plane containing both endpoints.
Uncertainties in the pole location for our newly found streams were estimated by assuming a $20\%$ uncertainty on the heliocentric distance before converting to Galactocentric coordinates. 
The poles for the new and previously discovered streams are shown in \figref{stream_poles}.

The distribution of DES stream poles shown in the left panel of \figref{stream_poles} is clearly non-uniform.
We do not find any strong association of stream poles coinciding with the proposed Vast Polar Structure \citep[VPOS;][]{Pawlowski:2012,Pawlowski:2015}.
Specifically, in the left panel of \figref{stream_poles} we plot the VPOS+new pole  \citep[Table 1 of ][]{Pawlowski:2015} transformed into Galactocentric coordinates assuming a distance of $100\kpc$.
However, the limited sky coverage of the DES footprint will bias the observable distribution of stream poles.
To estimate this bias, we generate a uniform random sample of Galactocenric great-circle orbits with a radius of 25 kpc. 
We calculate the fraction of each great-circle orbit contained within the DES footprint as a function of Galactocentric orbital pole and show this in the right panel of \figref{stream_poles}.  
We find that the observed distribution of stream poles is consistent with the predictions from our simple simulation.
The DES footprint is clearly biased against detecting streams having poles with $\psi_{\rm pole} < 15\degrees$ and $240\degrees < \phi_{\rm pole} < 320\degrees$.
We find qualitatively similar results for random samples of orbits with Galactocentric radii of 15 kpc and 50 kpc.

\begin{figure}[t]
\centering
\includegraphics[width=0.95\columnwidth]{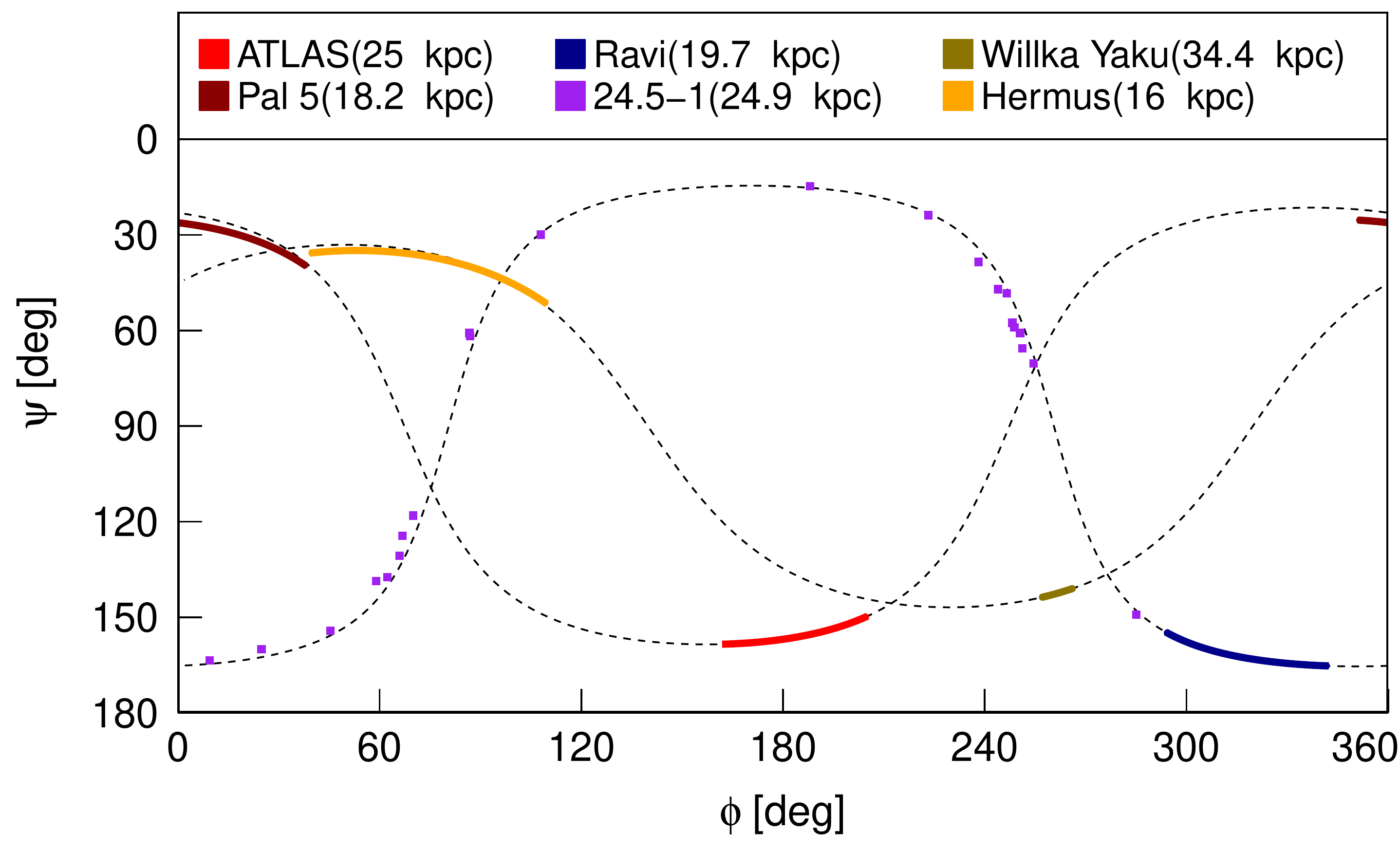}
\caption{ Great circle orbital models for 3 of the closest associations between DES and previously detected streams are plotted in Galactocentric coordinates ($\phi,\psi$). The dotted lines are great circle orbits derived from the DES stream endpoints. The solid colored lines show the locations of the streams themselves, the purple squares show the RR Lyrae \citep{Mateu:2017} that make up the stream 24.5-1.
}
\label{fig:stream_assoc}
\end{figure}

To investigate potential associations for the new DES streams, we plot Galactocentric great-circle orbits for the DES streams and other known streams with similar orbital poles (\figref{stream_assoc}). 
Full phase-space information is necessary to definitively match between streams systems; however, we do note several tentative associations based on the photometrically measured properties of the DES streams.
\figref{stream_assoc} shows a strong correspondence between Ravi and the tentative candidate RR Lyrae stream 24.5-1 \citep{Mateu:2017}.
The poles of these two stream candidates match within \CHECK{$3\degrees$} while their distance moduli differ by \CHECK{$\Delta(\mM) \sim 0.2 \magn$} (well within the systematic error associated with our isochrone fitting).
Such an association supports the robustness of tentative candidates identified below the conservative $>4\sigma$ significance threshold of \citet{Mateu:2017}.

The orbital pole of the Hermus stream \citep{Grillmair:2014} is only $3\fdg4$ from that of Willka Yaku, but the Galactocentric distances of the two streams differ by $\roughly 16 \kpc$.
It has previously been suggested that Hermus may be a northern extension of the Phoenix stream \citep{Grillmair:2016b}, which resides along a slightly different orbital plane, but is well-matched in distance.
Kinematic information would help to resolve this ambiguity.

Globular clusters and dwarf galaxies may provide possible progenitors for the newly discovered streams.
The globular cluster candidate with the smallest separation is IC 4499. 
It is within $1\fdg2$ of the great circle orbit of Turbio, and has a Galactocentric radius that differs by less than $2 \kpc$. IC 4499 is a moderate-mass, low-density cluster with signatures of an extra-tidal stellar halo \citep{Walker:2011}.
Cetus II, an ultra-faint dwarf galaxy \citep{Drlica-Wagner:2015}, is another candidate for association. It lies $1\fdg7$ off of the great circle orbit of Ailqa Uma, has a Galactocentric radius that is $1.1 \kpc$ larger, and is located $43 \degrees$ from the nearest endpoint of the stream. 
We note also that the tidal tails of Palomar 5 ($\Rgc = 18.2 \kpc$) have an orbital pole within $6\degrees$ of the ATLAS stream ($25 \kpc$). 

\citet{Agnello:2017} identified four stellar stream candidates overlapping the DES footprint using a WISE-Gaia multiple search.
It was proposed that two of these streams, WG3 and WG4, may have associated counterparts in DES \citep{Agnello:2017}. 
While WG3 and WG4 appear qualitatively similar to Indus and Jhelum, their absolute coordinates are offset by $\Delta \ra \sim -20\degrees$ (an angular separation of $\roughly 15\degrees$).
We do not see any significant stellar overdensities associated with the positions of WG3 or WG4 reported by \citet{Agnello:2017}, though their proximity to the Galactic plane and unknown distance makes it difficult to quantify the lack of a DES counterpart.
If we allow $\Delta \ra$ offsets of $\roughly 10\degree$, then we find a possible correspondence between WG1 and Wambelong.
WG1 is offset from Wambelong by $\Delta \ra \sim -8\degrees$ (angular separation of $\roughly 6 \degrees$) and extends both northeast and southwest along a similar path.
This observation provides circumstantial evidence in support of the longer extent of Wambelong proposed in \secref{wambelong}.
WG2 does not correspond to any of the high-significance stream candidates reported here.
However, WG2 appears qualitatively similar to a lower significance feature found to the southwest of NGC 1851 extending from $\ra,\dec = (56.1, -50.4)$ to $(78.5, -40.0)$ at a distance modulus of $\mM \sim 17.5$ (animation of \figref{isosel2}).
We expect Gaia DR2 to greatly improve the power of stellar stream searches using astrometric techniques.

\subsection{Milky Way Gravitational Potential}
\label{sec:flattening}

Stellar streams can be used to constrain the Milky Way gravitational potential \citep[e.g.,][]{Johnston:2005,Koposov:2010,Law:2010,Gibbons:2014,Bowden:2015,Kupper:2015,Bovy:2016}. 
Full potential modeling is beyond the scope of this work; however, we note that the streams discovered by DES span a wide range of Galactocentric radii and should be able to constrain how the Milky Way's density profile and shape evolves with radius. 
In this context, we expect that the ATLAS stream will be especially useful since it is long and does not lie on a great circle (see \figref{atlas}).

\begin{figure}
\includegraphics[width=\columnwidth]{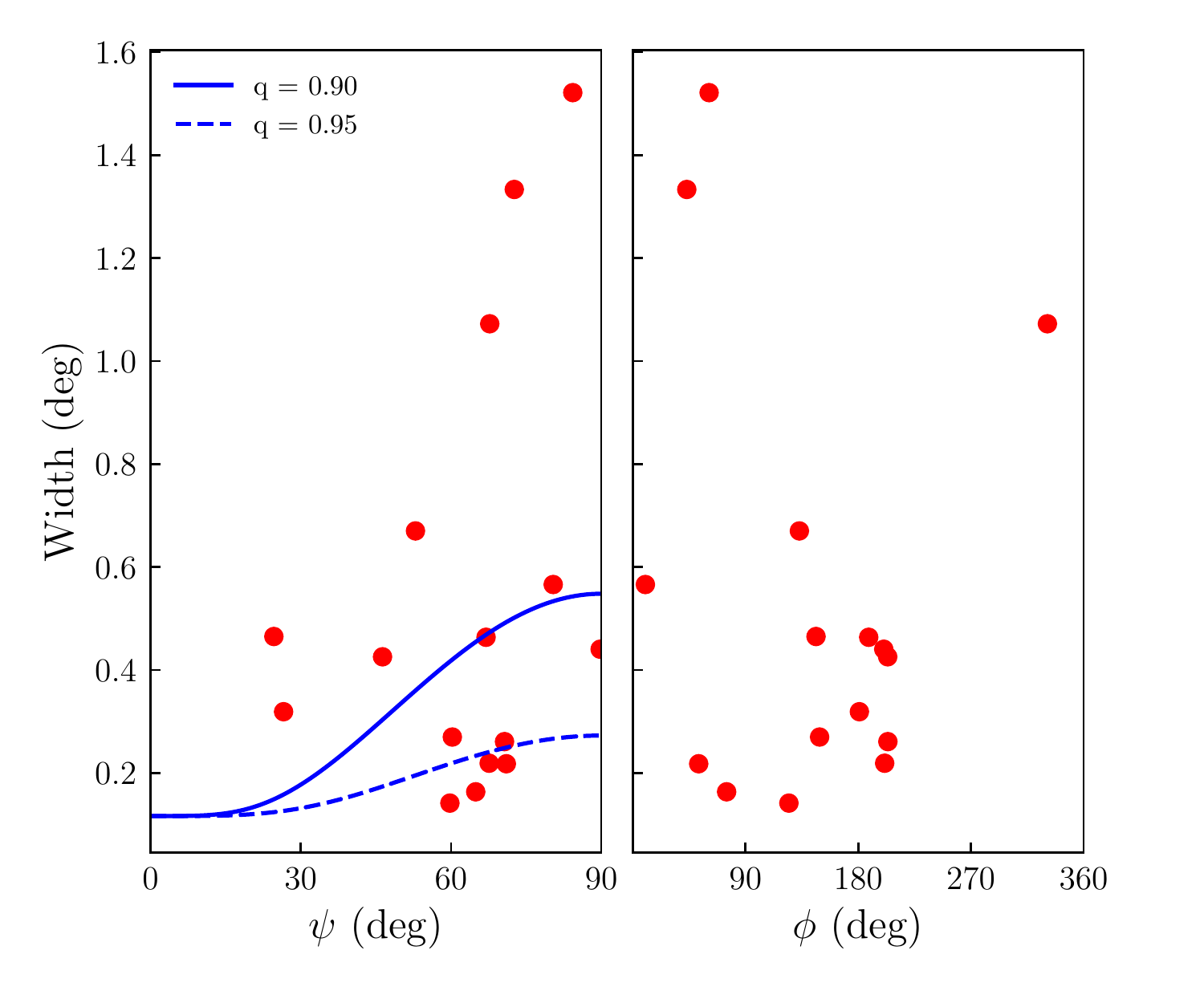}
\caption{Angular widths of stellar streams detected by DES (as perceived by an observer at the Galactic center) as a function of their Galactocentric pole orientation.
The left and right panels show the stream width as a function
of the Galactocentric polar and azimuthal angle, respectively. 
The curves show the expected stream width from \citet{Erkal:2016} for streams with age 4\Gyr in a potential with two different flattenings, $q = 0.9$ or $q = 0.95$.
These curves are produced assuming a progenitor with a mass of $10^{4.5}\Msun$ on an orbit with a pericenter of 15 kpc and an apocenter of 30 kpc in a logarithmic potential with a circular velocity of 220 km/s. 
The bunching of the polar angles of observed streams around low azimuthal angles is caused by the coverage of DES. 
Interestingly, the widest streams (Jhelum, Indus, Ravi, and Chenab) are close to polar orientations.
\label{fig:flattening}
}
\end{figure}

However, even without sophisticated modeling we can make some general observations in the context of the Milky Way potential. 
\citet{Erkal:2016} suggest that the connection between stream width and orbital inclination could provide an independent constraint on the symmetry axis and flattening of the Milky Way halo.
In \figref{flattening}, we plot the angular stream width (as would be observed from the Galactic center) against the Galactocentric polar angle, $\psi$, and the Galactocentric azimuthal angle, $\phi$. This figure shows a large scatter, which is to be expected from a heterogeneous population of progenitors. While it is interesting to note that the streams that appear the widest are also on nearly polar orbits, interpreting any trend in this figure is subject to a number of caveats.

In particular, inferences involving stream widths rely on the assumed mass, structural properties, orbit, and dynamical age of the stream and its progenitor.
Geometric effects can cause debris in the stream plane to contribute to the perceived width as calculated by a heliocentric observer.
Furthermore, the width can fluctuate along a stream due to the existence of nodes between the progenitor plane and the planes of the stream debris.
Nonetheless, it is interesting to note that \figref{flattening} is in general agreement with a model where the Galactic symmetry axis is perpendicular to the plane of the disk.\footnote{We note that Figure 16 in \citet{Erkal:2016} uses an inconsistent convention for the sign of the azimuthal coordinate $\phi$.}

\subsection{Stream progenitors}

In Table \ref{tab:deriv} we provided estimates of the stellar mass and progenitor mass \citep[based on the stream width, see][]{Erkal:2016} of each stream. These can be combined to give a mass-to-light ratio for the progenitor. Doing this, we find that five of the streams (Tucana III, ATLAS, Phoenix, Willka Yaku, and Turbio) have mass-to-light ratios less than $\sim \CHECK{30}$, while the other eight streams have significantly higher mass-to-light ratios. Given this seeming dichotomy, it is possible that the progenitors with low mass-to-light ratios are globular clusters while those with the higher ratios are dwarf galaxies. However, since the mass-to-light ratios are all higher than expected for a globular cluster (Phoenix stream has the lowest ratio of \CHECK{10}) we cannot make any firm 
conclusions. 

These higher than expected mass-to-light ratios could be due to a variety of reasons. As discussed in \ref{sec:isochrone}, the progenitor mass estimate is only approximate and only works on average for streams on eccentric orbits. Furthermore, if the stream widths have fanned out due to evolving in a non-spherical potential \citep[e.g.][]{Erkal:2016}, the inferred progenitor masses will be overestimated. Finally, the stellar masses will be underestimated for streams that are not fully contained within the DES footprint. 

\section{Conclusions}
\label{sec:conclusions}

We searched for Milky Way stellar streams by applying a matched-filter for old, metal-poor stellar populations to three years of data from DES.
The unprecedented photometric calibration, depth, and coverage area of the DES data allow us to detect stellar streams out to a distance of $> 50 \kpc$.
Our analysis recovers \NKNOWN narrow stellar streams previously identified within the DES footprint. 
In addition, we detect \NSTREAMS new stellar stream candidates.
\CHECK{In general, these newly detected streams are wider and lower surface brightness than those detected in previous surveys.}
We find several tentative associations of these new stream candidates with stellar structures detected in other regions of the sky.
In addition, we find evidence for extra-tidal stellar features around \NGLOB classical globular clusters.
The current analysis makes use of three years of DES data. 
We expect that additional DES observations, improved data reduction techniques, and improved stream detection algorithms will allow fainter and more distant streams to be detected in the near future.
While the DES data currently provide the most sensitive wide-area view of the southern sky, they are merely a precursor for larger sky coverage that can be achieved with DECam and, eventually, the Large Synoptic Survey Telescope (LSST).
LSST is expected to find $>100$ stellar streams with sensitivity out to the virial radius of the Milky Way \citep{LSST:2009}.
These wide-area photometric surveys will greatly expand our ability to probe the Milky Way stellar halo, providing unprecedented insights into Galactic archaeology and near-field cosmology.

\section*{Acknowledgments}

We thank the attendees at the ``Science at the Calyx'' event at the Royal Botanic Garden Sydney, as well as the students of St.\ John's School in North Ryde, Sydney, for their enthusiastic contributions to the process of naming our proposed stellar streams.  
We likewise acknowledge Aunty Maureen Sulter of the Gamilaraay nation and Mr.\ Drew Roberts of the Bundjalung nation for their gracious support of, and feedback on, this naming process.

We warmly thank Colegio Antonio Varas in Vicu\~{n}a, Chile, for organizing the search for aquatic terms from the native Quechua and Aymara cultures. In particular, we thank high school students D\'{a}nae Rojas and Emerson Carvajal, their teacher Yeimy Vargas, and dozens of enthusiastic kindergarteners and first graders who selected the Chilean stream names.

NS thanks the LSSTC Data Science Fellowship Program, her time as a Fellow has benefited this work.
EB acknowledges financial support from the European Research Council (ERC-StG-335936).

Funding for the DES Projects has been provided by the U.S. Department of Energy, the U.S. National Science Foundation, the Ministry of Science and Education of Spain, 
the Science and Technology Facilities Council of the United Kingdom, the Higher Education Funding Council for England, the National Center for Supercomputing 
Applications at the University of Illinois at Urbana-Champaign, the Kavli Institute of Cosmological Physics at the University of Chicago, 
the Center for Cosmology and Astro-Particle Physics at the Ohio State University,
the Mitchell Institute for Fundamental Physics and Astronomy at Texas A\&M University, Financiadora de Estudos e Projetos, 
Funda{\c c}{\~a}o Carlos Chagas Filho de Amparo {\`a} Pesquisa do Estado do Rio de Janeiro, Conselho Nacional de Desenvolvimento Cient{\'i}fico e Tecnol{\'o}gico and 
the Minist{\'e}rio da Ci{\^e}ncia, Tecnologia e Inova{\c c}{\~a}o, the Deutsche Forschungsgemeinschaft and the Collaborating Institutions in the Dark Energy Survey. 

The Collaborating Institutions are Argonne National Laboratory, the University of California at Santa Cruz, the University of Cambridge, Centro de Investigaciones Energ{\'e}ticas, 
Medioambientales y Tecnol{\'o}gicas-Madrid, the University of Chicago, University College London, the DES-Brazil Consortium, the University of Edinburgh, 
the Eidgen{\"o}ssische Technische Hochschule (ETH) Z{\"u}rich, 
Fermi National Accelerator Laboratory, the University of Illinois at Urbana-Champaign, the Institut de Ci{\`e}ncies de l'Espai (IEEC/CSIC), 
the Institut de F{\'i}sica d'Altes Energies, Lawrence Berkeley National Laboratory, the Ludwig-Maximilians Universit{\"a}t M{\"u}nchen and the associated Excellence Cluster Universe, 
the University of Michigan, the National Optical Astronomy Observatory, the University of Nottingham, The Ohio State University, the University of Pennsylvania, the University of Portsmouth, 
SLAC National Accelerator Laboratory, Stanford University, the University of Sussex, Texas A\&M University, and the OzDES Membership Consortium.

Based in part on observations at Cerro Tololo Inter-American Observatory, National Optical Astronomy Observatory, which is operated by the Association of Universities for Research in Astronomy (AURA) under a cooperative agreement with the National Science Foundation.

The DES data management system is supported by the National Science Foundation under Grant Numbers AST-1138766 and AST-1536171.
The DES participants from Spanish institutions are partially supported by MINECO under grants AYA2015-71825, ESP2015-88861, FPA2015-68048, SEV-2012-0234, SEV-2016-0597, and MDM-2015-0509, 
some of which include ERDF funds from the European Union. IFAE is partially funded by the CERCA program of the Generalitat de Catalunya.
Research leading to these results has received funding from the European Research
Council under the European Union's Seventh Framework Program (FP7/2007-2013) including ERC grant agreements 240672, 291329, and 306478.
We  acknowledge support from the Australian Research Council Centre of Excellence for All-sky Astrophysics (CAASTRO), through project number CE110001020.

This manuscript has been authored by Fermi Research Alliance, LLC under Contract No. DE-AC02-07CH11359 with the U.S. Department of Energy, Office of Science, Office of High Energy Physics. The United States Government retains and the publisher, by accepting the article for publication, acknowledges that the United States Government retains a non-exclusive, paid-up, irrevocable, world-wide license to publish or reproduce the published form of this manuscript, or allow others to do so, for United States Government purposes.

\facility{Blanco (DECam)} 
\software{\SExtractor \citep{Bertin:1996}, \healpix \citep{Gorski:2005}\footnote{\url{http://healpix.sourceforge.net}}, \code{astropy} \citep{Astropy:2013}, \code{matplotlib} \citep{Hunter:2007}, \code{numpy} \citep{numpy:2011}, \code{scipy} \citep{scipy:2001}, \code{healpy}\footnote{\url{https://github.com/healpy/healpy}}, \code{fitsio}\footnote{\url{https://github.com/esheldon/fitsio}}, \ngmix \citep{Sheldon:2014}\footnote{\url{https://github.com/esheldon/ngmix}}, \emcee \citep{Foreman_Mackey:2013}, \ugali \citep{Bechtol:2015}\footnote{\url{https://github.com/DarkEnergySurvey/ugali}},\code{galpot} \citep{Dehnen:1998}\footnote{\url{https://github.com/PaulMcMillan-Astro/GalPot}}}

\clearpage

\appendix
\numberwithin{figure}{section}
\numberwithin{table}{section}

\section{Stream fit configuration}
\label{app:fit}

\tabref{fits} contains the input specifications for the fits to each stream. The procedure for selecting on- and off-stream regions is described in \secref{isochrone}. These regions were derived in a variety of ways in order to optimize foreground subtraction and to avoid contamination by other resolved stellar populations, including other streams, globular clusters, and dwarf galaxies. For the ATLAS stream, when calculating $N_*$, we selected a region along the polynomial fit in \eqnref{atlas} to account for the curvature of the stream relative to a great circle. Due to the variation in region definitions, we list here the widths and separations of the selected regions for each stream. Additionally, we list the parameters that were fit for each stream. In many cases the data did not allow for a simultaneous fit of all parameters, so a subset of parameters were set to previously determined values or estimated by eye and held fixed.

\begin{deluxetable}{l ccc ccc}
\tablecolumns{13}
\tablewidth{0pt}
\tabletypesize{\scriptsize}
\tablecaption{Input specifications for stream fits \label{tab:fits}}
\tablehead{
Name & Width (on) & Width (off) & Separation & $\mM$ & Age & $Z$ \\
 & (deg) & (deg) & (deg) & & &
}
\startdata
Tucana III      & $0.8$ & $0.8$ & $\pm 0.8$ & fixed & fixed & fixed \\
ATLAS           & $1.2$ & $1.2$ & $\pm 1.2$ & free  & free  & free  \\
Molonglo        & $1.5$ & $3.0$ & $- 2.6$   & free  & fixed & fixed \\
Phoenix         & $0.7$ & $0.7$ & $\pm 0.8$ & free  & fixed & fixed \\
[+0.5em]\tableline\\[-1em]                           
Indus           & $3.3$ & $3.3$ & $\pm 4.1$ & free  & fixed & fixed \\
Jhelum          & $2.5$ & $1.2$ & $+ 2.0$   & free  & fixed & fixed \\
Ravi            & $2.9$ & $2.9$ & $\pm 3.6$ & free  & free  & free  \\
Chenab          & $4.0$ & $4.0$ & $\pm 4.7$ & free  & fixed & fixed \\
Elqui           & $2.0$ & $2.0$ & $\pm 2.2$ & free  & free  & free  \\
Aliqa Uma       & $0.6$ & $1.2$ & $- 1.5$   & fixed & fixed & fixed \\
Turbio          & $1.2$ & $2.4$ & $+ 2.0$   & free  & fixed & fixed \\
Willka Yaku     & $0.4$ & $0.8$ & $+ 0.8$& free  & free  & free  \\
Turranburra     & $2.4$ & $2.4$ & $\pm 3.0$ & free  & free  & free  \\
Wambelong       & $2.0$ & $2.0$ & $\pm 2.2$ & free  & free  & free  \\
[+0.5em]\tableline\\[-1em]                           
Palca           & \ldots & \ldots & \ldots & fixed  & fixed  & fixed  \\
\enddata
{\footnotesize \tablecomments{Input specifications for stellar stream fits. By default, the widths of the on- and off-stream regions are four times the Gaussian stream width. The separation between on and off regions is measured between the stream axis and the center of the off-stream region.}}
\end{deluxetable}

\clearpage

\bibliographystyle{aasjournal2}
\bibliography{main}

\begin{thebibliography}{}
\expandafter\ifx\csname natexlab\endcsname\relax\def\natexlab#1{#1}\fi
\providecommand{\url}[1]{\href{#1}{#1}}
\providecommand{\dodoi}[1]{doi:~\href{http://doi.org/#1}{\nolinkurl{#1}}}
\providecommand{\doeprint}[1]{\href{http://ascl.net/#1}{\nolinkurl{http://ascl.net/#1}}}
\providecommand{\doarXiv}[1]{\href{https://arxiv.org/abs/#1}{\nolinkurl{https://arxiv.org/abs/#1}}}

\bibitem[{{Agnello}(2017)}]{Agnello:2017}
{Agnello}, A. 2017, \mnras, 471, 2013, \dodoi{10.1093/mnras/stx1650}

\bibitem[{{Aihara} {et~al.}(2017){Aihara}, {Arimoto}, {Armstrong}, {Arnouts},
  {Bahcall}, {Bickerton}, {Bosch}, {Bundy}, {Capak}, {Chan}, {Chiba}, {Coupon},
  {Egami}, {Enoki}, {Finet}, {Fujimori}, {Fujimoto}, {Furusawa}, {Furusawa},
  {Goto}, {Goulding}, {Greco}, {Greene}, {Gunn}, {Hamana}, {Harikane},
  {Hashimoto}, {Hattori}, {Hayashi}, {Hayashi}, {He{\l}miniak}, {Higuchi},
  {Hikage}, {Ho}, {Hsieh}, {Huang}, {Huang}, {Ikeda}, {Imanishi}, {Inoue},
  {Iwasawa}, {Iwata}, {Jaelani}, {Jian}, {Kamata}, {Karoji}, {Kashikawa},
  {Katayama}, {Kawanomoto}, {Kayo}, {Koda}, {Koike}, {Kojima}, {Komiyama},
  {Konno}, {Koshida}, {Koyama}, {Kusakabe}, {Leauthaud}, {Lee}, {Lin}, {Lin},
  {Lupton}, {Mandelbaum}, {Matsuoka}, {Medezinski}, {Mineo}, {Miyama},
  {Miyatake}, {Miyazaki}, {Momose}, {More}, {More}, {Moritani}, {Moriya},
  {Morokuma}, {Mukae}, {Murata}, {Murayama}, {Nagao}, {Nakata}, {Niida},
  {Niikura}, {Nishizawa}, {Obuchi}, {Oguri}, {Oishi}, {Okabe}, {Okura}, {Ono},
  {Onodera}, {Onoue}, {Osato}, {Ouchi}, {Price}, {Pyo}, {Sako}, {Okamoto},
  {Sawicki}, {Shibuya}, {Shimasaku}, {Shimono}, {Shirasaki}, {Silverman},
  {Simet}, {Speagle}, {Spergel}, {Strauss}, {Sugahara}, {Sugiyama}, {Suto},
  {Suyu}, {Suzuki}, {Tait}, {Takata}, {Takada}, {Tamura}, {Tanaka}, {Tanaka},
  {Tanaka}, {Tanaka}, {Terai}, {Terashima}, {Toba}, {Toshikawa}, {Turner},
  {Uchida}, {Uchiyama}, {Umetsu}, {Uraguchi}, {Urata}, {Usuda}, {Utsumi},
  {Wang}, {Wang}, {Wong}, {Yabe}, {Yamada}, {Yamanoi}, {Yasuda}, {Yeh},
  {Yonehara}, \& {Yuma}}]{Aihara:2017}
{Aihara}, H., {Arimoto}, N., {Armstrong}, R., {et~al.} 2017, ArXiv e-prints.
\newblock \doarXiv{1704.05858}

\bibitem[{{An} {et~al.}(2015){An}, {Beers}, {Santucci}, {Carollo}, {Placco},
  {Lee}, \& {Rossi}}]{An:2015}
{An}, D., {Beers}, T.~C., {Santucci}, R.~M., {et~al.} 2015, \apjl, 813, L28,
  \dodoi{10.1088/2041-8205/813/2/L28}

\bibitem[{{An} {et~al.}(2013){An}, {Beers}, {Johnson}, {Pinsonneault}, {Lee},
  {Bovy}, {Ivezi{\'c}}, {Carollo}, \& {Newby}}]{An:2013}
{An}, D., {Beers}, T.~C., {Johnson}, J.~A., {et~al.} 2013, \apj, 763, 65,
  \dodoi{10.1088/0004-637X/763/1/65}

\bibitem[{{Astropy Collaboration} {et~al.}(2013){Astropy Collaboration},
  {Robitaille}, {Tollerud}, {Greenfield}, {Droettboom}, {Bray}, {Aldcroft},
  {Davis}, {Ginsburg}, {Price-Whelan}, {Kerzendorf}, {Conley}, {Crighton},
  {Barbary}, {Muna}, {Ferguson}, {Grollier}, {Parikh}, {Nair}, {Unther},
  {Deil}, {Woillez}, {Conseil}, {Kramer}, {Turner}, {Singer}, {Fox}, {Weaver},
  {Zabalza}, {Edwards}, {Azalee Bostroem}, {Burke}, {Casey}, {Crawford},
  {Dencheva}, {Ely}, {Jenness}, {Labrie}, {Lim}, {Pierfederici}, {Pontzen},
  {Ptak}, {Refsdal}, {Servillat}, \& {Streicher}}]{Astropy:2013}
{Astropy Collaboration}, {Robitaille}, T.~P., {Tollerud}, E.~J., {et~al.} 2013,
  \aap, 558, A33, \dodoi{10.1051/0004-6361/201322068}

\bibitem[{Balbinot \& Gieles(2017)}]{Balbinot:2018}
Balbinot, E., \& Gieles, M. 2017, \mnras, stx2708,
  \dodoi{10.1093/mnras/stx2708}

\bibitem[{{Balbinot} {et~al.}(2016){Balbinot}, {Yanny}, {Li}, {Santiago},
  {Marshall}, {Finley}, {Pieres}, {Abbott}, {Abdalla}, {Allam},
  {Benoit-L{\'e}vy}, {Bernstein}, {Bertin}, {Brooks}, {Burke}, {Carnero
  Rosell}, {Carrasco Kind}, {Carretero}, {Cunha}, {da Costa}, {DePoy}, {Desai},
  {Diehl}, {Doel}, {Estrada}, {Flaugher}, {Frieman}, {Gerdes}, {Gruen},
  {Gruendl}, {Honscheid}, {James}, {Kuehn}, {Kuropatkin}, {Lahav}, {March},
  {Martini}, {Miquel}, {Nichol}, {Ogando}, {Romer}, {Sanchez}, {Schubnell},
  {Sevilla-Noarbe}, {Smith}, {Soares-Santos}, {Sobreira}, {Suchyta}, {Tarle},
  {Thomas}, {Tucker}, {Walker}, \& {DES Collaboration}}]{Balbinot:2016}
{Balbinot}, E., {Yanny}, B., {Li}, T.~S., {et~al.} 2016, \apj, 820, 58,
  \dodoi{10.3847/0004-637X/820/1/58}

\bibitem[{{Bechtol} {et~al.}(2015){Bechtol}, {Drlica-Wagner}, {Balbinot},
  {Pieres}, {Simon}, {Yanny}, {Santiago}, {Wechsler}, {Frieman}, {Walker},
  {Williams}, {Rozo}, {Rykoff}, {Queiroz}, {Luque}, {Benoit-L{\'e}vy},
  {Tucker}, {Sevilla}, {Gruendl}, {da Costa}, {Fausti Neto}, {Maia}, {Abbott},
  {Allam}, {Armstrong}, {Bauer}, {Bernstein}, {Bernstein}, {Bertin}, {Brooks},
  {Buckley-Geer}, {Burke}, {Carnero Rosell}, {Castander}, {Covarrubias},
  {D'Andrea}, {DePoy}, {Desai}, {Diehl}, {Eifler}, {Estrada}, {Evrard},
  {Fernandez}, {Finley}, {Flaugher}, {Gaztanaga}, {Gerdes}, {Girardi},
  {Gladders}, {Gruen}, {Gutierrez}, {Hao}, {Honscheid}, {Jain}, {James},
  {Kent}, {Kron}, {Kuehn}, {Kuropatkin}, {Lahav}, {Li}, {Lin}, {Makler},
  {March}, {Marshall}, {Martini}, {Merritt}, {Miller}, {Miquel}, {Mohr},
  {Neilsen}, {Nichol}, {Nord}, {Ogando}, {Peoples}, {Petravick}, {Plazas},
  {Romer}, {Roodman}, {Sako}, {Sanchez}, {Scarpine}, {Schubnell}, {Smith},
  {Soares-Santos}, {Sobreira}, {Suchyta}, {Swanson}, {Tarle}, {Thaler},
  {Thomas}, {Wester}, {Zuntz}, \& {DES Collaboration}}]{Bechtol:2015}
{Bechtol}, K., {Drlica-Wagner}, A., {Balbinot}, E., {et~al.} 2015, \apj, 807,
  50, \dodoi{10.1088/0004-637X/807/1/50}

\bibitem[{{Bell} {et~al.}(2008){Bell}, {Zucker}, {Belokurov}, {Sharma},
  {et~al.}}]{Bell:2008}
{Bell}, E.~F., {Zucker}, D.~B., {Belokurov}, V., {Sharma}, S., {et~al.} 2008,
  \apj, 680, 295, \dodoi{10.1086/588032}

\bibitem[{{Belokurov} {et~al.}(2007{\natexlab{a}}){Belokurov}, {Evans},
  {Irwin}, {Lynden-Bell}, {et~al.}}]{Belokurov:2007b}
{Belokurov}, V., {Evans}, N.~W., {Irwin}, M.~J., {Lynden-Bell}, D., {et~al.}
  2007{\natexlab{a}}, \apj, 658, 337, \dodoi{10.1086/511302}

\bibitem[{{Belokurov} {et~al.}(2006{\natexlab{a}}){Belokurov}, {Zucker},
  {Evans}, {Wilkinson}, {Irwin}, {Hodgkin}, {Bramich}, {Irwin}, {Gilmore},
  {Willman}, {Vidrih}, {Newberg}, {Wyse}, {Fellhauer}, {Hewett}, {Cole},
  {Bell}, {Beers}, {Rockosi}, {Yanny}, {Grebel}, {Schneider}, {Lupton},
  {Barentine}, {Brewington}, {Brinkmann}, {Harvanek}, {Kleinman}, {Krzesinski},
  {Long}, {Nitta}, {Smith}, \& {Snedden}}]{2006ApJ...647L.111B}
{Belokurov}, V., {Zucker}, D.~B., {Evans}, N.~W., {et~al.} 2006{\natexlab{a}},
  \apjl, 647, L111, \dodoi{10.1086/507324}

\bibitem[{{Belokurov} {et~al.}(2006{\natexlab{b}}){Belokurov}, {Zucker},
  {Evans}, {Gilmore}, {Vidrih}, {Bramich}, {Newberg}, {Wyse}, {Irwin},
  {Fellhauer}, {Hewett}, {Walton}, {Wilkinson}, {Cole}, {Yanny}, {Rockosi},
  {Beers}, {Bell}, {Brinkmann}, {Ivezi{\'c}}, \& {Lupton}}]{Belokurov:2006}
---. 2006{\natexlab{b}}, \apjl, 642, L137, \dodoi{10.1086/504797}

\bibitem[{{Belokurov} {et~al.}(2007{\natexlab{b}}){Belokurov}, {Zucker},
  {Evans}, {Kleyna}, {Koposov}, {Hodgkin}, {Irwin}, {Gilmore}, {Wilkinson},
  {Fellhauer}, {Bramich}, {Hewett}, {Vidrih}, {De Jong}, {Smith}, {Rix},
  {Bell}, {Wyse}, {Newberg}, {Mayeur}, {Yanny}, {Rockosi}, {Gnedin},
  {Schneider}, {Beers}, {Barentine}, {Brewington}, {Brinkmann}, {Harvanek},
  {Kleinman}, {Krzesinski}, {Long}, {Nitta}, \& {Snedden}}]{Belokurov:2007}
---. 2007{\natexlab{b}}, \apj, 654, 897, \dodoi{10.1086/509718}

\bibitem[{{Belokurov} {et~al.}(2007{\natexlab{c}}){Belokurov}, {Evans}, {Bell},
  {Irwin}, {Hewett}, {Koposov}, {Rockosi}, {Gilmore}, {Zucker}, {Fellhauer},
  {Wilkinson}, {Bramich}, {Vidrih}, {Rix}, {Beers}, {Schneider}, {Barentine},
  {Brewington}, {Brinkmann}, {Harvanek}, {Krzesinski}, {Long}, {Pan},
  {Snedden}, {Malanushenko}, \& {Malanushenko}}]{Belokurov:2007c}
{Belokurov}, V., {Evans}, N.~W., {Bell}, E.~F., {et~al.} 2007{\natexlab{c}},
  \apjl, 657, L89, \dodoi{10.1086/513144}

\bibitem[{{Bernard} {et~al.}(2014){Bernard}, {Ferguson}, {Schlafly}, {Abbas},
  {Bell}, {Deacon}, {Martin}, {Rix}, {Sesar}, {Slater}, {Pe{\~n}arrubia},
  {Wyse}, {Burgett}, {Chambers}, {Draper}, {Hodapp}, {Kaiser}, {Kudritzki},
  {Magnier}, {Metcalfe}, {Morgan}, {Price}, {Tonry}, {Wainscoat}, \&
  {Waters}}]{Bernard:2014}
{Bernard}, E.~J., {Ferguson}, A.~M.~N., {Schlafly}, E.~F., {et~al.} 2014,
  \mnras, 443, L84, \dodoi{10.1093/mnrasl/slu089}

\bibitem[{{Bernard} {et~al.}(2016){Bernard}, {Ferguson}, {Schlafly}, {Martin},
  {Rix}, {Bell}, {Finkbeiner}, {Goldman}, {Mart{\'{\i}}nez-Delgado}, {Sesar},
  {Wyse}, {Burgett}, {Chambers}, {Draper}, {Hodapp}, {Kaiser}, {Kudritzki},
  {Magnier}, {Metcalfe}, {Wainscoat}, \& {Waters}}]{Bernard:2016}
---. 2016, \mnras, 463, 1759, \dodoi{10.1093/mnras/stw2134}

\bibitem[{{Bertin} \& {Arnouts}(1996)}]{Bertin:1996}
{Bertin}, E., \& {Arnouts}, S. 1996, \aaps, 117, 393

\bibitem[{{Bertin} {et~al.}(2002){Bertin}, {Mellier}, {Radovich}, {Missonnier},
  {Didelon}, \& {Morin}}]{Bertin:2002}
{Bertin}, E., {Mellier}, Y., {Radovich}, M., {et~al.} 2002, in Astronomical
  Society of the Pacific Conference Series, Vol. 281, Astronomical Data
  Analysis Software and Systems XI, ed. D.~A. {Bohlender}, D.~{Durand}, \&
  T.~H. {Handley}, 228

\bibitem[{{Besla} {et~al.}(2010){Besla}, {Kallivayalil}, {Hernquist}, {van der
  Marel}, {Cox}, \& {Kere{\v s}}}]{Besla:2010}
{Besla}, G., {Kallivayalil}, N., {Hernquist}, L., {et~al.} 2010, \apjl, 721,
  L97, \dodoi{10.1088/2041-8205/721/2/L97}

\bibitem[{{Besla} {et~al.}(2012){Besla}, {Kallivayalil}, {Hernquist}, {van der
  Marel}, {Cox}, \& {Kere{\v s}}}]{Besla:2012}
---. 2012, \mnras, 421, 2109, \dodoi{10.1111/j.1365-2966.2012.20466.x}

\bibitem[{{Blumenthal} {et~al.}(1984){Blumenthal}, {Faber}, {Primack}, \&
  {Rees}}]{Blumenthal:1984}
{Blumenthal}, G.~R., {Faber}, S.~M., {Primack}, J.~R., \& {Rees}, M.~J. 1984,
  \nat, 311, 517, \dodoi{10.1038/311517a0}

\bibitem[{{Bonaca} {et~al.}(2012){Bonaca}, {Geha}, \&
  {Kallivayalil}}]{Bonaca:2012}
{Bonaca}, A., {Geha}, M., \& {Kallivayalil}, N. 2012, \apjl, 760, L6,
  \dodoi{10.1088/2041-8205/760/1/L6}

\bibitem[{{Bonaca} {et~al.}(2014){Bonaca}, {Geha}, {K{\"u}pper}, {Diemand},
  {Johnston}, \& {Hogg}}]{Bonaca:2014}
{Bonaca}, A., {Geha}, M., {K{\"u}pper}, A.~H.~W., {et~al.} 2014, \apj, 795, 94,
  \dodoi{10.1088/0004-637X/795/1/94}

\bibitem[{{Bovy}(2014)}]{Bovy:2014}
{Bovy}, J. 2014, \apj, 795, 95, \dodoi{10.1088/0004-637X/795/1/95}

\bibitem[{{Bovy} {et~al.}(2016){Bovy}, {Bahmanyar}, {Fritz}, \&
  {Kallivayalil}}]{Bovy:2016}
{Bovy}, J., {Bahmanyar}, A., {Fritz}, T.~K., \& {Kallivayalil}, N. 2016, \apj,
  833, 31, \dodoi{10.3847/1538-4357/833/1/31}

\bibitem[{{Bovy} {et~al.}(2017){Bovy}, {Erkal}, \& {Sanders}}]{Bovy:2017}
{Bovy}, J., {Erkal}, D., \& {Sanders}, J.~L. 2017, \mnras, 466, 628,
  \dodoi{10.1093/mnras/stw3067}

\bibitem[{{Bowden} {et~al.}(2015){Bowden}, {Belokurov}, \&
  {Evans}}]{Bowden:2015}
{Bowden}, A., {Belokurov}, V., \& {Evans}, N.~W. 2015, \mnras, 449, 1391,
  \dodoi{10.1093/mnras/stv285}

\bibitem[{{Bullock} \& {Johnston}(2005)}]{Bullock:2005}
{Bullock}, J.~S., \& {Johnston}, K.~V. 2005, \apj, 635, 931,
  \dodoi{10.1086/497422}

\bibitem[{{Burke} {et~al.}(2018){Burke}, {Rykoff}, {Allam}, {Annis}, {Bechtol},
  {et~al.}}]{Burke:2017}
{Burke}, D., {Rykoff}, E.~S., {Allam}, S., {et~al.} 2018, \aj, 155, 41,
  \dodoi{10.3847/1538-3881/aa9f22}

\bibitem[{{Carballo-Bello} {et~al.}(2018){Carballo-Bello},
  {Mart{\'{\i}}nez-Delgado}, {et~al.}}]{Carballo-Bello:2018}
{Carballo-Bello}, J.~A., {Mart{\'{\i}}nez-Delgado}, D., {et~al.} 2018, \mnras,
  474, 683, \dodoi{10.1093/mnras/stx2767}

\bibitem[{{Carlberg}(2009)}]{Carlberg:2009}
{Carlberg}, R.~G. 2009, \apjl, 705, L223, \dodoi{10.1088/0004-637X/705/2/L223}

\bibitem[{{Carlberg}(2012)}]{Carlberg:2012}
---. 2012, \apj, 748, 20, \dodoi{10.1088/0004-637X/748/1/20}

\bibitem[{{Carlberg}(2016)}]{Carlberg:2016a}
---. 2016, \apj, 820, 45, \dodoi{10.3847/0004-637X/820/1/45}

\bibitem[{{Carollo} {et~al.}(2007){Carollo}, {Beers}, {Lee}, {Chiba}, {Norris},
  {Wilhelm}, {Sivarani}, {Marsteller}, {Munn}, {Bailer-Jones}, {Fiorentin}, \&
  {York}}]{Carollo:2007}
{Carollo}, D., {Beers}, T.~C., {Lee}, Y.~S., {et~al.} 2007, \nat, 450, 1020,
  \dodoi{10.1038/nature06460}

\bibitem[{{Carollo} {et~al.}(2010){Carollo}, {Beers}, {Chiba}, {Norris},
  {Freeman}, {Lee}, {Ivezi{\'c}}, {Rockosi}, \& {Yanny}}]{Carollo:2010}
{Carollo}, D., {Beers}, T.~C., {Chiba}, M., {et~al.} 2010, \apj, 712, 692,
  \dodoi{10.1088/0004-637X/712/1/692}

\bibitem[{{Chabrier}(2001)}]{Chabrier:2001}
{Chabrier}, G. 2001, \apj, 554, 1274, \dodoi{10.1086/321401}

\bibitem[{{Claydon} {et~al.}(2017){Claydon}, {Gieles}, \&
  {Zocchi}}]{Claydon:2016}
{Claydon}, I., {Gieles}, M., \& {Zocchi}, A. 2017, \mnras, 466, 3937,
  \dodoi{10.1093/mnras/stw3309}

\bibitem[{{Dambis}(2006)}]{Dambis:2006}
{Dambis}, A.~K. 2006, Astronomical and Astrophysical Transactions, 25, 185,
  \dodoi{10.1080/10556790600938628}

\bibitem[{{Das} \& {Binney}(2016)}]{Das:2016}
{Das}, P., \& {Binney}, J. 2016, \mnras, 460, 1725,
  \dodoi{10.1093/mnras/stw744}

\bibitem[{{de Grijs} \& {Bono}(2016)}]{deGrijs:2016}
{de Grijs}, R., \& {Bono}, G. 2016, \apjs, 227, 5,
  \dodoi{10.3847/0067-0049/227/1/5}

\bibitem[{{de Jong} {et~al.}(2010){de Jong}, {Yanny}, {Rix}, {Dolphin},
  {Martin}, \& {Beers}}]{Dejong:2010}
{de Jong}, J.~T.~A., {Yanny}, B., {Rix}, H.-W., {et~al.} 2010, \apj, 714, 663,
  \dodoi{10.1088/0004-637X/714/1/663}

\bibitem[{{Deason} {et~al.}(2011){Deason}, {Belokurov}, \&
  {Evans}}]{Deason:2011}
{Deason}, A.~J., {Belokurov}, V., \& {Evans}, N.~W. 2011, \mnras, 416, 2903,
  \dodoi{10.1111/j.1365-2966.2011.19237.x}

\bibitem[{{Dehnen} \& {Binney}(1998)}]{Dehnen:1998}
{Dehnen}, W., \& {Binney}, J. 1998, \mnras, 294, 429,
  \dodoi{10.1046/j.1365-8711.1998.01282.x}

\bibitem[{{DES Collaboration}(2005)}]{DES:2005}
{DES Collaboration}. 2005, ArXiv Astrophysics e-prints

\bibitem[{{DES Collaboration}(2016)}]{DES:2016}
---. 2016, \mnras, 460, 1270, \dodoi{10.1093/mnras/stw641}

\bibitem[{{DES Collaboration}(2017)}]{DES:2017}
---. 2017, ArXiv e-prints.
\newblock \doarXiv{1708.01530}

\bibitem[{{DES Collaboration}(2018)}]{DES:2018}
---. 2018, in preparation

\bibitem[{{Diehl} {et~al.}(2016){Diehl}, {Neilsen}, {Gruendl}, {Yanny},
  {et~al.}}]{Diehl:2016}
{Diehl}, H.~T., {Neilsen}, E., {Gruendl}, R., {Yanny}, B., {et~al.} 2016, in
  \procspie, Vol. 9910, Observatory Operations: Strategies, Processes, and
  Systems VI, 99101D

\bibitem[{{Diemand} {et~al.}(2008){Diemand}, {Kuhlen}, {Madau}, {Zemp},
  {Moore}, {Potter}, \& {Stadel}}]{Diemand:2008}
{Diemand}, J., {Kuhlen}, M., {Madau}, P., {et~al.} 2008, \nat, 454, 735,
  \dodoi{10.1038/nature07153}

\bibitem[{{Dinescu} {et~al.}(1997){Dinescu}, {Girard}, {van Altena}, {Mendez},
  \& {Lopez}}]{Dinescu:1997}
{Dinescu}, D.~I., {Girard}, T.~M., {van Altena}, W.~F., {Mendez}, R.~A., \&
  {Lopez}, C.~E. 1997, \aj, 114, 1014, \dodoi{10.1086/118532}

\bibitem[{{Dinescu} {et~al.}(1999){Dinescu}, {van Altena}, {Girard}, \&
  {L{\'o}pez}}]{Dinescu:1999}
{Dinescu}, D.~I., {van Altena}, W.~F., {Girard}, T.~M., \& {L{\'o}pez}, C.~E.
  1999, \aj, 117, 277, \dodoi{10.1086/300699}

\bibitem[{{Dotter} {et~al.}(2008){Dotter}, {Chaboyer}, {Jevremovi{\'c}},
  {Kostov}, {Baron}, \& {Ferguson}}]{Dotter:2008}
{Dotter}, A., {Chaboyer}, B., {Jevremovi{\'c}}, D., {et~al.} 2008, \apjs, 178,
  89, \dodoi{10.1086/589654}

\bibitem[{{Drlica-Wagner} {et~al.}(2017){Drlica-Wagner}, {Sevilla-Noarbe},
  {Rykoff}, {Gruendl}, {Yanny}, {et~al.}}]{Drlica-Wagner:2017}
{Drlica-Wagner}, A., {Sevilla-Noarbe}, I., {Rykoff}, E.~S., {et~al.} 2017,
  ArXiv e-prints.
\newblock \doarXiv{1708.01531}

\bibitem[{{Drlica-Wagner} {et~al.}(2015){Drlica-Wagner}, {Bechtol}, {Rykoff},
  {Luque}, {Queiroz}, {Mao}, {Wechsler}, {Simon}, {Santiago}, {Yanny},
  {Balbinot}, {Dodelson}, {Fausti Neto}, {James}, {Li}, {Maia}, {Marshall},
  {Pieres}, {Stringer}, {Walker}, {Abbott}, {Abdalla}, {Allam},
  {Benoit-L{\'e}vy}, {Bernstein}, {Bertin}, {Brooks}, {Buckley-Geer}, {Burke},
  {Carnero Rosell}, {Carrasco Kind}, {Carretero}, {Crocce}, {da Costa},
  {Desai}, {Diehl}, {Dietrich}, {Doel}, {Eifler}, {Evrard}, {Finley},
  {Flaugher}, {Fosalba}, {Frieman}, {Gaztanaga}, {Gerdes}, {Gruen}, {Gruendl},
  {Gutierrez}, {Honscheid}, {Kuehn}, {Kuropatkin}, {Lahav}, {Martini},
  {Miquel}, {Nord}, {Ogando}, {Plazas}, {Reil}, {Roodman}, {Sako}, {Sanchez},
  {Scarpine}, {Schubnell}, {Sevilla-Noarbe}, {Smith}, {Soares-Santos},
  {Sobreira}, {Suchyta}, {Swanson}, {Tarle}, {Tucker}, {Vikram}, {Wester},
  {Zhang}, {Zuntz}, \& {DES Collaboration}}]{Drlica-Wagner:2015}
{Drlica-Wagner}, A., {Bechtol}, K., {Rykoff}, E.~S., {et~al.} 2015, \apj, 813,
  109, \dodoi{10.1088/0004-637X/813/2/109}

\bibitem[{{Drlica-Wagner} {et~al.}(2016){Drlica-Wagner}, {Bechtol}, {Allam},
  {Tucker}, {Gruendl}, {Johnson}, {Walker}, {James}, {Nidever}, {Olsen},
  {Wechsler}, {Cioni}, {Conn}, {Kuehn}, {Li}, {Mao}, {Martin}, {Neilsen},
  {Noel}, {Pieres}, {Simon}, {Stringfellow}, {van der Marel}, \&
  {Yanny}}]{Drlica-Wagner:2016}
{Drlica-Wagner}, A., {Bechtol}, K., {Allam}, S., {et~al.} 2016, \apjl, 833, L5,
  \dodoi{10.3847/2041-8205/833/1/L5}

\bibitem[{{Erkal} \& {Belokurov}(2015{\natexlab{a}})}]{Erkal:2015}
{Erkal}, D., \& {Belokurov}, V. 2015{\natexlab{a}}, \mnras, 454, 3542,
  \dodoi{10.1093/mnras/stv2122}

\bibitem[{{Erkal} \& {Belokurov}(2015{\natexlab{b}})}]{Erkal:2015b}
---. 2015{\natexlab{b}}, \mnras, 454, 3542, \dodoi{10.1093/mnras/stv2122}

\bibitem[{{Erkal} {et~al.}(2016{\natexlab{a}}){Erkal}, {Belokurov}, {Bovy}, \&
  {Sanders}}]{Erkal:2016b}
{Erkal}, D., {Belokurov}, V., {Bovy}, J., \& {Sanders}, J.~L.
  2016{\natexlab{a}}, \mnras, 463, 102, \dodoi{10.1093/mnras/stw1957}

\bibitem[{{Erkal} {et~al.}(2017){Erkal}, {Koposov}, \&
  {Belokurov}}]{Erkal:2017}
{Erkal}, D., {Koposov}, S.~E., \& {Belokurov}, V. 2017, \mnras, 470, 60,
  \dodoi{10.1093/mnras/stx1208}

\bibitem[{{Erkal} {et~al.}(2016{\natexlab{b}}){Erkal}, {Sanders}, \&
  {Belokurov}}]{Erkal:2016}
{Erkal}, D., {Sanders}, J.~L., \& {Belokurov}, V. 2016{\natexlab{b}}, \mnras,
  461, 1590, \dodoi{10.1093/mnras/stw1400}

\bibitem[{{Fitzpatrick}(1999)}]{Fitzpatrick:1999}
{Fitzpatrick}, E.~L. 1999, \pasp, 111, 63, \dodoi{10.1086/316293}

\bibitem[{{Flaugher} {et~al.}(2015){Flaugher}, {Diehl}, {Honscheid}, {Abbott},
  {Alvarez}, {Angstadt}, {Annis}, {Antonik}, {Ballester}, {Beaufore},
  {Bernstein}, {Bernstein}, {Bigelow}, {Bonati}, {Boprie}, {Brooks},
  {Buckley-Geer}, {Campa}, {Cardiel-Sas}, {Castander}, {Castilla}, {Cease},
  {Cela-Ruiz}, {Chappa}, {Chi}, {Cooper}, {da Costa}, {Dede}, {Derylo},
  {DePoy}, {de Vicente}, {Doel}, {Drlica-Wagner}, {Eiting}, {Elliott}, {Emes},
  {Estrada}, {Fausti Neto}, {Finley}, {Flores}, {Frieman}, {Gerdes},
  {Gladders}, {Gregory}, {Gutierrez}, {Hao}, {Holland}, {Holm}, {Huffman},
  {Jackson}, {James}, {Jonas}, {Karcher}, {Karliner}, {Kent}, {Kessler},
  {Kozlovsky}, {Kron}, {Kubik}, {Kuehn}, {Kuhlmann}, {Kuk}, {Lahav}, {Lathrop},
  {Lee}, {Levi}, {Lewis}, {Li}, {Mandrichenko}, {Marshall}, {Martinez},
  {Merritt}, {Miquel}, {Mu{\~n}oz}, {Neilsen}, {Nichol}, {Nord}, {Ogando},
  {Olsen}, {Palaio}, {Patton}, {Peoples}, {Plazas}, {Rauch}, {Reil}, {Rheault},
  {Roe}, {Rogers}, {Roodman}, {Sanchez}, {Scarpine}, {Schindler}, {Schmidt},
  {Schmitt}, {Schubnell}, {Schultz}, {Schurter}, {Scott}, {Serrano}, {Shaw},
  {Smith}, {Soares-Santos}, {Stefanik}, {Stuermer}, {Suchyta}, {Sypniewski},
  {Tarle}, {Thaler}, {Tighe}, {Tran}, {Tucker}, {Walker}, {Wang}, {Watson},
  {Weaverdyck}, {Wester}, {Woods}, {Yanny}, \& {DES
  Collaboration}}]{Flaugher:2015}
{Flaugher}, B., {Diehl}, H.~T., {Honscheid}, K., {et~al.} 2015, \aj, 150, 150,
  \dodoi{10.1088/0004-6256/150/5/150}

\bibitem[{{Foreman-Mackey} {et~al.}(2013){Foreman-Mackey}, {Hogg}, {Lang}, \&
  {Goodman}}]{Foreman_Mackey:2013}
{Foreman-Mackey}, D., {Hogg}, D.~W., {Lang}, D., \& {Goodman}, J. 2013, \pasp,
  125, 306, \dodoi{10.1086/670067}

\bibitem[{{Gibbons} {et~al.}(2014){Gibbons}, {Belokurov}, \&
  {Evans}}]{Gibbons:2014}
{Gibbons}, S.~L.~J., {Belokurov}, V., \& {Evans}, N.~W. 2014, \mnras, 445,
  3788, \dodoi{10.1093/mnras/stu1986}

\bibitem[{{Gillessen} {et~al.}(2009){Gillessen}, {Eisenhauer}, {Trippe},
  {Alexander}, {Genzel}, {Martins}, \& {Ott}}]{Gillessen:2009}
{Gillessen}, S., {Eisenhauer}, F., {Trippe}, S., {et~al.} 2009, \apj, 692,
  1075, \dodoi{10.1088/0004-637X/692/2/1075}

\bibitem[{{Gnedin} \& {Ostriker}(1997)}]{Gnedin:1997}
{Gnedin}, O.~Y., \& {Ostriker}, J.~P. 1997, \apj, 474, 223,
  \dodoi{10.1086/303441}

\bibitem[{{G{\'o}rski} {et~al.}(2005){G{\'o}rski}, {Hivon}, {Banday},
  {Wandelt}, {Hansen}, {Reinecke}, \& {Bartelmann}}]{Gorski:2005}
{G{\'o}rski}, K.~M., {Hivon}, E., {Banday}, A.~J., {et~al.} 2005, \apj, 622,
  759, \dodoi{10.1086/427976}

\bibitem[{{Grillmair}(2006)}]{Grillmair:2006}
{Grillmair}, C.~J. 2006, \apjl, 645, L37, \dodoi{10.1086/505863}

\bibitem[{{Grillmair}(2014)}]{Grillmair:2014}
---. 2014, \apjl, 790, L10, \dodoi{10.1088/2041-8205/790/1/L10}

\bibitem[{{Grillmair}(2017)}]{Grillmair:2017}
---. 2017, \apj, 847, 119, \dodoi{10.3847/1538-4357/aa8872}

\bibitem[{{Grillmair} \& {Carlberg}(2016)}]{Grillmair:2016b}
{Grillmair}, C.~J., \& {Carlberg}, R.~G. 2016, \apjl, 820, L27,
  \dodoi{10.3847/2041-8205/820/2/L27}

\bibitem[{{Grillmair} {et~al.}(1995){Grillmair}, {Freeman}, {Irwin}, \&
  {Quinn}}]{Grillmair:1995}
{Grillmair}, C.~J., {Freeman}, K.~C., {Irwin}, M., \& {Quinn}, P.~J. 1995, \aj,
  109, 2553, \dodoi{10.1086/117470}

\bibitem[{{Grillmair} {et~al.}(2004){Grillmair}, {Jarrett}, \&
  {Ha}}]{Grillmair:2004}
{Grillmair}, C.~J., {Jarrett}, T.~H., \& {Ha}, A.~C. 2004, in Astronomical
  Society of the Pacific Conference Series, Vol. 327, Satellites and Tidal
  Streams, ed. F.~{Prada}, D.~{Martinez Delgado}, \& T.~J. {Mahoney}, 276

\bibitem[{{Hamilton} \& {Tegmark}(2004)}]{Hamilton:2004}
{Hamilton}, A.~J.~S., \& {Tegmark}, M. 2004, \mnras, 349, 115,
  \dodoi{10.1111/j.1365-2966.2004.07490.x}

\bibitem[{{Harris}(1996)}]{Harris:1996}
{Harris}, W.~E. 1996, \aj, 112, 1487, \dodoi{10.1086/118116}

\bibitem[{{Hattori} {et~al.}(2013){Hattori}, {Yoshii}, {Beers}, {Carollo}, \&
  {Lee}}]{Hattori:2013}
{Hattori}, K., {Yoshii}, Y., {Beers}, T.~C., {Carollo}, D., \& {Lee}, Y.~S.
  2013, \apjl, 763, L17, \dodoi{10.1088/2041-8205/763/1/L17}

\bibitem[{Hunter(2007)}]{Hunter:2007}
Hunter, J.~D. 2007, Computing In Science \& Engineering, 9, 90,
  \dodoi{10.1109/MCSE.2007.55}

\bibitem[{{Ibata} {et~al.}(2002){Ibata}, {Lewis}, {Irwin}, \&
  {Quinn}}]{Ibata:2002}
{Ibata}, R.~A., {Lewis}, G.~F., {Irwin}, M.~J., \& {Quinn}, T. 2002, \mnras,
  332, 915, \dodoi{10.1046/j.1365-8711.2002.05358.x}

\bibitem[{{Jarvis} {et~al.}(2016){Jarvis}, {Sheldon}, {Zuntz}, {Kacprzak},
  {Bridle}, {et~al.}}]{Jarvis:2015}
{Jarvis}, M., {Sheldon}, E., {Zuntz}, J., {et~al.} 2016, \mnras, 460, 2245,
  \dodoi{10.1093/mnras/stw990}

\bibitem[{{Jethwa} {et~al.}(2017){Jethwa}, {Torrealba}, {Navarrete},
  {et~al.}}]{Jethwa:2017}
{Jethwa}, P., {Torrealba}, G., {Navarrete}, C., {et~al.} 2017, submitted to
  MNRAS.
\newblock \doarXiv{1711.09103}

\bibitem[{{Johnston}(1998)}]{Johnston:1998}
{Johnston}, K.~V. 1998, \apj, 495, 297, \dodoi{10.1086/305273}

\bibitem[{{Johnston} {et~al.}(2005){Johnston}, {Law}, \&
  {Majewski}}]{Johnston:2005}
{Johnston}, K.~V., {Law}, D.~R., \& {Majewski}, S.~R. 2005, \apj, 619, 800,
  \dodoi{10.1086/426777}

\bibitem[{{Johnston} {et~al.}(2002){Johnston}, {Spergel}, \&
  {Haydn}}]{Johnston:2002}
{Johnston}, K.~V., {Spergel}, D.~N., \& {Haydn}, C. 2002, \apj, 570, 656,
  \dodoi{10.1086/339791}

\bibitem[{Jones {et~al.}(2001)Jones, Oliphant, Peterson, {et~al.}}]{scipy:2001}
Jones, E., Oliphant, T., Peterson, P., {et~al.} 2001, {SciPy}: Open source
  scientific tools for {Python}.
\newblock \url{http://www.scipy.org/}

\bibitem[{{Juri{\'c}} {et~al.}(2008){Juri{\'c}}, {Ivezi{\'c}}, {Brooks},
  {Lupton}, {Schlegel}, {Finkbeiner}, {Padmanabhan}, {Bond}, {Sesar},
  {Rockosi}, {Knapp}, {Gunn}, {Sumi}, {Schneider}, {Barentine}, {Brewington},
  {Brinkmann}, {Fukugita}, {Harvanek}, {Kleinman}, {Krzesinski}, {Long},
  {Neilsen}, {Nitta}, {Snedden}, \& {York}}]{Juric:2008}
{Juri{\'c}}, M., {Ivezi{\'c}}, {\v Z}., {Brooks}, A., {et~al.} 2008, \apj, 673,
  864, \dodoi{10.1086/523619}

\bibitem[{{Kafle} {et~al.}(2013){Kafle}, {Sharma}, {Lewis}, \&
  {Bland-Hawthorn}}]{Kafle:2013}
{Kafle}, P.~R., {Sharma}, S., {Lewis}, G.~F., \& {Bland-Hawthorn}, J. 2013,
  \mnras, 430, 2973, \dodoi{10.1093/mnras/stt101}

\bibitem[{{Kim} \& {Jerjen}(2015)}]{Kim:2015}
{Kim}, D., \& {Jerjen}, H. 2015, \apjl, 808, L39,
  \dodoi{10.1088/2041-8205/808/2/L39}

\bibitem[{{Kirby} {et~al.}(2013){Kirby}, {Cohen}, {Guhathakurta}, {Cheng},
  {Bullock}, \& {Gallazzi}}]{Kirby:2013}
{Kirby}, E.~N., {Cohen}, J.~G., {Guhathakurta}, P., {et~al.} 2013, \apj, 779,
  102, \dodoi{10.1088/0004-637X/779/2/102}

\bibitem[{{Koposov} {et~al.}(2015){Koposov}, {Belokurov}, {Torrealba}, \&
  {Evans}}]{Koposov:2015}
{Koposov}, S., {Belokurov}, V., {Torrealba}, G., \& {Evans}, W. 2015, IAU
  General Assembly, 22, 2256759

\bibitem[{{Koposov} {et~al.}(2014){Koposov}, {Irwin}, {Belokurov},
  {Gonzalez-Solares}, {Yoldas}, {Lewis}, {Metcalfe}, \&
  {Shanks}}]{Koposov:2014}
{Koposov}, S.~E., {Irwin}, M., {Belokurov}, V., {et~al.} 2014, \mnras, 442,
  L85, \dodoi{10.1093/mnrasl/slu060}

\bibitem[{{Koposov} {et~al.}(2010){Koposov}, {Rix}, \& {Hogg}}]{Koposov:2010}
{Koposov}, S.~E., {Rix}, H.-W., \& {Hogg}, D.~W. 2010, \apj, 712, 260,
  \dodoi{10.1088/0004-637X/712/1/260}

\bibitem[{{Koposov} {et~al.}(2012){Koposov}, {Belokurov}, {Evans}, {Gilmore},
  {Gieles}, {Irwin}, {Lewis}, {Niederste-Ostholt}, {Pe{\~n}arrubia}, {Smith},
  {Bizyaev}, {Malanushenko}, {Malanushenko}, {Schneider}, \&
  {Wyse}}]{Koposov:2012}
{Koposov}, S.~E., {Belokurov}, V., {Evans}, N.~W., {et~al.} 2012, \apj, 750,
  80, \dodoi{10.1088/0004-637X/750/1/80}

\bibitem[{{K{\"u}pper} {et~al.}(2015{\natexlab{a}}){K{\"u}pper}, {Balbinot},
  {Bonaca}, {Johnston}, {Hogg}, {Kroupa}, \& {Santiago}}]{Kuepper:2015}
{K{\"u}pper}, A.~H.~W., {Balbinot}, E., {Bonaca}, A., {et~al.}
  2015{\natexlab{a}}, \apj, 803, 80, \dodoi{10.1088/0004-637X/803/2/80}

\bibitem[{{K{\"u}pper} {et~al.}(2015{\natexlab{b}}){K{\"u}pper}, {Balbinot},
  {Bonaca}, {et~al.}}]{Kupper:2015}
---. 2015{\natexlab{b}}, \apj, 803, 80, \dodoi{10.1088/0004-637X/803/2/80}

\bibitem[{{K{\"u}pper} {et~al.}(2012){K{\"u}pper}, {Lane}, \&
  {Heggie}}]{Kuepper:2012}
{K{\"u}pper}, A.~H.~W., {Lane}, R.~R., \& {Heggie}, D.~C. 2012, \mnras, 420,
  2700, \dodoi{10.1111/j.1365-2966.2011.20242.x}

\bibitem[{{Kuzma} {et~al.}(2018){Kuzma}, {Da Costa}, \& {Mackey}}]{Kuzma:2018}
{Kuzma}, P.~B., {Da Costa}, G.~S., \& {Mackey}, A.~D. 2018, \mnras, 473, 2881,
  \dodoi{10.1093/mnras/stx2353}

\bibitem[{{Law} {et~al.}(2005){Law}, {Johnston}, \& {Majewski}}]{Law:2005}
{Law}, D.~R., {Johnston}, K.~V., \& {Majewski}, S.~R. 2005, \apj, 619, 807,
  \dodoi{10.1086/426779}

\bibitem[{{Law} \& {Majewski}(2010)}]{Law:2010}
{Law}, D.~R., \& {Majewski}, S.~R. 2010, \apj, 714, 229,
  \dodoi{10.1088/0004-637X/714/1/229}

\bibitem[{{Leon} {et~al.}(2000){Leon}, {Meylan}, \& {Combes}}]{Leon:2000}
{Leon}, S., {Meylan}, G., \& {Combes}, F. 2000, \aap, 359, 907

\bibitem[{{Li} {et~al.}(2016){Li}, {Balbinot}, {Mondrik}, {Marshall}, {Yanny},
  {Bechtol}, {Drlica-Wagner}, {Oscar}, {Santiago}, {Simon}, {Vivas}, {Walker},
  {Wang}, {Abbott}, {Abdalla}, {Benoit-L{\'e}vy}, {Bernstein}, {Bertin},
  {Brooks}, {Burke}, {Carnero Rosell}, {Carrasco Kind}, {Carretero}, {da
  Costa}, {DePoy}, {Desai}, {Diehl}, {Doel}, {Estrada}, {Finley}, {Flaugher},
  {Frieman}, {Gruen}, {Gruendl}, {Gutierrez}, {Honscheid}, {James}, {Kuehn},
  {Kuropatkin}, {Lahav}, {Maia}, {March}, {Martini}, {Ogando}, {Plazas},
  {Reil}, {Romer}, {Roodman}, {Sanchez}, {Scarpine}, {Schubnell},
  {Sevilla-Noarbe}, {Smith}, {Soares-Santos}, {Sobreira}, {Suchyta}, {Swanson},
  {Tarle}, {Tucker}, {Zhang}, \& {DES Collaboration}}]{Li:2016}
{Li}, T.~S., {Balbinot}, E., {Mondrik}, N., {et~al.} 2016, \apj, 817, 135,
  \dodoi{10.3847/0004-637X/817/2/135}

\bibitem[{{Li} {et~al.}(2018)}]{Li:2018}
{Li}, T.~S., {et~al.} 2018, in prep.

\bibitem[{{LSST Science Collaboration}(2009)}]{LSST:2009}
{LSST Science Collaboration}. 2009, ArXiv e-prints.
\newblock \doarXiv{0912.0201}

\bibitem[{{Luque} {et~al.}(2017{\natexlab{a}}){Luque}, {Pieres}, {Santiago},
  {Yanny}, {Vivas}, {et~al.}}]{Luque:2017}
{Luque}, E., {Pieres}, A., {Santiago}, B., {et~al.} 2017{\natexlab{a}}, \mnras,
  468, 97, \dodoi{10.1093/mnras/stx405}

\bibitem[{{Luque} {et~al.}(2016){Luque}, {Queiroz}, {Santiago}, {Pieres},
  {Balbinot}, {et~al.}}]{Luque:2016}
{Luque}, E., {Queiroz}, A., {Santiago}, B., {et~al.} 2016, \mnras, 458, 603,
  \dodoi{10.1093/mnras/stw302}

\bibitem[{{Luque} {et~al.}(2017{\natexlab{b}}){Luque}, {Santiago}, {Pieres},
  {Marshall}, {Pace}, {et~al.}}]{Luque:2017b}
{Luque}, E., {Santiago}, B., {Pieres}, A., {et~al.} 2017{\natexlab{b}}, ArXiv
  e-prints.
\newblock \doarXiv{1709.05689}

\bibitem[{{Majewski} {et~al.}(2003){Majewski}, {Skrutskie}, {Weinberg}, \&
  {Ostheimer}}]{Majewski:2003}
{Majewski}, S.~R., {Skrutskie}, M.~F., {Weinberg}, M.~D., \& {Ostheimer}, J.~C.
  2003, \apj, 599, 1082, \dodoi{10.1086/379504}

\bibitem[{{Majewski} {et~al.}(2004){Majewski}, {Kunkel}, {Law}, {Patterson},
  {Polak}, {Rocha-Pinto}, {Crane}, {Frinchaboy}, {Hummels}, {Johnston}, {Rhee},
  {Skrutskie}, \& {Weinberg}}]{Majewski:2004}
{Majewski}, S.~R., {Kunkel}, W.~E., {Law}, D.~R., {et~al.} 2004, \aj, 128, 245,
  \dodoi{10.1086/421372}

\bibitem[{{Mateu} {et~al.}(2017){Mateu}, {Read}, \& {Kawata}}]{Mateu:2017}
{Mateu}, C., {Read}, J.~I., \& {Kawata}, D. 2017, ArXiv e-prints.
\newblock \doarXiv{1711.03967}

\bibitem[{{McMillan}(2017)}]{McMillan:2017}
{McMillan}, P.~J. 2017, \mnras, 465, 76, \dodoi{10.1093/mnras/stw2759}

\bibitem[{{Morganson} {et~al.}(2018)}]{Morganson:2018}
{Morganson}, E., {et~al.} 2018, {\rm in prep.}

\bibitem[{{Newberg} \& {Carlin}(2016)}]{Newberg:2016}
{Newberg}, H.~J., \& {Carlin}, J.~L., eds. 2016, Astrophysics and Space Science
  Library, Vol. 420, {Tidal Streams in the Local Group and Beyond}

\bibitem[{{Newberg} {et~al.}(2002){Newberg}, {Yanny}, {Rockosi}, {Grebel},
  {Rix}, {Brinkmann}, {Csabai}, {Hennessy}, {Hindsley}, {Ibata}, {Ivezi{\'c}},
  {Lamb}, {Nash}, {Odenkirchen}, {Rave}, {Schneider}, {Smith}, {Stolte}, \&
  {York}}]{Newberg:2002}
{Newberg}, H.~J., {Yanny}, B., {Rockosi}, C., {et~al.} 2002, \apj, 569, 245,
  \dodoi{10.1086/338983}

\bibitem[{{Ngan} \& {Carlberg}(2014)}]{Ngan:2014}
{Ngan}, W.~H.~W., \& {Carlberg}, R.~G. 2014, \apj, 788, 181,
  \dodoi{10.1088/0004-637X/788/2/181}

\bibitem[{{Nidever} {et~al.}(2008){Nidever}, {Majewski}, \& {Butler
  Burton}}]{Nidever:2008}
{Nidever}, D.~L., {Majewski}, S.~R., \& {Butler Burton}, W. 2008, \apj, 679,
  432, \dodoi{10.1086/587042}

\bibitem[{{Nidever} {et~al.}(2017){Nidever}, {Olsen}, {Walker}, {Vivas},
  {Blum}, {Kaleida}, {Choi}, {Conn}, {Gruendl}, {Bell}, {Besla}, {Mu{\~n}oz},
  {Gallart}, {Martin}, {Olszewski}, {Saha}, {Monachesi}, {Monelli}, {de Boer},
  {Johnson}, {Zaritsky}, {Stringfellow}, {van der Marel}, {Cioni}, {Jin},
  {Majewski}, {Martinez-Delgado}, {Monteagudo}, {No{\"e}l}, {Bernard},
  {Kunder}, {Chu}, {Bell}, {Santana}, {Frechem}, {Medina}, {Parkash},
  {Ser{\'o}n Navarrete}, \& {Hayes}}]{Nidever:2017}
{Nidever}, D.~L., {Olsen}, K., {Walker}, A.~R., {et~al.} 2017, \aj, 154, 199,
  \dodoi{10.3847/1538-3881/aa8d1c}

\bibitem[{{Odenkirchen} {et~al.}(2001){Odenkirchen}, {Grebel}, {Rockosi},
  {Dehnen}, {Ibata}, {Rix}, {Stolte}, {Wolf}, {Anderson}, {Bahcall},
  {Brinkmann}, {Csabai}, {Hennessy}, {Hindsley}, {Ivezi{\'c}}, {Lupton},
  {Munn}, {Pier}, {Stoughton}, \& {York}}]{Odenkirchen:2001}
{Odenkirchen}, M., {Grebel}, E.~K., {Rockosi}, C.~M., {et~al.} 2001, \apjl,
  548, L165, \dodoi{10.1086/319095}

\bibitem[{{Olszewski} {et~al.}(2009){Olszewski}, {Saha}, {Knezek},
  {Subramaniam}, {de Boer}, \& {Seitzer}}]{Olszewski:2009}
{Olszewski}, E.~W., {Saha}, A., {Knezek}, P., {et~al.} 2009, \aj, 138, 1570,
  \dodoi{10.1088/0004-6256/138/6/1570}

\bibitem[{{Pawlowski} {et~al.}(2015){Pawlowski}, {McGaugh}, \&
  {Jerjen}}]{Pawlowski:2015}
{Pawlowski}, M.~S., {McGaugh}, S.~S., \& {Jerjen}, H. 2015, \mnras, 453, 1047,
  \dodoi{10.1093/mnras/stv1588}

\bibitem[{{Pawlowski} {et~al.}(2012){Pawlowski}, {Pflamm-Altenburg}, \&
  {Kroupa}}]{Pawlowski:2012}
{Pawlowski}, M.~S., {Pflamm-Altenburg}, J., \& {Kroupa}, P. 2012, \mnras, 423,
  1109, \dodoi{10.1111/j.1365-2966.2012.20937.x}

\bibitem[{{Peebles}(1965)}]{Peebles:1965}
{Peebles}, P.~J.~E. 1965, \apj, 142, 1317, \dodoi{10.1086/148417}

\bibitem[{{Piatti}(2018)}]{Piatti:2018}
{Piatti}, A.~E. 2018, \mnras, 473, 492, \dodoi{10.1093/mnras/stx2471}

\bibitem[{{Pieres} {et~al.}(2016){Pieres}, {Santiago}, {Balbinot}, {Luque},
  {Queiroz}, {et~al.}}]{Pieres:2016}
{Pieres}, A., {Santiago}, B., {Balbinot}, E., {et~al.} 2016, \mnras, 461, 519,
  \dodoi{10.1093/mnras/stw1260}

\bibitem[{{Pieres} {et~al.}(2017){Pieres}, {Santiago}, {Drlica-Wagner},
  {Bechtol}, {Marel}, {et~al.}}]{Pieres:2017}
{Pieres}, A., {Santiago}, B.~X., {Drlica-Wagner}, A., {et~al.} 2017, \mnras,
  468, 1349, \dodoi{10.1093/mnras/stx507}

\bibitem[{{Press} \& {Schechter}(1974)}]{Press:1974}
{Press}, W.~H., \& {Schechter}, P. 1974, \apj, 187, 425, \dodoi{10.1086/152650}

\bibitem[{{Price-Whelan} {et~al.}(2014){Price-Whelan}, {Hogg}, {Johnston}, \&
  {Hendel}}]{Price-Whelan:2014}
{Price-Whelan}, A.~M., {Hogg}, D.~W., {Johnston}, K.~V., \& {Hendel}, D. 2014,
  \apj, 794, 4, \dodoi{10.1088/0004-637X/794/1/4}

\bibitem[{{Rocha-Pinto} {et~al.}(2004){Rocha-Pinto}, {Majewski}, {Skrutskie},
  {Crane}, \& {Patterson}}]{Rocha-Pinto:2004}
{Rocha-Pinto}, H.~J., {Majewski}, S.~R., {Skrutskie}, M.~F., {Crane}, J.~D., \&
  {Patterson}, R.~J. 2004, \apj, 615, 732, \dodoi{10.1086/424585}

\bibitem[{{Rockosi} {et~al.}(2002){Rockosi}, {Odenkirchen}, {Grebel}, {Dehnen},
  {Cudworth}, {Gunn}, {York}, {Brinkmann}, {Hennessy}, \&
  {Ivezi{\'c}}}]{Rockosi:2002}
{Rockosi}, C.~M., {Odenkirchen}, M., {Grebel}, E.~K., {et~al.} 2002, \aj, 124,
  349, \dodoi{10.1086/340957}

\bibitem[{{Rozo} {et~al.}(2016){Rozo}, {Rykoff}, {Abate}, {Bonnett}, {Crocce},
  {et~al.}}]{Rozo:2016}
{Rozo}, E., {Rykoff}, E.~S., {Abate}, A., {et~al.} 2016, \mnras, 461, 1431,
  \dodoi{10.1093/mnras/stw1281}

\bibitem[{{Sanders}(2014)}]{Sanders:2014}
{Sanders}, J.~L. 2014, \mnras, 443, 423, \dodoi{10.1093/mnras/stu1159}

\bibitem[{{Sanders} {et~al.}(2016){Sanders}, {Bovy}, \& {Erkal}}]{Sanders:2016}
{Sanders}, J.~L., {Bovy}, J., \& {Erkal}, D. 2016, \mnras, 457, 3817,
  \dodoi{10.1093/mnras/stw232}

\bibitem[{{Sanderson} {et~al.}(2016){Sanderson}, {Vera-Ciro}, {Helmi}, \&
  {Heit}}]{Sanderson:2016}
{Sanderson}, R.~E., {Vera-Ciro}, C., {Helmi}, A., \& {Heit}, J. 2016, ArXiv
  e-prints.
\newblock \doarXiv{1608.05624}

\bibitem[{{Sandford} {et~al.}(2017){Sandford}, {K{\"u}pper}, {Johnston}, \&
  {Diemand}}]{Sandford:2017}
{Sandford}, E., {K{\"u}pper}, A.~H.~W., {Johnston}, K.~V., \& {Diemand}, J.
  2017, \mnras, 470, 522, \dodoi{10.1093/mnras/stx1268}

\bibitem[{{Schlafly} \& {Finkbeiner}(2011)}]{Schlafly:2011}
{Schlafly}, E.~F., \& {Finkbeiner}, D.~P. 2011, \apj, 737, 103,
  \dodoi{10.1088/0004-637X/737/2/103}

\bibitem[{{Schlegel} {et~al.}(1998){Schlegel}, {Finkbeiner}, \&
  {Davis}}]{Schlegel:1998}
{Schlegel}, D.~J., {Finkbeiner}, D.~P., \& {Davis}, M. 1998, \apj, 500, 525,
  \dodoi{10.1086/305772}

\bibitem[{{Sch{\"o}nrich} {et~al.}(2010){Sch{\"o}nrich}, {Binney}, \&
  {Dehnen}}]{Schonrich:2010}
{Sch{\"o}nrich}, R., {Binney}, J., \& {Dehnen}, W. 2010, \mnras, 403, 1829,
  \dodoi{10.1111/j.1365-2966.2010.16253.x}

\bibitem[{{Sheldon}(2014)}]{Sheldon:2014}
{Sheldon}, E.~S. 2014, \mnras, 444, L25, \dodoi{10.1093/mnrasl/slu104}

\bibitem[{{Simon} {et~al.}(2017){Simon}, {Li}, {Drlica-Wagner}, {Bechtol},
  {Marshall}, {James}, {Wang}, {Strigari}, {Balbinot}, {Kuehn}, {Walker},
  {Abbott}, {Allam}, {Annis}, {Benoit-L{\'e}vy}, {Brooks}, {Buckley-Geer},
  {Burke}, {Carnero Rosell}, {Carrasco Kind}, {Carretero}, {Cunha}, {D'Andrea},
  {da Costa}, {DePoy}, {Desai}, {Doel}, {Fernandez}, {Flaugher}, {Frieman},
  {Garc{\'{\i}}a-Bellido}, {Gaztanaga}, {Goldstein}, {Gruen}, {Gutierrez},
  {Kuropatkin}, {Maia}, {Martini}, {Menanteau}, {Miller}, {Miquel}, {Neilsen},
  {Nord}, {Ogando}, {Plazas}, {Romer}, {Rykoff}, {Sanchez}, {Santiago},
  {Scarpine}, {Schubnell}, {Sevilla-Noarbe}, {Smith}, {Sobreira}, {Suchyta},
  {Swanson}, {Tarle}, {Whiteway}, {Yanny}, \& {DES Collaboration}}]{Simon:2017}
{Simon}, J.~D., {Li}, T.~S., {Drlica-Wagner}, A., {et~al.} 2017, \apj, 838, 11,
  \dodoi{10.3847/1538-4357/aa5be7}

\bibitem[{{Springel} {et~al.}(2008){Springel}, {Wang}, {Vogelsberger},
  {Ludlow}, {Jenkins}, {Helmi}, {Navarro}, {Frenk}, \& {White}}]{Springel:2008}
{Springel}, V., {Wang}, J., {Vogelsberger}, M., {et~al.} 2008, \mnras, 391,
  1685, \dodoi{10.1111/j.1365-2966.2008.14066.x}

\bibitem[{{Swanson} {et~al.}(2008){Swanson}, {Tegmark}, {Hamilton}, \&
  {Hill}}]{Swanson:2008}
{Swanson}, M.~E.~C., {Tegmark}, M., {Hamilton}, A.~J.~S., \& {Hill}, J.~C.
  2008, \mnras, 387, 1391, \dodoi{10.1111/j.1365-2966.2008.13296.x}

\bibitem[{{Van Der Walt} {et~al.}(2011){Van Der Walt}, {Colbert}, \&
  {Varoquaux}}]{numpy:2011}
{Van Der Walt}, S., {Colbert}, S.~C., \& {Varoquaux}, G. 2011, Computing in
  Science \& Engineering, 13, 22, \dodoi{10.1109/MCSE.2011.37}

\bibitem[{{Walker} {et~al.}(2011){Walker}, {Kunder}, {Andreuzzi},
  {et~al.}}]{Walker:2011}
{Walker}, A.~R., {Kunder}, A.~M., {Andreuzzi}, G., {et~al.} 2011, \mnras, 415,
  643, \dodoi{10.1111/j.1365-2966.2011.18736.x}

\bibitem[{{Willman} {et~al.}(2005{\natexlab{a}}){Willman}, {Blanton}, {West},
  {Dalcanton}, {Hogg}, {Schneider}, {Wherry}, {Yanny}, \&
  {Brinkmann}}]{Willman:2005a}
{Willman}, B., {Blanton}, M.~R., {West}, A.~A., {et~al.} 2005{\natexlab{a}},
  \aj, 129, 2692, \dodoi{10.1086/430214}

\bibitem[{{Willman} {et~al.}(2005{\natexlab{b}}){Willman}, {Dalcanton},
  {Martinez-Delgado}, {West}, {Blanton}, {Hogg}, {Barentine}, {Brewington},
  {Harvanek}, {Kleinman}, {Krzesinski}, {Long}, {Neilsen}, {Nitta}, \&
  {Snedden}}]{Willman:2005b}
{Willman}, B., {Dalcanton}, J.~J., {Martinez-Delgado}, D., {et~al.}
  2005{\natexlab{b}}, \apjl, 626, L85, \dodoi{10.1086/431760}

\bibitem[{{Yanny} {et~al.}(2003){Yanny}, {Newberg}, {Grebel}, {Kent},
  {Odenkirchen}, {Rockosi}, {Schlegel}, {Subbarao}, {Brinkmann}, {Fukugita},
  {Ivezic}, {Lamb}, {Schneider}, \& {York}}]{Yanny:2003}
{Yanny}, B., {Newberg}, H.~J., {Grebel}, E.~K., {et~al.} 2003, \apj, 588, 824,
  \dodoi{10.1086/374220}

\bibitem[{{Yoon} {et~al.}(2011){Yoon}, {Johnston}, \& {Hogg}}]{Yoon:2011}
{Yoon}, J.~H., {Johnston}, K.~V., \& {Hogg}, D.~W. 2011, \apj, 731, 58,
  \dodoi{10.1088/0004-637X/731/1/58}

\bibitem[{{York} {et~al.}(2000){York}, {Adelman}, {Anderson}, {Anderson},
  {Annis}, {Bahcall}, {Bakken}, {Barkhouser}, {Bastian}, {Berman}, {Boroski},
  {Bracker}, {Briegel}, {Briggs}, {Brinkmann}, {Brunner}, {Burles}, {Carey},
  {Carr}, {Castander}, {Chen}, {Colestock}, {Connolly}, {Crocker}, {Csabai},
  {Czarapata}, {Davis}, {Doi}, {Dombeck}, {Eisenstein}, {Ellman}, {Elms},
  {Evans}, {Fan}, {Federwitz}, {Fiscelli}, {Friedman}, {Frieman}, {Fukugita},
  {Gillespie}, {Gunn}, {Gurbani}, {de Haas}, {Haldeman}, {Harris}, {Hayes},
  {Heckman}, {Hennessy}, {Hindsley}, {Holm}, {Holmgren}, {Huang}, {Hull},
  {Husby}, {Ichikawa}, {Ichikawa}, {Ivezi{\'c}}, {Kent}, {Kim}, {Kinney},
  {Klaene}, {Kleinman}, {Kleinman}, {Knapp}, {Korienek}, {Kron}, {Kunszt},
  {Lamb}, {Lee}, {Leger}, {Limmongkol}, {Lindenmeyer}, {Long}, {Loomis},
  {Loveday}, {Lucinio}, {Lupton}, {MacKinnon}, {Mannery}, {Mantsch}, {Margon},
  {McGehee}, {McKay}, {Meiksin}, {Merelli}, {Monet}, {Munn}, {Narayanan},
  {Nash}, {Neilsen}, {Neswold}, {Newberg}, {Nichol}, {Nicinski}, {Nonino},
  {Okada}, {Okamura}, {Ostriker}, {Owen}, {Pauls}, {Peoples}, {Peterson},
  {Petravick}, {Pier}, {Pope}, {Pordes}, {Prosapio}, {Rechenmacher}, {Quinn},
  {Richards}, {Richmond}, {Rivetta}, {Rockosi}, {Ruthmansdorfer}, {Sandford},
  {Schlegel}, {Schneider}, {Sekiguchi}, {Sergey}, {Shimasaku}, {Siegmund},
  {Smee}, {Smith}, {Snedden}, {Stone}, {Stoughton}, {Strauss}, {Stubbs},
  {SubbaRao}, {Szalay}, {Szapudi}, {Szokoly}, {Thakar}, {Tremonti}, {Tucker},
  {Uomoto}, {Vanden Berk}, {Vogeley}, {Waddell}, {Wang}, {Watanabe},
  {Weinberg}, {Yanny}, {Yasuda}, \& {SDSS Collaboration}}]{York:2000}
{York}, D.~G., {Adelman}, J., {Anderson}, Jr., J.~E., {et~al.} 2000, \aj, 120,
  1579, \dodoi{10.1086/301513}

\bibitem[{{Zucker} {et~al.}(2006{\natexlab{a}}){Zucker}, {Belokurov}, {Evans},
  {Kleyna}, {Irwin}, {Wilkinson}, {Fellhauer}, {Bramich}, {Gilmore}, {Newberg},
  {Yanny}, {Smith}, {Hewett}, {Bell}, {Rix}, {Gnedin}, {Vidrih}, {Wyse},
  {Willman}, {Grebel}, {Schneider}, {Beers}, {Kniazev}, {Barentine},
  {Brewington}, {Brinkmann}, {Harvanek}, {Kleinman}, {Krzesinski}, {Long},
  {Nitta}, \& {Snedden}}]{Zucker:2006}
{Zucker}, D.~B., {Belokurov}, V., {Evans}, N.~W., {et~al.} 2006{\natexlab{a}},
  \apjl, 650, L41, \dodoi{10.1086/508628}

\bibitem[{{Zucker} {et~al.}(2006{\natexlab{b}}){Zucker}, {Belokurov}, {Evans},
  {Wilkinson}, {Irwin}, {Sivarani}, {Hodgkin}, {Bramich}, {Irwin}, {Gilmore},
  {Willman}, {Vidrih}, {Fellhauer}, {Hewett}, {Beers}, {Bell}, {Grebel},
  {Schneider}, {Newberg}, {Wyse}, {Rockosi}, {Yanny}, {Lupton}, {Smith},
  {Barentine}, {Brewington}, {Brinkmann}, {Harvanek}, {Kleinman}, {Krzesinski},
  {Long}, {Nitta}, \& {Snedden}}]{2006ApJ...643L.103Z}
---. 2006{\natexlab{b}}, \apjl, 643, L103, \dodoi{10.1086/505216}

\end{thebibliography}

\end{document}